\documentclass{kasper}
\usepackage{relsize}
\usepackage[bottom]{footmisc}
\usepackage{slashed}

\newcommand{\teight}[1]{t_8^{(#1)}}
\newcommand{\eom}[1]{{\cal E}(#1)}
\newcommand{\bareom}[1]{\bar{\cal E}(#1)}
\newcommand{\norder}[1]{:\!#1\!:}
\newcommand{\specialcolon}{\mathrel{\mathop:}}
\newcommand{\defas}{\specialcolon =}

\newdimen\tableauside\tableauside=.8ex   
\newdimen\tableaurule\tableaurule=.32pt   
\newdimen\tableaustep
\def\phantomhrule#1{\hbox{\vbox to0pt{\hrule height\tableaurule width#1\vss}}}
\def\phantomvrule#1{\vbox{\hbox to0pt{\vrule width\tableaurule height#1\hss}}}
\def\sqr{\vbox{%
  \phantomhrule\tableaustep
  \hbox{\phantomvrule\tableaustep\kern\tableaustep\phantomvrule\tableaustep}%
  \hbox{\vbox{\phantomhrule\tableauside}\kern-\tableaurule}}}
\def\squares#1{\hbox{\count0=#1\noindent\loop\sqr
  \advance\count0 by-1 \ifnum\count0>0\repeat}}
\def\tableau#1{\vcenter{\offinterlineskip
  \tableaustep=\tableauside\advance\tableaustep by-\tableaurule
  \kern\normallineskip\hbox
    {\kern\normallineskip\vbox
      {\gettableau#1 0 }%
     \kern\normallineskip\kern\tableaurule}%
  \kern\normallineskip\kern\tableaurule}}
\def\gettableau#1 {\ifnum#1=0\let\next=\null\else
  \squares{#1}\let\next=\gettableau\fi\next}

\numberwithin{equation}{section}

\begin{document}
\preprintnumber{DAMTP-2000-82\\ NORDITA-2000/72 HE\\ 
SPHT-T00/116\\ hep-th/0010167}
\title{Supersymmetric higher-derivative actions in ten and eleven dimensions,
the associated superalgebras and their formulation in superspace}
\author{Kasper Peeters$^{1}$, Pierre Vanhove$^{2}$ and Anders Westerberg$^{3}$}
\address{1}{CERN\\
TH-division\\
1211 Geneva 23\\
Switzerland}
\address{2}{SPT\\
Orme des Merisiers\\
CEA/Saclay\\
91191 Gif-sur-Yvette Cedex\\
France}
\address{3}{NORDITA \\
Blegdamsvej 17\\
DK-2100 Copenhagen \O\\
Denmark}
\date{October 19th, 2000}
\email{k.peeters, p.vanhove@damtp.cam.ac.uk, a.westerberg@nordita.dk}

\maketitle

\begin{abstract}
  Higher-derivative terms in the string and M-theory effective actions
  are strongly constrained by supersymmetry.  Using a mixture of
  techniques, involving both string-amplitude calculations and an
  analysis of supersymmetry requirements, we determine the
  supersymmetric completion of the $R^4$ action in eleven dimensions
  to second order in the fermions, in a form compact enough for
  explicit further calculations.  Using these results, we obtain the
  modifications to the field transformation rules and determine the
  resulting field-dependent modifications to the coefficients in the
  supersymmetry algebra.  We then make the link to the superspace
  formulation of the theory and discuss the mechanism by which
  higher-derivative interactions lead to modifications to the
  supertorsion constraints. For the particular interactions under
  discussion we find that no such modifications are induced.
\end{abstract}
\bigskip

\vfill\eject\setcounter{page}{2}\pagestyle{plain}
\hrule
\tableofcontents
\bigskip\bigskip
\hrule

\begin{sectionunit}
\title{Introduction}
\maketitle

The effective field theory actions describing the dynamics of the
massless modes of the various string theories contain, in addition to the
well-known supergravity terms, an infinite number of higher-derivative
corrections. These terms encode the dynamics of the massive modes and
the extended character of strings. For the first few orders in the
string slope $\alpha'$ and the string coupling $g_s$, many of these
terms are known explicitly from either sigma-model anomaly
computations or from string amplitudes.  The best-known one
is perhaps the Green-Schwarz Lorentz anomaly cancellation
term~\cite{Green:1984sg} in the heterotic string effective action,
appearing at genus one:
\begin{equation}
\label{e:GS}
S_{\text{GS}} = \int\! B\wedge t_8 R^4\, .
\end{equation}
The $t_8$ tensor (whose precise definition will be recalled below) 
contracts the eight flat indices of the Riemann
tensors. As was first pointed out by \dcite{vafa7}, this term arises
also in the non-chiral type IIA theory, where a duality argument
relates it to the five-brane Lorentz anomaly (see \dcite{duff4}).
Another higher-derivative term which has received considerable
attention is the famous
\begin{equation}
\label{e:R4}
S_{R^4} = \int\!{\rm d}^{10}x\,e\, t_8 t_8 R^4\, .
\end{equation}
This term arises in all string theories, and not only at one-loop
order but also at tree level, when multiplied with the appropriate 
power of the dilaton\footnote{The results of \dcite{Bern:1998ug} and
\dcite{Dunbar:1999nj} show that it does not get corrections beyond one
loop. There are, however, non-perturbative contributions weighted
with an appropriate power of the dilaton.}.  Upon
compactification to four dimensions, the action~\eqn{e:R4} is
responsible for corrections to the hypermultiplet geometry; see for
instance \dcite{Strominger:1998eb} and \dcite{Antoniadis:1997eg}.
Moreover, both \eqn{e:GS} and \eqn{e:R4} contribute to the equations
of motion and are therefore expected to modify the supergravity
brane solutions.  In addition, several other explicit higher-derivative
terms are known or conjectured to exist by duality arguments (see
e.g.~\dcite{Berkovits:1998ex}).

Although one can in principle determine higher-derivative terms
through explicit string scattering calculations, it is expected that
their structure is strongly constrained by the symmetries of string
theory. In particular, suitable linear combinations of the bosonic
terms discussed above should be part of a full supersymmetric action
when amended with quadratic and higher-order fermion terms.
Ultimately one may perhaps hope that for every string theory (i.e.\ 
for every string model with a given field content) and for every genus
(i.e.\ for every given power of the dilaton), supersymmetry will be
enough to fix the structure of the action at the corresponding order
of the string loop expansion. One can argue (see also
section~\ref{s:fieldredefinitions}) that supersymmetry does not mix
the different string loop orders.  However, in some cases additional
symmetries may be enough to bridge even the genus gap; an explicit
example is the SL(2,${\mathbb Z}$) symmetry of type IIB string theory, 
as shown by \dcite{Green:1998by}.  \medskip

Supersymmetry on the component level often provides us with an elegant
underlying explanation of the rather complicated form of
higher-derivative actions (the present paper will exhibit many examples 
of this fact). But if we are interested in world-volume actions for 
branes embedded in target-space supergravity backgrounds, it is necessary 
to also understand how one can describe such background theories in their 
\emph{superspace} formulations.  Indeed, supersymmetric brane actions 
are known only in formulations with manifest target-space supersymmetry. 
One expects, based on previous experience with string models, that there 
will be a close link between world-volume quantum effects and target-space
corrections, although such a link is obviously much more difficult to
establish given the complications that arise in quantising kappa-symmetric 
actions.

The relation between the two formalisms is easiest to understand when
one realises that superspace provides a geometric rationale for the
algebraic structure of the component-space theory. Therefore, one should 
focus on possible modifications to the supersymmetry algebra induced by
higher-derivative terms in the action. In a component language, one
finds that, when higher-derivative terms are included, one also has to
modify the supersymmetry transformation rules of the fields.  In other
words, supersymmetry for these actions means that the invariance of
the action is expressed by
\begin{equation}
\Big(\delta_0 + \sum_n (\alpha')^n \delta_n\Big)\,\Big( S_0 + \sum_{n}
(\alpha')^n S_n\Big)=0\, . 
\end{equation}
Due to these modifications, the field-dependent coefficients of the
supersymmetry algebra can pick up corrections as well. As a result, the
geometrical structure of superspace is modified, which is reflected in
modified supertorsion constraints. We will return to these issues in
more detail later; it suffices to say at this point that our ultimate
goal is to derive the modifications to the supersymmetry
transformation rules and use these to make contact with a manifestly
supersymmetric formulation of the theory.
\medskip

The task of constructing supersymmetric string effective actions using
supersymmetry as the only input is horrendously complicated. Already
in pure gauge theories without gravity such an approach is rather
difficult. The sub-leading terms were constructed by
\dcite{Metsaev:1987by} and later extended to the non-abelian case by
\dcite{Bergshoeff:1987jm}. Much less is known about the higher-derivative 
supergravity invariants. Explicit results are only known for the
\emph{heterotic} effective action, while there are some expectations
about the form of higher-derivative invariants for other theories
(based on string calculations) which, however, have so far defied a
supersymmetry analysis.

It turns out that the heterotic theory admits two distinct types of
higher-derivative superinvariants. The first type is related to the
shift of the Neveu-Schwarz three-form field strength by the Lorentz
Chern-Simons term as required by anomaly cancellation.  In order to
restore supersymmetry, which is broken by this mechanism, additional
terms are required at genus zero. The $\alpha' R^2$ terms were
first analysed by \dcite{Romans:1986xd}, but there are corrections at
any order in the string slope, of which the $(\alpha')^3 R^4$ ones
have been computed explicitly by \dcite{berg8}.  Of course, none of
these terms are expected to play a role in the other string theories
or in eleven dimensions.  The second set of superinvariants, which
\emph{are} known to be relevant for the other string theories as well
(by virtue of the sigma-model computations of \dcite{gris11} and
several string-amplitude calculations) has $(\alpha')^3 R^4$ terms at
the lowest order. These invariants have been studied in an impressive
paper by \dcite{dero3} and reported in more detail in the thesis by
\dcite{suel1}.  These authors were able to pin down the exact form (up
to and including fermion bilinears) of three actions, each of which is
separately invariant under supersymmetry within the limitations of
their analysis.

These limitations arise as follows. The supergravity theories in ten
dimensions involve many fields, so one often restricts to checking
supersymmetry only with respect to a subset of the transformation
rules.  Furthermore, the four-fermi terms of higher-derivative actions
are extremely difficult to compute and are therefore in practice
always ignored.  Because of these restrictions, the analysis of
\dcite{dero3} has provided us with three candidate building blocks for
invariants, for which invariance has been checked, but only for a
subset of the full transformation rules. In addition, they 
constructed a generalisation of the Yang-Mills invariant which
includes the couplings of lowest order in the gravity fields. Their
results can be transcribed in a form adapted to string theory, in
which case the bosonic terms of the gravitational invariants are
expressed as (the precise form of the index contractions is not
important at this point)
\begin{equation}
\label{e:basicinvariants}
\begin{aligned}
I_X     &= t_8t_8 R^4 + \tfrac{1}{2}\varepsilon_{10} t_8 B R^4 + {\cal O}(\alpha')\, ,\\[.5ex]
I_{Y_1} &= (t_8 + \tfrac{1}{2}\varepsilon_{10} B) (\text{tr} R^2)^2  + 
 4\,HR^2 DR + {\cal O}(\alpha')\, ,\\[.5ex]
I_{Y_2} &= (t_8 + \tfrac{1}{2}\varepsilon_{10} B)  \text{tr} R^4 
+  HR^2 DR + {\cal O}(\alpha')  \, ,\\[.5ex]
I_{Z}   &= -\varepsilon_{10}\varepsilon_{10} R^4 + 4\, \varepsilon_{10} t_8 B
R^4 + {\cal O}(\alpha')\, .
\end{aligned}
\end{equation}
(The first three invariants are related via the identity $I_X = 24\,
I_{Y_2}- 6\, I_{Y_1}$, and so they comprise only two linearly
invariant combinations.) The terms of higher order in the string slope
involve higher powers of the fields as well as derivatives thereof.
Only one particular linear combination of the above invariants is
completely independent of the anti-symmetric tensor field, namely $I_X
- \tfrac{1}{8}I_Z$.  Because of gauge invariance for the $B$-field,
this is the only invariant that can appear in string theory at
arbitrary loop order, and in particular at tree level.  \medskip

The fact that these particular combinations can be made supersymmetric
is, as expected, consistent with explicit calculations of the
heterotic string effective action. The bosonic parts of the tree-level
contributions have been evaluated by \dcite{cai1} and
\dcite{Gross:1987mw} (three- and four-point amplitudes) while the
bosonic one-loop terms were computed by \dcite{Sakai:1987bi},
\ddcite{Abe:1988ud}{Abe:1988cq} and \dcite{Ellis:1988dc} 
(four-point amplitudes) and \dcite{Lerche:1987sg} (five-point amplitudes
involving the Neveu-Schwarz tensor field). Ignoring normalisation factors,
the result reads
\begin{equation}
\label{e:heteroticinvariant}
{\cal L}_{\text{heterotic}}\Big|_{(\alpha')^3} = 
 e^{-2\phi^H} \big( I_{\text{BdR}} + I_X - \tfrac{1}{8} I_Z\big) 
+ \big(I_{FR} + I_X  \big)\, .
\end{equation}
The term $I_{\text{BdR}}$ is the $(\alpha')^3$ piece of the invariant
constructed by \dcite{berg8}, and $I_{FR}$ is the invariant of
\dcite{dero3} which describes the coupling to the Yang-Mills fields.
When the transformations of the dilaton and the tensor field are fully
taken into account, the linear combinations appearing above are
expected to be completely fixed by supersymmetry, although this has to
our knowledge never been checked. Because the relative normalisation
(suppressed above) between the tree-level and one-loop terms is known
to contain a transcendental $\zeta(3)$ factor, it is impossible to get
further constraints from supersymmetry that relate these two
contributions.

The heterotic invariants discussed above seem to have some relevance for 
the other theories as well. The absence of Yang-Mills fields makes these 
actions considerably simpler (when higher-rank form fields are excluded). 
The four-point amplitudes for the IIA theory were computed by
\ddcite{Green:1982xx}{Green:1982ya} and \dcite{Kiritsis:1997em};
we already commented on the five-point analysis by \dcite{vafa7}. The
result is that
\begin{equation}
\label{e:L_IIA}
{\cal L}_{\text{IIA}}\Big|_{(\alpha')^3} = e^{-2\phi^A} \big( I_X - \tfrac{1}{8} I_Z \big)
+ \big( I_X + \tfrac{1}{8} I_Z \big)\, .
\end{equation}
Alternatively, the sigma-model results of \dcite{gris11} can be used to 
determine this action.  The type IIB theory~\cite{schw2,Schwarz:1983qr,howe1} 
can, due to its modular invariance, be written in terms of a single linear 
combination of invariants at any loop order, multiplied with an overall 
factor in terms of the complexified coupling constant 
$\Omega=C^{(0)} + ie^{-\phi}$ \cite{Green:1997tv,Green:1998by},
\begin{equation}
\label{e:Rfourdef}
{\cal L}_{\text{IIB}}\Big|_{(\alpha')^3} =
 f(\Omega,\bar\Omega) \big(  I_X-\tfrac{1}{8}I_Z \big) \,.
\end{equation}
It should be stressed that none of these type-II results have any
solid backing from a supersymmetry analysis; the bosonic terms have
been computed directly in string theory and their fermionic completion
is as of yet unknown.
\medskip

The above summarises the current knowledge about the form of the
higher-derivative actions in string theory. Given the considerable
technical difficulties that arise when one wants to analyse the
supersymmetric completions of these actions, and the even bigger
obstacles one faces when deriving the required modifications to the
supersymmetry transformation rules, we will in this paper follow a
slightly different path. Using information from various sources,
namely string amplitudes, supersymmetry requirements as well as an
interesting parallel between super-Maxwell theory and supergravity, we
set out to systematically analyse string-based higher-derivative
supergravity theories. One of our main goals is to obtain enough
information to construct the superinvariant in eleven dimensions,
write down the full set of modifications to the supersymmetry
transformation rules and determine the implications of these
modifications for the field-dependent coefficients in the
supersymmetry algebra. Using the latter, it will then be possible to
study the interplay of kappa symmetry with higher-derivative target
space and world-volume corrections.  However, there are many steps
involved before we reach this goal.
\medskip

In the first part of this paper we re-analyse the higher-derivative
$F^4$ invariants for the simpler super-Maxwell model and the coupling
to an external supergravity background. We show how the various
fermionic terms arise from string-amplitude calculations and use
supersymmetry to fix their relative normalisation. This provides a
very compact form of the results reported in the thesis by
\dcite{suel1}.

We then show, using observations made earlier by
\dcite{Bellucci:1988ff} and \dcite{berg8}, how the super-Maxwell
results can be used to bootstrap the construction of the
higher-derivative $I_X$ invariant for $N=1$ supergravity. Since all
our terms will be organised, right from the beginning, in a form that
is adapted to string theory, we will again find a very compact form of
the fermionic bilinears and the modified transformation rules.  More
importantly, since one can check superinvariance of this action
\emph{by hand}, it becomes feasible to try to understand the way in
which generalisations to the type IIA and IIB models can be
implemented.

We will leave such considerations for later and instead focus on the
extension to eleven dimensions. This poses additional problems. First
of all, supersymmetry is highly dimension dependent, so it is not at
all clear whether the analysis of \dcite{dero3} can be extended in a
straightforward way.  Another puzzle is the appearance of the $t_8$
tensors.  In string theory, they arise from the integration of eight
world-sheet fermion zero modes in the even spin-structure sector,
while the $\varepsilon_{10}$ comes from the odd spin structure. As
this can happen for left- and right-movers separately, the structure
of the contractions in \eqn{e:basicinvariants} is very natural from a
string point of view.  Clearly, such a split is not expected to
survive in eleven dimensions.  Our analysis of the fermionic terms
sheds considerable light on this issue. As we shall see, the part of
the action bilinear in the fermions does not exhibit a complete
$t_8\,t_8$ structure, even in ten dimensions.  In addition, it will
become clear that there is an inherent dimension dependence in these
terms which, when lifted to eleven dimensions, forces us to give up
many of the nice $t_8$ tensors in exchange for more complicated
couplings. One of the main results of our paper, namely the very
compact form of the eleven-dimensional superinvariant (in the absence
of gauge fields) is exhibited in~\eqn{e:elevendimI_X}.  The new
tensorial structures found there may perhaps be explained by looking
at a manifestly covariant eleven-dimensional superparticle or membrane
loop calculation using the vertex operators recently constructed by
\dcite{Dasgupta:2000df}, although we have not yet attempted to do so.
\medskip

With these results at hand, we are able to return to our original
motivation for this project. In section~\ref{s:modsusyrules}
and~\ref{s:modtorsion} we compute the supersymmetry algebra generated by the
modified transformation rules.  In general, the supersymmetry algebra of
supergravity theories takes the form 
\begin{equation}
{}[\delta^{\text{susy}}_1, \delta^{\text{susy}}_2 ] = \delta^{\text{translation}} 
+ \delta^{\text{susy}}
+ \delta^{\text{gauge}}
+ \delta^{\text{Lorentz}}\, .
\end{equation}
The coefficients on the right-hand side are, however, not necessarily
simply functions of the supersymmetry parameters $\epsilon_1$ and
$\epsilon_2$ appearing on the left-hand side; rather, they can (in an
on-shell formulation without auxiliary fields) depend explicitly on
the supergravity fields.  With the $\alpha'$ corrections in place,
each of these coefficients can pick up correction terms as well.
Focussing on the translation part, we have for instance
\begin{equation}
{}[ \bar\epsilon_1 Q, \bar\epsilon_2 Q ] e_\mu{}^r =
\xi^\nu \partial_\nu\, e_\mu{}^r+ \cdots\, ,
\end{equation}
where the translation parameter is
$\xi^\nu=2(\bar\epsilon_1\Gamma^\nu\epsilon_2)+\cdots$, the dots
indicating field-dependent $\alpha'$ correction terms.  In the
superspace formulation, the algebra is generated by supercovariant
derivatives,
\begin{equation}
\{ {D_a}, D_b \} = T_{ab}{}^r\, D_r\, .
\end{equation}
In other words, the structure of the bundle tangent to the
supermanifold reflects the com\-po\-nent-field supersymmetry algebra,
and the link is provided by constraints on the superspace
torsions. Consequently, corrections to the algebra are in direct
correspondence with modifications to the classical superspace
supergravity constraints.  Indeed, as has been shown by~\dcite{howe7},
such corrections are expected to be necessary in order to describe
modifications of the eleven-dimensional supergravity theory. Since our
analysis provides us with the full set of modified field
transformation rules, we can \emph{derive} the required superspace
torsion constraints.

Within the limitations of our analysis (we do not consider variations
proportional to the gauge field), we find that there are, however,
\emph{no} corrections to the supersymmetry algebra arising from the
lifted $I_X$ action. Consequently, the modifications to the superspace
torsion constraints anticipated by \dcite{howe7} and
\ddcite{Cederwall:2000ye}{Cederwall:2000fj} will have to follow from a
more elaborate analysis involving also gauge-field dependent terms in
the superinvariant.  In the last part of this paper we discuss
possible explanations for and implications of this result.  Our
techniques can be used to include gauge-field terms in the analysis as
well, and work along these directions is in progress.  \medskip

For reference, we have included an appendix on the first- and
second-order formulations of the $N=1$ supergravity theories in ten and 
eleven dimensions. Most of it is not new, but the reader may find it 
helpful to have the entire derivation of these standard actions spelled 
out in one place, together with an explanation of the origin of various
normalisation factors which have been quite crucial in our analysis.
A second appendix collects details on the expansion of the $t_8$
tensor, a number of useful gamma-matrix identities and the conventions
used in this paper.

\end{sectionunit}

\begin{sectionunit}
\title{First step: stringy construction of the $F^4$ action}
\maketitle
\label{s:supermaxwell}
\begin{sectionunit}
\title{Effective actions and field redefinitions}
\maketitle
\label{s:fieldredefinitions}

As the first step in our construction of the supersymmetric
$(\alpha')^3$ corrections to the supergravity actions, we will tackle
a related problem, namely the construction of the $(\alpha')^2$
corrections to the super-Maxwell action in ten dimensions. This
invariant has been constructed before by \dcite{Metsaev:1987by}, and the
non-abelian case (which we will not address) was subsequently worked out
by \dcite{Bergshoeff:1987jm}.\footnote{We should also mention that a
supersymmetric version of the ten-dimensional Born-Infeld action has
been constructed  by \dcite{Aganagic:1996nn} by gauge-fixing the 
kappa-symmetric D9-brane action.} We will also discuss the coupling to
supergravity which was worked out by \dcite{suel1}. However, as we will be
using string input, we are able to rederive these results in a much
simpler way, and at the same time this approach allows us to see how
to generalise the invariants to the supergravity $W^4$ actions. 
In the present section we  discuss a number of general issues, while
the following two sections are concerned with the string analysis and the
supersymmetry constraints respectively.

The field content of the on-shell super-Maxwell theory consists of an
abelian vector $A_\mu$ and a negative-chirality Majorana-Weyl spinor
$\chi$. We also consider the interaction with the vielbein 
$e_\mu{}^r$, the negative-chirality Majorana-Weyl gravitino $\psi_\mu$ 
and the two-form $B_{\mu\nu}$ of the supergravity multiplet. 
At the lowest order, the super-Maxwell action coupled to background 
supergravity is given by (we have normalised the fields in
such a way as to make comparison with the supergravity calculations
using the conventions of appendix~\ref{s:sugra_appendix} simple)
\begin{equation}
\label{e:superF2}
S_{F^2} = \int\!{\rm d}^{10}x \,e \big[ -\tfrac{1}{4}
F_{\mu\nu}F^{\mu\nu} - 8\,\bar\chi \slashed{D}(\omega) \chi
+2\,\bar\chi\Gamma^\mu \Gamma^{\nu\rho} \psi_\mu\, F_{\nu\rho}\, \big]
\end{equation}
(the coupling of the gauge field to the gaugino is identically zero in
the abelian case due to the fact that the gaugino is Majorana).  This
action is invariant under local supersymmetry, under which the fields
transform according to the following rules:
\begin{equation}
\label{e:YMtrafo}
\begin{aligned}
\delta A_\mu &= -4\,\bar\epsilon \Gamma_\mu \chi\, ,\\[.5ex] \delta
\chi &= \tfrac{1}{8}\Gamma^{\mu\nu}\epsilon F_{\mu\nu}\, ,\\[.5ex]
\delta F_{\mu\nu} &= -8\,\big(D_{[\mu}(\omega)\bar\epsilon\big) \Gamma_{\nu]} \chi +
8\,\bar\epsilon \Gamma_{[\mu} D_{\nu]}(\omega) \chi\, .
\end{aligned}
\end{equation}
We have not displayed the gravitational part of the action (see
appendix~\ref{s:sugra_appendix} for details and references). The only
coupling there consists of the usual shift of the field strength
$H_{\mu\nu\rho}$ of the two-form $B_{\mu\nu}$ by the Yang-Mills
Chern-Simons form. The contribution of that term to the equation of
motion is irrelevant in the variations of the terms that we 
consider below. However, in order for local supersymmetry to work out 
we still have to consider the transformations of the supergravity
multiplet fields:
\begin{equation}
\label{e:SYMtrafos}
\begin{aligned}
\delta e_\mu{}^r &= 2\, \bar\epsilon\Gamma^r\psi_\mu \, ,\\[1ex]
\delta \psi_\mu &= D_\mu(\omega) \epsilon + \cdots\,, \\[1ex]
\delta B_{\mu\nu} &= 2\, \bar\epsilon\Gamma_{[\mu}
\psi_{\nu]}\, .
\end{aligned}
\end{equation}
The dots represent terms proportional to the field strength
$H_{\mu\nu\rho}$ of the two-form $B_{\mu\nu}$ which again will not be
needed for our calculation below. We have suppressed the dilaton as it
only transforms into the dilatino and we do not consider variations
proportional to the latter. The transformation rules above are given 
in the string frame.

The equations of motion that follow from the above action are
\begin{subequations}
\label{e:SYMeoms}
\begin{align}
\label{e:SYMeomA}
\eom{A}^\mu &= \frac{1}{e}\frac{\delta S_{F^2}}{\delta
A_\mu} = D_\lambda F^{\lambda\mu} - 4\,D_\nu \big( \bar\chi \Gamma^\lambda
\Gamma^{\nu\mu} \psi_\lambda\big)\, ,\\[1em]
\label{e:SYMeomchi}
\eom{\bar\chi} &= \frac{1}{e}\frac{\delta S_{F^2}}{\delta
\bar\chi} =  16\,\Gamma^\mu \big( D_\mu(\omega)\chi -
\tfrac{1}{8} \Gamma^{rs} \psi_\mu F_{rs} \big) =: 16\,\Gamma^\mu \hat \chi_\mu \, .
\end{align}
\end{subequations}
The derivative on the gaugino is not supercovariant, i.e.~it picks up
a derivative of the supersymmetry parameter $\epsilon$, in contrast
to the hatted object appearing in the equation of motion for the
gaugino. In the higher-derivative action, the identity
\begin{equation}
\label{e:chimu_id}
\slashed{D}(\omega) \chi = \tfrac{1}{8} \Gamma^\mu \Gamma^{rs} \psi_\mu F_{rs}
+ \tfrac{1}{16} \eom{\bar\chi}
\end{equation}
can be used to trade derivatives on the gaugino for terms proportional
to the lowest-order equations of motion plus terms proportional to the
gravitino, and will be used repeatedly below. We stress that it is an
expression which is non-linear in the fields, and therefore, when used
in the amplitudes, can produce a five-point interaction from a
four-point string amplitude.

There is considerable ambiguity in the structure of the on-shell
higher-derivative effective action, due to the fact that field
redefinitions can be used to remove higher-order $\alpha'$ terms
proportional to the field equations; see for
instance~\dcite{Tseytlin:1986zz} for an overview of this problem from
various calculational points of view. As the field equations involve
both gauge fields as well as fermions, we have to be very careful
about the field-redefinition freedom. If the action at higher order in
some coupling constant $g$ contains terms proportional to the
equations of motion obtained from the lowest-order action, as in
\begin{equation}
S[\phi] = S_0[\phi] + g \int\!{\rm d}x\,
\frac{\delta S_0[\phi]}{\delta \phi(x)}\,R(\phi(x),\partial\phi(x)) \, ,
\end{equation}
these can be removed by a field redefinition:
\begin{equation}
\begin{aligned}
\phi \rightarrow \phi - g\,R(\phi,\partial\phi) \quad \Rightarrow 
\quad S[\phi] \rightarrow S_0[\phi] + {\cal O}(g^2)\, .
\end{aligned}
\end{equation}
Similarly, higher-order terms in the variation of the action under some
transformation $\delta_\lambda\phi$ which are proportional to the 
lowest-order equations of motion,
\begin{equation}
\delta_\lambda S[\phi] = V_0[\phi] + 
g \int\!{\rm d}x\, \frac{\delta S_0[\phi]}{\delta\phi(x)}\, 
V_1(\phi(x),\partial\phi(x);\lambda)\,,
\end{equation}
can be removed by a modification of the field transformation rules:
\begin{equation}
\delta_\lambda \phi \rightarrow \delta_\lambda\phi - 
g V_1(\phi,\partial\phi;\lambda)\quad
\Rightarrow\quad
\delta_\lambda S[\phi] \rightarrow V_0[\phi] + 
{\cal O}(g^2)\, .
\end{equation}
In addition to these two mechanisms, we will also encounter terms
that are zero in a specific gauge. In particular, terms proportional to
the spin-$1/2$ part of the gravitino or terms involving the de~Donder gauge
condition of the graviton are identically zero in string theory.
Such terms can, however, not be removed by a procedure of the type sketched
above, and they have to be kept for the supersymmetric completion.

We should finally comment on the field redefinition freedom involving
rescaling by powers of the dilaton (see also
appendix~\ref{s:sugra_appendix}). In~\eqn{e:SYMtrafos} we wrote
transformation rules which do not involve the dilaton on the
right-hand side. The existence of a particular frame in which the
transformations take such a simple form is related to the scale
invariance of the heterotic (see the work of \dcite{Kallosh:1985eh}
and \dcite{Kallosh:1986cd}) and type~II supergravity theories. In
this frame---the string frame---the classical supersymmetry
transformations only involve derivatives of the dilaton (as was
observed by~\dcite{berg8}).  Since the string genus expansion is
ordered by powers of exponentials of the dilaton, classical
supersymmetry thus does not mix string loop orders.  Moreover, the
higher-order modifications to the supersymmetry transformations that
we derive in this paper do not contain the dilaton, and thus also
respect the genus expansion.
\end{sectionunit}

\begin{sectionunit}
\title{Tensor structures from string amplitudes}
\maketitle

In order to get a covariant result, which is by far the most
convenient starting point for the supersymmetry analysis, we
employ the covariant formalism to compute the string amplitudes.
Moreover, the tensor structures arise here in a very simple way from
the operator products of world-sheet fermions, whereas a manifestly
space-time supersymmetric approach would require complicated fermion
zero-mode integrals to be performed. We will comment further on such
alternative approaches in the conclusions.

There are few detailed accounts of string amplitudes with external
fermions in the literature (for exceptions see
\ddcite{Green:1982xx}{Green:1982ya} for a light-cone approach and
\dcite{Atick:1987rs} and
\ddcite{Pasquinucci:1997gv}{Pasquinucci:1998mu} for covariant
calculations). Let us therefore first review some of the formalism and
present the required technical details. For other texts on this
subject the reader is referred to the book by \dcite{b_lues1} and the
review by \dcite{D'Hoker:1988ta}. Some of the subtleties have, however,
only been discussed in the papers cited below.

In order to find the higher-derivative action we analyse string
amplitudes with two external fermions. The relevant vertex operators
are given by
\begin{equation}
\label{e:SYMvertexoperators}
\begin{aligned}
V_A^{(0)}(k) =\,\,& \int\!{\rm d}^2 z\, A_\mu \norder{(i\partial X^\mu + 
\tfrac{\alpha'}{2} k\cdot \Psi\,\Psi^\mu) \, e^{i k \cdot X}}\, ,\\[1ex]
V_\chi^{(-1/2)}(k) =\,\,& \int\!{\rm d}^2 z\, \bar\chi \norder{S^+_{\rm L} \, e^{-\phi/2} \, 
e^{i k\cdot X}}\,,\\[1ex]
V^{(-1/2)}_{\psi}(k) =\,\,& \int\!{\rm d}^2 z\, \bar\psi_\nu 
\norder{S^+_{\rm L}  e^{-\phi/2} \bar\partial X^\nu e^{ik\cdot X}}\, , \\[1ex]
V^{(-1)}_B(k) =\,\,&\int\!{\rm d}^2 z\, B_{\mu\nu} 
\norder{e^{-\phi} \Psi^\mu \bar\partial X^\nu e^{ik\cdot X}} \, .
\end{aligned}
\end{equation}
for the abelian gauge field, the abelian gaugino, the gravitino and
two-form gauge potential, respectively (as we will not determine
precise normalisations of our amplitudes, we have ignored any overall
factors in the vertex operators listed above).
The right-moving sector carrying the abelian gauge 
degree of freedom has been omitted. The world-sheet bosons are denoted 
by $X^\mu$, while $\Psi^\mu$ are the world-sheet fermions and $\phi$ is a 
bosonic field representing the ghosts \cite{frie1}. The spin field $S^+_{\rm L}$ 
has positive space-time chirality and arises purely from world-sheet
fermions of one chirality, associated to the left-moving sector after
imposing the equations of motion.  

All of the vertex operators above are taken in their canonical ghost
picture, as deduced by linearisation of the string action. The ghost 
charges are balanced by inserting a sufficient number of copies of the 
picture-changing operator
\begin{equation}
Y(w) \defas T_F e^{\phi}(w) = i \sqrt{\tfrac{2}{\alpha'}}\, \partial X^\mu \Psi_\mu\, e^{\phi}(w)
\end{equation}
at arbitrary points in the correlators (in the odd spin-structure
sector, one always gets at least one power of the picture-changing
operator from integration over the odd supermoduli). The end result
can be shown to be independent of these arbitrary locations (see
\dcite{frie1}, \ddcite{verl1}{Verlinde:1988tx} and in particular
also section 3.2 of \dcite{pasq1} for an explicit example). One may be
tempted to use the vertex operators in different pictures by taking
the insertion points of $Y(\omega)$ to coincide with those of the
vertex operators. As was shown by \dcite{Green:1988qu}, this requires
very careful treatment of terms in the vertex operators that are
proportional to the world-sheet equations of motion. These extra
pieces have been interpreted as vertex operators for the
$N=1$ auxiliary fields of the gauge multiplets \cite{atic1}.  This
phenomenon, specific to $N=1$ superstring models (the heterotic and
the open string), does not occur for type~II superstrings. Since the
origin of the tensorial structures for the type~II invariants are the
same up to modifications we will explain later, we adopt a more
symmetric procedure, similar to the computations
of~\dcite{Atick:1987rs}.

\begin{table}
\smaller\smaller
\begin{align*}
\intertext{\bf even spin structure:}
\left.\begin{aligned}
{}& V_A^{(0)} &&\quad V_A^{(0)} &&\quad V_A^{(0)} &&\quad V_A^{(0)} \\
{}& \norder{k\Psi\Psi} &&\quad \norder{k\Psi\Psi} &&\quad \norder{k\Psi\Psi} &&\quad \norder{k\Psi\Psi} 
\end{aligned}\quad\right\}& \quad t_8 F^4 \\[2ex]
\left.\begin{aligned}
{}& T_F && \quad V_A^{(0)} && &&V_\chi^{(-1/2)}\!\!&& &&\quad V_\chi^{(-1/2)}\!\!
&& &&\quad V_A^{(0)} \\
{}& \partial X\Psi && \quad \norder{k\Psi\Psi} &&
\stackrel{\Gamma^{[1]}S^+\Psi^2}{\longleftrightarrow} &&\quad S^+  &&
\stackrel{\Gamma^{[1]}\Psi}{\longleftrightarrow}
&&\quad\quad  S^+ &&
\stackrel{\Gamma^{[1]}S^+\Psi^2}{\longleftrightarrow} &&\quad \norder{ k\Psi\Psi} 
\end{aligned}\quad\right\}&\quad
t_8 (\bar\chi  \Gamma_r \eta_{rr} D_r\chi) F_{rr}F_{rr}\\[2ex]
\left.\begin{aligned}
{}& T_F && \quad V_\chi^{-1/2} && &&\quad V_\chi^{-1/2} &&\quad V_A^{0} &&\quad V_A^{0}\\
{}& \partial X\Psi && \quad   S^+         && \stackrel{\Gamma^{[1]}\Psi}{\longleftrightarrow} && \quad
 S^+ && \quad\norder{k\Psi\Psi} && \quad \norder{k\Psi\Psi} 
\end{aligned}\quad\right\} & \quad (\bar\chi \Gamma_m D_n \chi) F^2{}_{mn}\\[2ex]
\left.\begin{aligned}
{}& T_F && \quad V^{(-1/2)}_\psi\, && &&\quad V^{(-1/2)}_\chi\, &&\quad V^{(0)}_A &&\quad V^{(0)}_A &&\quad V^{(0)}_A \\
{}& \partial X\Psi && \quad S^+ &&\stackrel{\Gamma^{[1]}\Psi}{\longleftrightarrow} &&\quad  S^+ &&\quad \norder{k\Psi \Psi} 
&&\quad \norder{k\Psi\Psi} &&\quad \norder{k\Psi\Psi}\\
{}& &&\quad \bar\partial X^\nu 
\end{aligned}\quad\right\}& \quad t_8(\bar\psi_r \Gamma_r \chi) F_{rr}F_{rr}F_{rr}\\[3ex]
\intertext{\bf odd spin structure:}
\left.\begin{aligned}
{}& V_\chi^{(-1/2)} && &&\quad V_\chi^{(-1/2)} && &&\quad T_F &&\quad V_A^{(0)}
&& \quad V_A^{(0)}\\
{}& S^+ && \stackrel{\Gamma^{[5]}\Psi^5}{\longleftrightarrow} && \quad
S^+ \exp(ikX) && \quad\leftrightarrow &&\quad \norder{\partial X\Psi} &&\quad 
\norder{k\Psi\Psi} && \quad \norder{k\Psi\Psi}
\end{aligned}\quad\right\}& \quad \varepsilon_{10} (\bar\chi \Gamma_{rrrrr} D_r\chi) F_{rr}F_{rr}\\[2ex]
\left.\begin{aligned}
{}& \quad T_F && \quad V_B^{(-1)} &&\quad V_A^{(0)} &&\quad V_A^{(0)} &&\quad V_A^{(0)}
&& \quad V_A^{(0)}\\
{}& \partial X \Psi && \quad \Psi^\mu &&\quad \norder{k\Psi\Psi} &&\quad
\norder{k\Psi\Psi} &&\quad \norder{k\Psi\Psi} && \quad \norder{k\Psi\Psi} \\
{}&  && \quad \bar\partial X^\nu
\end{aligned}\quad\right\}& \quad \varepsilon_{10} B F^4\\[2ex]
\left.\begin{aligned}
{}& T_F &&\quad V_\psi^{(-1/2)} && &&\quad V_\chi^{(-1/2)} &&\quad V_A^{(0)}
&& \quad V_A^{(0)} && \quad V_A^{(0)} \\
{}& \partial X \Psi && \quad S^+ && \stackrel{\Gamma^{[3]}\Psi^3}{\longleftrightarrow} && \quad
S^+ &&\quad \norder{k\Psi\Psi} && \quad \norder{k\Psi\Psi} &&\quad \norder{k\Psi\Psi}\\
{}& && \quad \bar\partial X && && && && && 
\end{aligned}\quad\right\}& \quad \varepsilon^{(r)}_{10}  (\bar\psi_{r}
\Gamma_{rrr} \chi) F_{rr} F_{rr} F_{rr}\\[2ex]
\end{align*}
\caption{Vertex operator products for the heterotic string leading to
the supersymmetric completion of the $F^4$ effective action. The top
 row shows the vertex operators, while the second and third row
 exhibit the relevant fields in the left- and right-moving sector
respectively. The given products are those relevant for one-loop
 amplitudes and Abelian gauge group; the first column shows the relevant terms from each
vertex operator in the product, while the second column shows which
terms in the effective action are generated by the amplitude on the
left. The spin structure from which the various terms originate has
been indicated. The gauge part and the ghost factors, being trivial,
have been omitted. Plane-wave exponentials are only displayed whenever
they are non-trivially contracted. The precise index structure of the
terms in the last column is exhibited in~\eqn{e:SYMstring}.}
\label{t:SYMvops}
\end{table}

In order to compute the operator product expansions, we need the
following building blocks: 
{\allowdisplaybreaks[4]
\begin{align}
\Psi_\mu(z) \Psi_\nu(w)            &= \frac{\eta_{\mu\nu}}{z-w} + \text{finite}\, ,\\[1ex]
X^\mu(z,\bar z) X^\nu(w,\bar w)    &=-\tfrac{1}{2}\alpha'\, \eta^{\mu\nu} \ln|z-w| +\text{finite}\, ,\\[1ex]
\partial X^\mu (z,\bar z) \bar\partial X^\nu(w,\bar w) &= \frac{\pi\alpha'}{2}\,
\eta^{\mu\nu} \delta^2(z-w)+ \text{finite}\, ,\\[1ex]
X^\mu(z) e^{ik\cdot X} &= -\tfrac{1}{2}i\alpha'\, k^\mu e^{ik\cdot X}
\ln(z) + \text{finite}\, .
\end{align}}
For the ghost we have
\begin{equation}
e^{q_1\phi(z)} e^{q_2\phi(0)} = \frac{1}{z^{q_1q_2}} e^{(q_1+q_2)\phi(z)} +
\text{finite}\, .
\end{equation}
The operator products of the spin operators are slightly more complicated,
as we need some finite parts as well. One finds 
\begin{equation}
\label{e:OPE}
\begin{aligned}
S^+(z,\bar z)\otimes S^+(0)     &= 2^{-4}\,\sum_{n=1,3,\ldots}\tfrac{1}{n!} ({\cal P}^+{\cal C}\Gamma^{\mu_1\cdots \mu_n}{\cal P}^+)
                            \norder{\Psi_{\mu_1}\cdots\Psi_{\mu_n}} z^{-5/4+n/2}\,, \\[1ex]
S^+(z,\bar z)\otimes S^-(0)     &=2^{-4}\, \sum_{n=0,2,4,\ldots}\!\!\!
                            \tfrac{1}{n!} ({\cal P}^+{\cal C}\Gamma^{\mu_1\cdots \mu_n}{\cal P}^-)
                            \norder{\Psi_{\mu_1}\cdots\Psi_{\mu_n}} z^{-5/4+n/2}\, ,\\[1ex]
S^{\pm}(z,\bar z)\Psi^\mu(0) &= ({\cal C}\Gamma^\mu) S^{\mp}  z^{-1/2} -
                           i({\cal C}\Gamma^{\nu}) \norder{S^{\mp}
                            \Psi^{\mu}\Psi_\nu} z^{1/2} 
+\text{finite}\, .
\end{aligned}
\end{equation}
The objects ${\cal P}^\pm$ are projectors onto positive and negative
space-time chirality. These identities hold for both left- and 
right-handed world-sheet sectors.  We refer to appendix~\ref{s:conventions}
for further details about our conventions.

There are two strong constraints that have to be satisfied for a
particular vertex operator product to give a non-vanishing expectation
value: the fermionic $\Psi$ zero-mode integrals (when present) have to
be saturated and the total ghost number has to cancel the ghost charge
of the vacuum (or in other words the ghost zero modes have to be
saturated as well).  When integrating over the world-sheet fermions
$\Psi$ there are ten fermionic zero-mode integrals for odd spin
structure. This implies that in this sector the integral picks out the
part of the vertex operator product that depends on ten $\Psi$s. The
result is an $\varepsilon_{10}$ tensorial structure. For the even spin
structure there are no restrictions on the number of world-sheet
fermions. An integral over eight $\Psi$s with anti-periodic
boundary conditions can be rewritten (through the usual bosonisation
and refermionisation procedure) in terms of an integral over a
space-time fermion $\theta$ with periodic boundary conditions. The
resulting tensorial structure corresponds to the $t_8$ tensor. For
smaller numbers of fermions, other tensorial structures will appear.
Concerning the ghost number cancellation, recall that the Riemann-Roch
theorem yields $2g-2$ for the difference between the number of $\beta$
zero modes minus the number of $\gamma$ zero modes. For our one-loop
considerations, we therefore have to consider vertex operator products
with total ghost-charge zero. We refer the reader to section~12.6 of
\dcite{b_polc2} and to \dcite{Verlinde:1988tx} for more details.

The first corrections to the action~\eqn{e:superF2} come in at order
$(\alpha')^2$ relative to the classical action. As the $F^4$ term
arises from a four-point function in string theory, one may expect
that the fermion bilinears which are related to it by linear
supersymmetry can also be obtained by considering four-point functions
only. This turns out to be true when considering only global
supersymmetry invariance, but local supersymmetry forces us to
consider both four- and five-point functions (see in this context also the
work by \ddcite{Grisaru:1977vm}{Grisaru:1977px} which discusses
a related but not quite identical issue). Table~\ref{t:SYMvops}
lists all the various operator products that occur. Many of them are
straightforward to obtain so we will here just comment on the more 
subtle ones.

In the even spin-structure sector, there is a term which receives
contributions from only six world-sheet fermions. Only the charge
conjugation matrix is kept in the product of the spin field with a
world-sheet fermion, and using the first term of the third identity 
in~\eqn{e:OPE} one obtains
\begin{equation}
 S^+\otimes S^+ \, \Psi^\mu \, \Psi^{\nu_1} \Psi^{\nu_2} \,
 \Psi^{\nu_3} \Psi^{\nu_4}  \to \Gamma_\nu
 \times \eta^{\nu\nu_1} \eta^{\nu_2\nu_3} \eta^{\nu_4\mu}\, .
\end{equation}
The $\partial X^\mu$ from the picture-changing operator is contracted
with the plane-wave factors of the fermion vertex operators. Since we
are only interested in this leading contribution the remaining
correlation function between the plane-wave factors is approximated to
one
\begin{equation}
\left\langle \prod_{i=1}^4 \, e^{i k^{(i)}\cdot X} \right\rangle
\simeq 1 + \mathcal{O}(k^2) \,.
\end{equation}
This contribution has the correct (anti-)symmetry in the labels of the
external (fermions) bosons. 

Another tricky term in the even spin-structure sector is the one on
the second line of table~\ref{t:SYMvops}. Here one has to keep, for
each of the operator products of a spin field with a world-sheet
fermion, the term $({\cal C}\Gamma^m) \Psi^m$. The result is
\begin{equation}
S^+\otimes  S^+ \, \Psi^\mu \, \Psi^{\nu_1} \Psi^{\nu_2} \,
 \Psi^{\nu_4} \Psi^{\nu_4}  \to
\Gamma_\nu\Gamma_\rho\Gamma_\lambda \,\times
 t_8^{\nu\rho\lambda \mu \nu_1\nu_2 \nu_3\nu_4}\, .
\end{equation}
The fermions $\Psi$ have been contracted in the usual way
\cite{b_polc2} to obtain the $t_8$-structure, while the plane-wave part 
has been treated as before. This contribution to the amplitude is already
symmetric in the labels of the external gauge bosons, and the
anti-symmetry in the labels of the external gaugini forbids
the appearance of a $\Gamma_{\nu\rho\lambda}$ term. The final
form of this contribution is therefore
\begin{equation}
\label{e:etagamma1}
t_8^{(r)} \left( \bar\chi \left[\Gamma_{r_1r_2} \Gamma_{r_3} -
  \Gamma_{r_3} \Gamma_{r_1r_2}\right] D_{r_4} \chi\right)\, F_{r_5r_6}
  F_{r_7r_8}\, .
\end{equation}
The other products are straightforward. In the next section we will
use supersymmetry to determine the precise coefficients of these
terms.
\end{sectionunit}

\begin{sectionunit}
\title{Supersymmetry of the higher-derivative effective action}
\maketitle
\label{s:maxwellsusy}

The analysis in the previous section has provided us with all the
distinct tensorial structures of the fermion bilinears appearing in
the supersymmetric completion of the $F^4$ action. Since we did not
compute the full amplitudes, we have not obtained the normalisations
of these terms.  However, the normalisations can easily be fixed using
supersymmetry.  As we will show in detail in this section, the correct
combination of terms is given by the following action:
\begin{equation}
\label{e:SYMstring}
\begin{aligned}
S_{F^4} = \frac{(\alpha')^2}{32}\int\! {\rm d}^{10}x \Big[ &\,\tfrac{1}{6}
e\, \teight{r} F_{r_1r_2}\cdots F_{r_7r_8}
 + \tfrac{1}{12} \varepsilon_{10}^{(r)} B_{r_1r_2}
F_{r_3r_4} \cdots F_{r_9r_{10}} \\[1ex]
& - \tfrac{32}{5} e\, \teight{r} \eta_{r_2r_3} (\bar\chi \Gamma_{r_1}
\chi_{r_4})  F_{r_5r_6}F_{r_7r_8}\\[1ex]
& + \tfrac{12\cdot32}{5} e (\bar\chi\Gamma_{r_1}\chi_{r_2})
F^{r_1m}F_m{}^{r_2}  \\[1ex]
&  -\tfrac{16}{5!} \varepsilon_{10}^{(r)} (\bar\chi \Gamma_{r_1\cdots r_5}
\chi_{r_6}) F_{r_7r_8} F_{r_9r_{10}} \\[1ex]
&  + \tfrac{16}{3}e\,\teight{r} (\bar\psi_{r_1} \Gamma_{r_2}\chi)
F_{r_3r_4}F_{r_5r_6}F_{r_7r_8} \\[1ex]
& + \tfrac{8}{3}e\, (\bar\psi_{m}\Gamma^{mr_1\cdots r_6} \chi) F_{r_1
r_2}\cdots F_{r_5r_6} \Big] \,.
\end{aligned}
\end{equation}
Here $\teight{r}\defas t_8^{r_1\cdots r_8}$ and
$\varepsilon_{10}^{(r)}\defas \varepsilon^{r_1\cdots r_{10}}$.  The pure
gauge-field terms in this action taken together with \eqn{e:superF2}
agree with the expansion of the Born-Infeld action to the
corresponding order (and we used this correspondence to fix the
overall normalisation of~\eqn{e:SYMstring}). Our result is minimal in
the sense that we have not included any terms proportional to the
lowest-order equations of motion. Indeed, this is perhaps the most
natural thing to do, given our on-shell string results.  Anyhow, such
terms can always be eliminated by field redefinitions.

The action above displays several surprising features. The most
striking one is perhaps the explicit dependence on the spacetime
dimension arising from the contraction of $\teight{r}$ indices in the
term on the second line (the tensor $\eta_{r_2r_3}$ in this term
originates from the gamma matrix product in \eqn{e:etagamma1}).  The
dimension dependence makes it clear already at this point that some of
the stringy $\teight{r}$ structure will be lost when going to eleven
dimensions; we will address this point in more detail later.
Secondly, one observes that there is a term which does not have the
$\teight{r}$ at all, due to the fact that it only involves six
world-sheet fermion zero modes.


In order to compare the result \eqn{e:SYMstring} to equation~(C.9) of
\dcite{suel1}, one has use the gaugino equation of
motion~\eqn{e:chimu_id} wherever possible in order to produce
additional gravitino terms. In this way, one arrives at the following
Lagrangian (from here on we will write just $D$ 
instead of $D(\omega)$):
\begin{equation}
\label{e:expandedSYMstring}
\begin{aligned}
%
%
(\alpha')^{-2} {\cal L}_{\Gamma^{[0]}} =\,\, &-&\tfrac{1}{32}e\,&( (F^2)^2 - 4 F^4 ) \\[1ex]
&+&\tfrac{1}{384}&\varepsilon^{t_1t_2r_1\cdots r_8} B_{t_1t_2}
F_{r_1r_2}\cdots F_{r_7r_8}\, , \\[1ex]
%
%
(\alpha')^{-2} {\cal L}_{\Gamma^{[1]}} =\,\, &-&4e\, &(\bar\chi \Gamma^{r_1} D^{r_2} \chi)
F^2_{r_1r_2} \\[1ex]
&+&\tfrac{1}{4}e\,&(\bar\psi^{r_1}\Gamma^{r_2}\chi) F_{r_1r_2}F^2  \\[1ex]
&-&3e\,&(\bar\psi^{r_1}\Gamma^{r_2}\chi) F^3_{r_1r_2} \, , \\[1ex]
%
%
(\alpha')^{-2} {\cal L}_{\Gamma^{[3]}} =\,\, &+& 2e\,& (\bar\chi\Gamma^{r_1r_2r_3} D^{r_4}
\chi) F_{r_1r_2}F_{r_3r_4}  \\[1ex]
 &-&\tfrac{1}{8}e\,&(\bar\psi_m \Gamma^{mr_1r_2}\chi)
F_{r_1r_2} F^2 \\[1ex] 
 &+&\tfrac{1}{2}e\, &(\bar\psi_m \Gamma^{mr_1r_2} \chi) F^3_{r_1r_2}  \\[1ex]
 &+&e\,&(\bar\psi^{r_4} \Gamma^{r_1r_2r_3}\chi) F_{r_1r_2}
F^2_{r_3r_4}\, , \\[1ex]
%
%
(\alpha')^{-2} {\cal L}_{\Gamma^{[5]}} =\,\, &+&\tfrac{1}{8}e \,
  &(\bar\psi^{r_6} \Gamma^{r_1\cdots r_5}\chi)
  F_{r_1r_2}F_{r_3r_4} F_{r_5r_6}  \, , \\[1ex]
%
%
(\alpha')^{-2} {\cal L}_{\Gamma^{[7]}} =\,\,  &+&\tfrac{1}{48}e \,
  &(\bar\psi_m \Gamma^{m r_1\cdots r_6}\chi)
  F_{r_1r_2}F_{r_3r_4} F_{r_5r_6}  \, .
\end{aligned}
\end{equation}
(Here $F^2_{mn}\defas F_{mp}F^p{}_n$, $F^2\defas F_{mn}F^{nm}$, etc.)
The terms ${\cal L}^{2}_{\Gamma[1]}$ and ${\cal L}^{3}_{\Gamma[1]}$
receive contributions from both the four- and five-point functions;
the coefficients arise as $\tfrac{1}{4}=-\tfrac{3}{4}+1$ and $-3=1-4$
respectively. 

We should comment on the fact that we dropped the terms proportional
to the equations of motion when going from~\eqn{e:SYMstring}
to~\eqn{e:expandedSYMstring}. The underlying idea is that string
theory cannot give us any information about such terms, so from this
perspective the two actions are equally good. When we discuss
supersymmetry of the action, and the required modifications to the
transformation rules, we will always start
from~\eqn{e:expandedSYMstring}. The transformation rules of the fields
which appear in~\eqn{e:SYMstring} pick up additional modifications due
to the fact that they are related to the fields
in~\eqn{e:expandedSYMstring} by a field redefinition.

Let us now exhibit supersymmetry in detail.  For pedagogical purposes,
we write the variation of the above action in a form which makes it
easy to read off the global supersymmetry invariance when the
gravitino is set to zero identically. Repeated use of the Bianchi
identity and the lowest-order equations of motion for the gauge
potential leads to the following result for the variation of the terms
derived from a four-point calculation:
\begin{subequations}
\begin{equation}
\label{e:maxwellhorror}
\begin{aligned}
%
%
(\alpha')^{-2} \delta {\cal L}^1_{\Gamma^{[0]}} \stackrel{\delta
F}{\rightarrow}
\,\, &+&e\,&(\bar\chi\Gamma^{r_1}\epsilon) D^{r_2}(F^2 F_{r_1r_2} -4
F^3_{r_1r_2}) && \quad
X_1+X_2\\[1ex]
%
%
(\alpha')^{-2} \delta {\cal L}^1_{\Gamma^{[1]}} =\,\,& -&e\,
&(\bar\chi\Gamma^{r_1}\Gamma^{r_3r_4}
D^{r_2}\epsilon) F_{r_3r_4} F^2_{r_1r_2} && \quad Z_1\\[1ex]
 & +&2e\, &(\bar\chi\Gamma^{r_3}\Gamma^{r_1r_4} \epsilon) 
F^2_{r_1r_2} D_{r_3}F^{r_2}{}_{r_4} &&\quad G_1\\[1ex]
 & -&\tfrac{1}{8}e\,& (\bar\chi\Gamma^{r_1}\Gamma^{r_3r_4}\epsilon) 
 D_{r_1} ( F_{r_3r_4} F^2 ) &&\quad G_2\\[1ex]
 & -&e\,& (\bar\chi\Gamma^{r_1}\epsilon)D^{r_2}(F^2 F_{r_1r_2}-4
F^3_{r_1r_2}) && \quad X_1+X_2\\[1ex]
 & +&\tfrac{1}{2}e\,&(\bar\chi\Gamma^{r_1r_2r_3} \epsilon) F_{r_1r_2}
 F_{r_3}{}^{r_4}\, \eom{A}_{r_4}\\[1ex]
 & +&e\,&(\bar\chi\Gamma^{r_1}\epsilon) F^2_{r_1}{}^{r_2}\, \eom{A}_{r_2}\\[1ex]
 & -&\tfrac{3}{4}e\,&(\bar\chi\Gamma^{r_1}\epsilon) F^2\,\eom{A}_{r_1}\,,\\[1ex]
%
%
(\alpha')^{-2} \delta {\cal L}^1_{\Gamma^{[3]}} =\,\,& +&\tfrac{1}{48}e\,
 &(\bar\chi\Gamma^m\Gamma^{r_1\cdots r_6} \epsilon) D_{m}(
 F_{r_1r_2}\cdots F_{r_5r_6}) && \quad Z_2\\[1ex]
 & -&e\,&(\bar\chi\Gamma^{r_1}\Gamma^{r_6r_2}\epsilon) F^2_{r_6r_4}
 D_{r_1} F^{r_4}{}_{r_2} &&\quad G_1\\[1ex]
 & +&\tfrac{1}{2}e\,&(\bar\chi\Gamma^{r_1}\Gamma^{r_3r_6}\epsilon)
 F_{r_3r_4}(D_{r_1}F^{r_4}{}_{r_2})F^{r_2}{}_{r_6} &&\quad G_1\\[1ex]
 &+&\tfrac{1}{8}e\,& (\bar\chi \Gamma^{r_1\cdots r_5} \epsilon)
 F_{r_1r_2} F_{r_3r_4}\,\eom{A}_{r_5}\\[1ex]
 &-&\tfrac{1}{2}e\,&(\bar\chi\Gamma^{r_1r_2r_3}\epsilon) F_{r_1r_2}
 F_{r_3}{}^{r_4}\,\eom{A}_{r_4}\\[1ex]
 & +&\tfrac{1}{2}e\,&(\bar\chi\Gamma^{r_1}\epsilon) F^2\, \eom{A}_{r_1}\, .
\end{aligned}
\end{equation}
The arrow in $\delta L_{\Gamma[0]}^1$ indicates that we have not yet
considered the transformation of the vielbein. The column on the 
right-hand side lists symbolic names for terms which still have to be
cancelled. The $\eom{\cdot}$ lines indicate terms proportional to the
equations of motion; see~\eqn{e:SYMeomA} and~\eqn{e:SYMeomchi}.  The
quickest way to arrive at the above result is to integrate away from
the gaugino, leave all terms with derivatives on epsilon alone, and
then write all $\Gamma^{[3]}$ terms as $\Gamma^{[1]}\Gamma^{[2]}$
products. After that, the derivative index can be made to contract
onto the single gamma matrix by Bianchi cycling, yielding gaugino
equations of motion.  The remaining terms are single gamma terms.

We should stress that in the above variation, it never happened that we 
got a dimension-dependent factor (which could in principle occur when
contracting gamma matrices). This fact will be very important later on in
the supergravity generalisation.

The $G_i$ terms in \eqn{e:maxwellhorror} are zero in global susy by
using the gaugino equation of motion; note that their sum can be recast
in $t_8$ form as
\begin{equation*}
G_1+G_2 = -\tfrac{1}{48} e \teight{r}(\bar\chi\Gamma^m\Gamma_{r_7r_8}
\epsilon)D_m(F_{r_1r_2}\cdots F_{r_5r_6}).
\end{equation*}
The $Z_i$ terms are zero identically
in global susy. The above shows that the action without gravitini is
invariant under global susy.

For local susy, we are now left with the following terms in the
variation of terms coming from the four-point functions:
\begin{equation}
\begin{aligned}
\label{e:maxwellhorror_intermediate}
\delta {\cal L}_{\text{four-point}} =\,\, &+&\tfrac{1}{2}e\, &(\bar\epsilon\Gamma^{r_1r_2r_3}\chi) D_{r_1}
(F^3_{r_2r_3}) &&\quad X_6\\[1ex]
  &+& e\, &(\bar\epsilon\Gamma^{r_3}\chi) D^{r_2} F^3_{r_3r_2}\,
,&& \quad X_2\\[3ex]
 &-&\tfrac{1}{8}e\, &(\bar\epsilon\Gamma^{r_1r_2r_3} \chi) D_{r_1}( 
F_{r_2r_3}F^2) && \quad X_7\\[1ex]
  &-& \tfrac{1}{4}e\, &(\bar\epsilon\Gamma^{r_1}\chi) D^{r_2}(F_{r_1r_2}F^2)
\, , &&\quad X_1\\[3ex]
  &+&e\, &(\bar\epsilon\Gamma^{r_1r_3r_4}D^{r_2}\chi)
  F_{r_3r_4}F^2_{r_1r_2} && \quad X_{11}\\[1ex]
 &+& e\,& (\bar\epsilon\Gamma^{r_1r_3r_4}\chi) D^{r_2}(
  F_{r_3r_4}F^2_{r_1r_2} ) &&\quad X_8\\[1ex]
 &+& 2e\, &(\bar\epsilon\Gamma^{r_4}D^{r_2}\chi) F^3_{r_4r_2} &&
\quad X_{12}\\[1ex]
 &+& 2e\, &(\bar\epsilon\Gamma^{r_4}\chi)
D^{r_2}(F^3_{r_4r_2}) && \quad X_2\\[4ex]
 &+& \tfrac{1}{8}e\,& (\bar\epsilon\Gamma^{r_1\cdots
r_5}\chi) D^{r_6} (F_{r_1r_2}\cdots F_{r_5r_6}) \, . && \quad X_5
\end{aligned}
\end{equation}
The four different blocks come from $G_1$, $G_2$, $Z_1$ and
$Z_2$ respectively. These terms will have to cancel against the
variation of the terms derived from a five-point calculation in string
theory, namely the terms proportional to the gravitino.
For those five-point terms, we integrate away from the supersymmetry
parameter and use the identity~\eqn{e:chimu_id} whenever possible. The
upshot is:
\begin{equation}
\label{e:maxwellhorror2}
\begin{aligned}
%
%
(\alpha')^{-2} \delta {\cal L}^2_{\Gamma^{[1]}} =\,\,
  &+&\tfrac{1}{4}e\,&(\bar\epsilon\Gamma^{r_1} D^{r_2}\chi) F_{r_1r_2}
  F^2 && \quad Y_1\\[1ex]
 &+& \tfrac{1}{4}e\, &(\bar\epsilon\Gamma^{r_1} \chi) D^{r_2}
  F_{r_1r_2} F^2 && \quad X_1\\[1ex]
 &+& \tfrac{1}{32}e\,& (\bar\psi^{r_1}\Gamma^{r_2r_3r_4}\epsilon)
  F_{r_1r_2}F_{r_3r_4} F^2 && \\[1ex]
 &+& \tfrac{1}{16}e\, &(\bar\psi^{r_1} \Gamma^{r_2} \epsilon)  F^2_{r_1r_2} F^2
 \\[1ex]
%
%
(\alpha')^{-2} \delta {\cal L}^3_{\Gamma^{[1]}} =\,\, 
  &-&3e\,& (\bar\epsilon\Gamma^{r_1}D^{r_2}\chi) F^3_{r_1r_2} && \quad X_{12}\\[1ex]
 &-& 3e\,& (\bar\epsilon\Gamma^{r_1} \chi) D^{r_2} F^3_{r_1r_2}
  && \quad X_2 \\[1ex]
 &-&\tfrac{3}{8}e\,& (\bar\psi^{r_1} \Gamma^{r_2r_3r_4} \epsilon)
  F^3_{r_1r_2} F_{r_3r_4}\\[1ex]
 &-&\tfrac{3}{4}e\,& (\bar\psi^{r_1}\Gamma^{r_2} \epsilon) F^4_{r_1r_2}
\end{aligned}
\end{equation}
{\renewcommand{\theequation}{\ref{e:maxwellhorror2} cont.}
\addtocounter{equation}{-1}
\begin{equation}
\begin{aligned}
%
%
(\alpha')^{-2} \delta {\cal L}^2_{\Gamma^{[3]}} =\,\, 
 &+&\tfrac{1}{64}e\,&
 (\bar\epsilon\Gamma^{r_1r_2}\Gamma^m\Gamma^{r_3r_4}\psi_m)
 F_{r_1r_2}F_{r_3r_4} F^2 - \tfrac{1}{64}e\, (\epsilon\leftrightarrow
 \psi_m)\\[1ex]
 &+& \tfrac{1}{128}e\,& (\bar\epsilon\Gamma^{r_1r_2}\eom{\bar\chi}) F_{r_1r_2} F^2 \\[1ex]
 &+& \tfrac{1}{32} e\,& (\bar\psi^{r_2}\Gamma^{r_1}\Gamma^{r_3r_4}\epsilon)
 F_{r_1r_2} F_{r_3r_4} F^2 \\[1ex]
 &-& \tfrac{1}{4}e\,& (\bar\epsilon\Gamma^{r_1}D^{r_2}\chi)
 F_{r_1r_2} F^2 && \quad Y_1\\[1ex]
 &+& \tfrac{1}{8}e\, &(\bar\epsilon\Gamma^{mr_1r_2}\chi) D_m
 (F_{r_1r_2} F^2) && \quad X_7 \\[1ex]
%
%
(\alpha')^{-2} \delta {\cal L}^3_{\Gamma^{[3]}} =\,\, &-&\tfrac{1}{16}e\, &(\bar\epsilon
  \Gamma^{r_1r_2} \Gamma^{m} \Gamma^{r_3r_4} \psi_m ) F_{r_3r_4}
  F^3_{r_1r_2} + \tfrac{1}{16}e\,(\epsilon\leftrightarrow\psi_m)
  \\[1ex]
 &-& \tfrac{1}{32}e\,& (\bar\epsilon\Gamma^{r_1r_2}\eom{\bar\chi}) F^3_{r_1r_2}\\[1ex]
 &-& \tfrac{1}{8}e\, &
  (\bar\psi^{r_2}\Gamma^{r_1}\Gamma^{r_3r_4}\epsilon) 
 F_{r_1r_2}^3 F_{r_3r_4} \\[1ex]
 &+& e\, & (\bar\epsilon\Gamma^{r_1}D^{r_2}\chi)
  F^3_{r_1r_2} && \quad X_{12}\\[1ex]
 &-& \tfrac{1}{2}e\,& (\bar\epsilon\Gamma^{r_1r_2r_3} \chi) D_{r_1}
  F^3_{r_2r_3} && \quad X_6\\[1ex]
%
%
(\alpha')^{-2} \delta {\cal L}^4_{\Gamma^{[3]}} =\,\, 
  &-& e\, &(\bar\epsilon\Gamma^{r_1r_2r_3} D^{r_4}\chi)
  F_{r_1r_2}F^2_{r_3r_4} && \quad X_{11}\\[1ex]
 &-& e\,&(\bar\epsilon\Gamma^{r_1r_2r_3}\chi) 
 D^{r_4}( F_{r_1r_2}F^2_{r_3r_4}) && \quad X_8\\[1ex]
 &+& \tfrac{1}{8}e\,& (\bar\psi^{r_6}\Gamma^{r_1r_2r_3}\Gamma^{r_4r_5}\epsilon) 
 F_{r_1r_2} F_{r_4r_5}  F^2_{r_3r_6}\\[1ex]
%
%
(\alpha')^{-2} \delta {\cal L}_{\Gamma^{[5]}} =\,\, 
 &-& \tfrac{1}{8}e\,&(\bar\epsilon\Gamma^{r_1\cdots r_5} D^{r_6}\chi)
 F_{r_1r_6} F_{r_2r_3} F_{r_4r_5} && \quad Y_3\\[1ex]
 &-& \tfrac{1}{8}e\,& (\bar\epsilon\Gamma^{r_1\cdots
 r_5} \chi) D^{r_6} (F_{r_1r_6} F_{r_2r_3} F_{r_4r_5}) && \quad X_5\\[1ex]
 &-& \tfrac{1}{64}e\,& (\bar\psi^m \Gamma^{r_1\cdots
 r_5}\Gamma^{r_6r_7}\epsilon)
 F_{m r_1} F_{r_2r_3} F_{r_4r_5} F_{r_6r_7} && \quad Y_4\\[1ex]
%
%
(\alpha')^{-2} \delta {\cal L}_{\Gamma^{[7]}} =\,\, &-&\tfrac{1}{384}e\,&
 (\bar\epsilon\Gamma^{r_1\cdots r_6}\Gamma^m\Gamma^{r_7r_8} \psi_m) 
 F_{r_1r_2}\cdots F_{r_7r_8} + \tfrac{1}{384}e\,(\epsilon\leftrightarrow\psi_m)&& \\[1ex]
 &-& \tfrac{1}{768}e\,& (\bar\epsilon\Gamma^{r_1\cdots r_6}\eom{\bar\chi}) F_{r_1r_2} 
\cdots F_{r_5r_6} \\[1ex]
 &+& \tfrac{1}{64}e\,& (\bar\psi^m \Gamma^{r_1\cdots r_5}
 \Gamma^{r_6r_7}\epsilon) F_{m r_1}
 F_{r_2r_3}\cdots F_{r_6r_7} &&\quad Y_4\\[1ex]
 &+& \tfrac{1}{8}e\,& (\bar\epsilon\Gamma^{r_1\cdots r_5} D^{r_6}
 \chi) F_{r_1r_2}\cdots F_{r_5r_6}\, . && \quad Y_3
\end{aligned}
\end{equation}}
The last step consists of showing that the $\bar\psi\Gamma\epsilon$
terms cancel, which makes contact with the remaining unused
transformation rule, namely the one of the vielbein. Expanding all
the gamma-matrix products in the expression above, we obtain
in this final step the variations
\begin{equation}
\label{e:maxwellhorror3}
\begin{aligned}
%
%
(\alpha')^{-2} \delta {\cal L}^1_{\Gamma^{[0]}} \stackrel{\delta e}{\rightarrow}\,\, 
 &-&\tfrac{1}{16}e\, &(\bar\epsilon \Gamma^m \psi_m) ((F^2)^2- 4
F^4) && \quad Y_{12} \\[1ex]
&+& \tfrac{1}{2} e\,& (\bar\epsilon \Gamma^{r_1} \psi^{r_2}) F^2{}_{r_1r_2} F^2 &&
\quad Y_{10} \\[1ex]
&-& 2 e\,& (\bar\epsilon \Gamma^{r_1} \psi^{r_2}) F^4{}_{r_1r_2}
&&\quad Y_{11}  \\[1ex]
%
%
(\alpha')^{-2} \delta {\cal L}^2_{\Gamma^{[0]}} =\,\, &+&\tfrac{1}{192}e\,& (\bar\epsilon
\Gamma^{r_1\cdots r_9} \psi_{r_9}) F_{r_1r_2} \cdots F_{r_7r_8}
&&\quad B \\[1ex]
%
%
(\alpha')^{-2} \delta{\cal L}_{\Gamma^{[1]}}^2 \rightarrow\,\, &+&\tfrac{1}{32}e\, &(\bar\psi^{r_1}\Gamma^{r_2r_3r_4}\epsilon)
  F_{r_1r_2}F_{r_3r_4} F^2 && \quad Y_5\\[1ex]
 &+& \tfrac{1}{16}e\,& (\bar\psi^{r_1} \Gamma^{r_2} \epsilon)  F^2_{r_1r_2} F^2
 &&\quad Y_{10}\\[1ex]
%
%
(\alpha')^{-2} \delta {\cal L}_{\Gamma^{[1]}}^3 \rightarrow\,\, &-&\tfrac{3}{8}e\,& (\bar\psi^{r_1} \Gamma^{r_2r_3r_4} \epsilon)
  F^3_{r_1r_2} F_{r_3r_4} && \quad Y_6\\[1ex]
 &+& \tfrac{3}{4}e\, &(\bar\epsilon\Gamma^{r_1}\psi^{r_2}) F^4_{r_1r_2} &&
  \quad Y_{11}
\end{aligned}
\end{equation}
{\renewcommand{\theequation}{\ref{e:maxwellhorror3} cont.}
\addtocounter{equation}{-1}
\begin{equation}
\begin{aligned}
%
%
(\alpha')^{-2} \delta{\cal L}_{\Gamma^{[3]}}^2 \rightarrow\,\,
  &+&\tfrac{1}{32}e\, &(\bar\epsilon\Gamma^{r_1\cdots r_5}\psi_{r_5})
  F_{r_1r_2}F_{r_3r_4} F^2 &&\quad Y_7\\[1ex]
  &+& \tfrac{1}{32}e\, & (\bar\epsilon\Gamma^{r_1r_2r_3}\psi^{r_4})
  F_{r_1r_2}F_{r_3r_4}F^2 && \quad Y_5\\[1ex]
  &-& \tfrac{3}{16}e\,& (\bar\epsilon\Gamma^{r_1}\psi^{r_2})
  F^2_{r_1r_2} F^2 && \quad Y_{10}\\[1ex]
  &+& \tfrac{1}{16}e\,& (\bar\epsilon\Gamma^m\psi_m) (F^2)^2 &&
  \quad Y_{12} \\[1ex]
%
%
(\alpha')^{-2} \delta{\cal L}_{\Gamma^{[3]}}^3 \rightarrow\,\, 
  &-&\tfrac{1}{8}e\, & (\bar\epsilon\Gamma^{r_1\cdots r_5}\psi_{r_5})
  F_{r_3r_4}F^3_{r_1r_2} &&\quad Y_8\\[1ex]
  &-& \tfrac{1}{8}e\,& (\bar\epsilon\Gamma^{r_1r_2r_3}\psi^{r_4})
  F_{r_1r_2} F^3_{r_3r_4}&& \quad Y_6 \\[1ex]
  &+& \tfrac{3}{4}e\,& (\bar\epsilon\Gamma^{r_1}\psi^{r_2}) F^4_{r_1r_2} &&
  \quad Y_{11}\\[1ex]
  &-& \tfrac{1}{4}e\,&(\bar\epsilon\Gamma^m\psi_m) F^4 &&\quad Y_{12}\\[1ex]
%
%
(\alpha')^{-2} \delta{\cal L}_{\Gamma^{[3]}}^4 \rightarrow \,\,
  &-&\tfrac{1}{8}e\,&(\bar\epsilon\Gamma^{r_1\cdots r_5}\psi^{r_6})
  F_{r_1r_2}F_{r_3r_4} F^2_{r_5r_6} && \quad Y_9\\[1ex]
  &-& \tfrac{1}{4}e\,&(\bar\epsilon\Gamma^{r_1r_2r_3}\psi^{r_4}) F_{r_1r_2}
  F^3_{r_3r_4} && \quad Y_6\\[1ex]
  &+& \tfrac{1}{2}e\,&(\bar\epsilon\Gamma^{r_1}\psi^{r_2}) F^4_{r_1r_2} &&
  \quad Y_{11}\\[1ex]
  &-& \tfrac{1}{4}e\,&(\bar\epsilon\Gamma^{r_1}\psi^{r_2}) F^2_{r_1r_2} F^2
   && \quad Y_{10}\\[1ex]
%
%
(\alpha')^{-2} \delta{\cal L}_{\Gamma^{[7]}} \rightarrow \,\,
  &-&\tfrac{1}{192}e\,& (\bar\epsilon\Gamma^{r_1\cdots r_9}\psi_{r_9})
  F_{r_1r_2}\cdots F_{r_7r_8} && \quad B\\[1ex]
  &-& \tfrac{1}{32}e\,& (\bar\epsilon\Gamma^{r_1\cdots
  r_5}\psi_{r_5}) F_{r_1r_2}F_{r_3r_4} F^2 && \quad Y_7\\[1ex]
  &+& \tfrac{1}{8}e\,& (\bar\epsilon\Gamma^{r_1\cdots r_5}
  \psi_{r_5}) F_{r_1r_2}F^3_{r_3r_4} &&\quad Y_8\\[1ex]
  &+& \tfrac{1}{8}e\,& (\bar\epsilon\Gamma^{r_1\cdots r_5} 
  \psi^{r_6}) F_{r_1r_2}F_{r_3r_4} F^2_{r_5r_6}\, . && \quad Y_9
\end{aligned}
\end{equation}}
\end{subequations}
All terms cancel which proves supersymmetry of the action~\eqn{e:SYMstring}.

If we had not started with the input from string theory, but had tried to
obtain the fermionic terms directly using supersymmetry, things would
have been much more difficult. While it is definitely possible to find
the higher-order $\Gamma$ terms in~\eqn{e:expandedSYMstring}, the
lower-order terms are hard to guess due to the fact that there is no
$\teight{r}$ tensor present. On the other hand, obtaining the unexpanded
form of the action~\eqn{e:SYMstring} is also complicated, as it relies
on the subtle mixing between terms with four and terms with five powers
of the fields. Altogether, we have found the string input crucial to
understand the supersymmetric completion, even though supersymmetry has
played an essential role in finding the `non-standard' five-point 
contact terms.

\end{sectionunit}

\begin{sectionunit}
\title{Modifications to the transformation rules}
\maketitle
\label{s:SYMmodsusy}

We have so far not discussed the terms in the variation that are
proportional to the equations of motion. As was explained in
section~\ref{s:fieldredefinitions}, these terms can all be absorbed by
modifications to the field transformation rules at order
$(\alpha')^2$. For the globally supersymmetric action, one obtains
from~\eqn{e:maxwellhorror} that the new transformation rules are
\begin{equation}
\label{e:modtrafoYM}
\begin{aligned}
\delta A_\mu &= -4\,\bar\epsilon \Gamma_\mu \chi - (\alpha')^2\Big[ \tfrac{1}{4}(\bar\epsilon\Gamma_\mu\chi)
F^2 - (\bar\epsilon\Gamma^m\chi) F^2_{m\mu} 
- \tfrac{1}{8}(\bar\epsilon\Gamma^{r_1\cdots r_4}{}_\mu\chi) F_{r_1r_2}F_{r_3r_4}
\Big]\, ,\\[1ex]
\delta \chi &= \tfrac{1}{8}\Gamma^{\mu\nu}\epsilon F_{\mu\nu}
+ \tfrac{1}{768}(\alpha')^2\Big[ \big(\teight{r} \Gamma_{r_7r_8}\epsilon 
- \Gamma^{r_1\cdots r_6}\epsilon\big) F_{r_1r_2}F_{r_3r_4} F_{r_5r_6}\Big]\, .
\end{aligned}
\end{equation}
Using these modified transformation rules one can now verify that the
modified gaugino equation of motion is again supercovariant.  This
equation can be read off from the action~\eqn{e:expandedSYMstring} in
a straightforward way and one indeed finds that
\begin{equation}
\big(\delta^{(\alpha')^0} + \delta^{(\alpha')^2}\big) \frac{\delta
\big(S_{F^2}+S_{F^4}\big)}{\delta\bar\chi} = \text{independent of
$D\epsilon$}\, .
\end{equation}
This constitutes an independent check of our results and is required
for the equations of motion to sit in a supermultiplet.

We can now compute the commutator of two supersymmetry
transformations (though only on the gauge field; the commutator on the
gaugino requires knowledge about higher-order fermi terms in the
modified transformation rules). It turns out that the structure
coefficients of the algebra are unmodified as compared to the 
lowest-order ones:
\begin{equation}
{}\big[ \delta_{\epsilon_1}^{(\alpha')^0} + \delta_{\epsilon_1}^{(\alpha')^2},
\delta_{\epsilon_2}^{(\alpha')^0} + \delta_{\epsilon_2}^{(\alpha')^2} \big] A_\mu =
{}\big[ \delta_{\epsilon_1}^{(\alpha')^0}, \delta_{\epsilon_2}^{(\alpha')^0} \big] A_\mu +
{\cal O}\big((\alpha')^4\big)\, .
\end{equation}
Note that all of the modifications in~\eqn{e:modtrafoYM} are needed to
make the $(\alpha')^2$ contributions cancel.  The above result was
obtained previously by~\dcite{Metsaev:1987by} and generalized to the 
non-Abelian $t_8{\rm tr}(F^4)$ action by \dcite{Bergshoeff:1987jm} with
the same conclusions. When we extend their result to include the coupling 
to the gravitino background, \eqn{e:maxwellhorror} no longer produces 
equation-of-motion terms for the gaugino (because one no longer 
performs a partial integration of the $G_1+G_2$ terms). However, one finds 
that the same modifications to the gaugino transformation rules are now 
obtained from the equation-of-motion terms in~\eqn{e:maxwellhorror2} 
instead.  In order to determine the algebra, one would of course need to 
have information also about the gravity sector which have not addressed 
so far.

Before we proceed to the analysis of the supergravity invariants, let
us note that the fact that the structure coefficients of the
supersymmetry algebra are unchanged for super-Maxwell is compatible
with the observation that standard superspace (i.e.~superspace for
which the canonical dimension-zero torsion constraint
$T_{ab}{}^r=2\,(\Gamma^r)_{ab}$ is imposed) is sufficient to describe
the super-Born-Infeld action. This is most explicit in the
construction of the $N=1$ action in four dimensions, which can be done
in a power-series expansion in the standard vector superfield; see in
particular section~3 of \dcite{Bagger:1997wp}. Of course, an analogous
formulation in terms of an unconstrained superfield does not exist in
ten dimensions, but our results nevertheless show that the superspace
geometry does not have to be changed in this case either.

Let us now turn to the supergravity case.

\end{sectionunit}

\end{sectionunit}

\begin{sectionunit}
\title{Second step: completion of the $W^4$ action}
\maketitle
\label{s:W4complete}
\begin{sectionunit}
\title{From super-Maxwell to supergravity}
\maketitle 
\label{s:max2sugra}

In the previous section we derived the compact form \eqn{e:SYMstring}
of the leading higher-derivative string correction to the
super-Maxwell action in ten dimensions, including the fermion
bilinears required for supersymmetry and the coupling to the
supergravity background.  As should be clear from the lengthy
supersymmetry analysis, the information we used from string scattering
amplitudes was crucial to enable us to organise the terms in a
systematic way. Such additional string input is also helpful
in finding the supersymmetric completion of the higher-derivative
corrections to the supergravity action, although for this case we
fortunately do not have to start the calculations from scratch.  
In fact, with the string-inspired higher-derivative super-Maxwell
action~\eqn{e:SYMstring} at hand, it turns out that no additional
string analysis is actually needed to arrive at the supersymmetric
completion of the invariant $I_X$ due to the existence of a very
close formal similarity between the super-Maxwell supersymmetry
structure and its supergravity analogue.  As a consequence, most
results derived in the previous section can be mapped in a
straightforward way to the gravity case, with only a few
easy-to-handle exceptions; the details of the procedure are given
below.  Once the construction of the supergravity invariant for the
$N=1$ case is completed, we will comment on extensions to the type~II
theories and then show explicitly how to lift the analysis to eleven
dimensions.

The main ingredient in the construction of the supergravity invariant
for the $N=1$ model in ten dimensions is the strong parallel between
the super-Maxwell field transformation rules~\eqn{e:YMtrafo} and the
on-shell, lowest-order supergravity ones obtained
from~\eqn{e:hatpsi2_trafo} and~\eqn{e:deltaomega_full_new}, which we
redisplay here for the convenience of the reader in a form that
highlights the similarities:
\begin{equation}
\begin{aligned}
\label{e:trafo_comp}
\delta\chi &= \tfrac{1}{8}\Gamma^{\mu\nu}\epsilon \, F_{\mu\nu}\,, 
&\qquad\qquad   
\delta\psi_{rs} &= \tfrac{1}{8} \Gamma^{\mu\nu}\epsilon \, R_{\mu\nu rs}
+ \cdots \,,\\[1ex]
\delta F_{\mu\nu} &= - 8 D_{[\mu} (\bar\epsilon\Gamma_{\nu]}\chi) \,,  &
\delta R_{\mu\nu}{}^{rs} &= -4 D_{[\mu}(\bar\epsilon\Gamma_{\nu]}\psi^{rs}
-2\,\bar\epsilon\Gamma^{[r}\psi^{s]}{}_{\nu]}) + \cdots \,.
\end{aligned}
\end{equation}
(In addition, the transformation laws for the vielbein, the gravitino and
the two-form potential as given in~\eqn{e:SYMtrafos} apply in both cases.)
Crucial for the correspondence is also the fact that the super-Maxwell 
identity~\eqn{e:chimu_id} has a direct analogue in~\eqn{e:Dslashedpsi2} on 
the supergravity side:
\begin{equation}
\label{e:identity_comp}
\slashed{D}\chi = \tfrac{1}{8} \Gamma^m \Gamma^{\mu\nu} \psi_m\,F_{\mu\nu} 
+ \cdots\,, \qquad
\slashed{D}\psi_{rs} = \tfrac{1}{8}\Gamma^m\Gamma^{\mu\nu}\psi_m\,
R_{\mu\nu rs} + \cdots \,.
\end{equation}
In both~\eqn{e:trafo_comp} and~\eqn{e:identity_comp} the dots indicate 
terms of higher order in the fermionic fields and/or terms proportional to 
the lowest-order equations of motion. For later use, it is important
to observe that the supergravity results quoted above are dimension
independent, essentially because of the fact that we did not include
the dilaton or any gauge-field dependent terms. More information can
be found in appendix~\ref{s:sugra_appendix}.

In the transition from the Maxwell case to gravity, one is led by the above
equations to make the tentative substitutions 
\begin{equation}
\label{e:SYM2SUGRA}
\begin{aligned}
F_{r_1r_2}  &\rightarrow R_{r_1r_2s_1s_2} \, ,\\[1ex]
\chi        &\rightarrow \psi_{s_1s_2} \, ,\\[1ex]
D_r\chi &\rightarrow D_r\psi_{s_1s_2}\,, \\[1ex]
\end{aligned}
\end{equation}
in the action~\eqn{e:SYMstring}, while inserting at the same time an
additional $\teight{s}$ tensor to saturate the extra vector indices
introduced in the process.  The origin of this `trick'---which indeed
turns out to be most useful in spite of the imperfect match between
the transformation rules for the gauge-field strength and the Riemann
tensor---becomes evident when one compares the string amplitudes
involving the super-Maxwell multiplet with those involving the
supergravity fields, as we do in the next subsection.

However, let us first explain in more detail how the differences between 
the super-Maxwell and the supergravity cases arise from a supersymmetry
perspective. One source for these differences is the mismatch between the 
gauge-potential and the spin-connection transformation rules, carrying 
over to the transformation rules for the corresponding field
strengths (in other words, the map $F_{r_1r_2}\rightarrow R_{r_1r_2s_1s_2}$ does not 
commute with supersymmetry). Rewriting, trivially, the Riemann tensor 
transformation rule in~\eqn{e:trafo_comp} as
\begin{equation}
\label{e:riem_trafo_split}
\begin{aligned}
\delta R_{r_1r_2}{}^{s_1s_2} = \,\,&- 8 D_{[r_1}(\bar\epsilon\Gamma_{r_2]}\psi^{s_1s_2}) \\[1ex]
 &+ 4 D_{[r_1}(\bar\epsilon\Gamma_{r_2]}\psi^{s_1s_2}
+2\,\bar\epsilon\Gamma^{[s_1}\psi^{s_2]}{}_{r_2]}) + \cdots \,,
\end{aligned}
\end{equation}
the first line is what the transformation rule would have looked like, 
had the naive matching with the super-Maxwell case been sufficient. 
Putting it differently, given the mapping~\eqn{e:SYM2SUGRA} and the
accompanying introduction of $\teight{s}$ applied to the higher-derivative
super-Maxwell action~\eqn{e:SYMstring}, this part of the transformation 
acting on the $R^4$ term thus generated will conspire with the variations of
all the remaining terms in the `naive' action to produce a vanishing result.  
We are then faced with the puzzle that there is nothing left to cancel 
the contribution from the second line of~\eqn{e:riem_trafo_split} to the
variation of the $R^4$ term, which after partial integration can be written
as:
\begin{equation}
\label{e:SUGRA_var_remainder}
6\cdot32\,(\alpha')^{-3}\Delta \defas 48\, e\,\teight{r}\teight{s}\, 
(\bar\epsilon\Gamma_{r_7}\psi_{s_7s_8} 
-2\,\bar\epsilon\Gamma_{s_8}\psi_{r_7s_7})\,(D_{r_8}R_{r_1r_2s_1s_2})
R_{r_3r_4s_3s_4}R_{r_5r_6s_5s_6}\,.
\end{equation}
A key observation that points towards the resolution of this puzzle, is that 
the second term above exhibits non-trivial mixing between the left-moving
$r$-indices and the right-moving $s$-indices. This fact, in turn, leads
us to examine the effects of such mixing on the naive gravitino bilinears; 
indeed, using the cyclic identity~\eqn{e:cyclicpsi2} on the gravitino 
curvature, one can show that 
\begin{equation}
\begin{aligned}
\delta\Big[ \alpha_1\,&e\, \teight{r} \teight{s}
(\bar\psi_{r_7}\Gamma_{r_8}\psi_{s_7s_8}) 
R_{r_1r_2s_1s_2} \cdots R_{r_5 r_6 s_5 s_6} +
2\alpha_2\, (r_8\leftrightarrow s_7) \Big] \\[1.5ex]
&= +\tfrac{1}{8}(\alpha_1+\alpha_2) e\, \teight{r} 
\teight{s} (\bar\psi_{r_7}\Gamma_{r_8}{}^{mn}\epsilon)
R_{r_1r_2s_1s_2}\cdots R_{s_7s_8mn} \\[1ex]
&\quad + \tfrac{1}{4}(\alpha_1+\alpha_2)e\, \teight{r} \teight{s} (\bar\epsilon\Gamma^n \psi_{r_7})
R_{n r_8 s_7s_8}R_{r_1r_2s_1s_2}\cdots R_{r_5r_6s_5s_6}\\[1ex]
&\quad + (\alpha_1+\alpha_2)e\, \teight{r} \teight{s}
(\bar\epsilon\Gamma_{r_7} D_{r_8}\psi_{s_7s_8}) R_{r_1r_2s_1s_2}\cdots
R_{r_5r_6s_5s_6} \\[1ex]
&\quad +3\alpha_1 e\, \teight{r} \teight{s} (\bar\epsilon
\Gamma_{r_7} \psi_{s_7s_8}) (D_{r_8} R_{r_1r_2s_1s_2})
R_{r_3r_4s_3s_4}
R_{r_5r_6s_5s_6} \\[1ex]
&\quad -6\alpha_2 e\, \teight{r} \teight{s} (\bar\epsilon
\Gamma_{s_8} \psi_{r_7s_7}) (D_{r_8} R_{r_1r_2s_1s_2})
R_{r_3r_4s_3s_4}
R_{r_5r_6s_5s_6}\, .
\end{aligned}
\end{equation}
Comparison with~\eqn{e:SUGRA_var_remainder} immediately shows that adding 
to the naive action terms of the above kind with relative weight such that 
$\alpha_1+\alpha_2=0$ produces a variation of precisely the kind that
we are looking to cancel. In particular, choosing the coefficients of these 
additional terms as $\alpha_1=-\tfrac{1}{32}(\alpha')^3{\cdot}\tfrac{8}{3}$ 
and $2\alpha_2=\tfrac{1}{32}(\alpha')^3{\cdot}\tfrac{16}{3}$, respectively,
leads to the supersymmetric completion of the $R^4$ invariant.

However, before we are ready to down write down the result, we need to
address the second point on which the super-Maxwell and the
supergravity cases differ. Since it is more natural from a string
theory perspective, we would like to present the action in a form in
which all terms proportional to the lowest-order equations of motion
have been subtracted. This amounts to making the substitution
\begin{equation}
\label{e:Riemann_subst}
R_{mn}{}^{pq} \rightarrow W_{mn}{}^{pq} - \frac{16}{d-2}
\delta_{[m}{}^{[p}(\bar\psi_{|r|}\Gamma^{|r|}\psi_{n]}{}^{q]}
- \bar\psi^{|r|}\Gamma^{q]}\psi_{n]r}) 
\end{equation}
wherever the Riemann tensor occurs in the action. Here we again made
use of the results~\eqn{e:RfromEOM} and~\eqn{e:RiccifromEOM} of
appendix~\ref{s:sugra_appendix}.  From~\eqn{e:Riemann_subst} it
immediately follows that the extraction of equation-of-motion terms
is, with one exception, achieved by simply replacing the Riemann
tensor by the Weyl tensor. The sole exception is the $R^4$ term, from
which a new term bilinear in the gravitino appears (strictly speaking
one would also get a new gravitino bilinear from the $BR^4$ term, but
that one does not contribute to any of the variations we considered,
so it falls outside the scope of our analysis).  This term, given on
the second line of the action below, lacks a partner on the
super-Maxwell side.  Let us also mention that when subtracting the
equation-of-motion terms we implicitly used the fact that the
supersymmetry variation of these terms is again proportional to the
equations of motion (see section 2.3 of \dcite{berg1} for more on this
issue).

Finally, we are in a position to write down the supersymmetric
completion of the $N=1$ invariant $I_X$ in ten dimensions:
\begin{equation}
\label{e:SUGRAstring}
\begin{aligned}
S_{X} = \frac{(\alpha')^3}{32}\int\!{\rm d}^{10}x \Big[ & \tfrac{1}{6}
e\,\teight{r}\teight{s} W_{r_1r_2s_1s_2}\cdots W_{r_7r_8s_7s_8}
 + \tfrac{1}{12} \varepsilon_{10}^{(r)} \teight{s} B_{r_9r_{10}} W_{r_1r_2s_1s_2}
 \cdots W_{r_7r_8 s_7s_8} \\[1ex]
&\quad + \tfrac{4}{3} e\,\teight{r}\teight{s} \eta_{r_8s_8} (\bar\psi^m
\Gamma_{r_7} \psi_{s_7m}) W_{r_1r_2s_1s_2}\cdots W_{r_5r_6s_5s_6}\\[1ex]
&\quad - \tfrac{32}{5} e\,\teight{r}\teight{s} \eta_{r_2r_3} (\bar\psi_{s_1s_2}
\Gamma_{r_1} D_{r_4}\psi_{s_3s_4})  W_{r_5r_6s_5s_6}W_{r_7r_8s_7s_8} \\[1ex]
&\quad +\tfrac{12{\cdot}32}{5}e\,\teight{s}
(\bar\psi_{s_1s_2}\Gamma_{r_1}D_{r_2}\psi_{s_3s_4})
W_{r_1m s_5s_6}W^m{}_{r_2 s_7s_8}\\[1ex]
 &\quad -\tfrac{16}{5!} \varepsilon_{10}^{(r)} \teight{s}
 (\bar\psi_{s_1s_2} \Gamma_{r_1\cdots r_5} D_{r_6}\psi_{s_3s_4}) 
 W_{r_7r_8s_5s_6} W_{r_9r_{10}s_7s_8} \\[1ex]
&\quad + \tfrac{8}{3}e\,\teight{r}\teight{s} (\bar\psi_{r_1} \Gamma_{r_2}\psi_{s_1s_2})
W_{r_3r_4s_3s_4} \cdots W_{r_7r_8s_7s_8}\\[1ex]
&\quad + \tfrac{16}{3}e\,\teight{r}\teight{s} (\bar\psi_{r_1} \Gamma_{s_1}\psi_{r_2s_2})
W_{r_3r_4s_3s_4} \cdots W_{r_7r_8s_7s_8}\\[1ex]
&\quad +
\tfrac{8}{3} e\,\teight{s}(\bar\psi^{m}\Gamma_{mr_1\cdots r_6} \psi_{s_7s_8}) W_{r_1
r_2s_1s_2}\cdots W_{r_5r_6s_5s_6} \Big]\, ,
\end{aligned}
\end{equation}
Notice the term on the penultimate line, which introduces left-right
mixing in the action. Note also that the coefficient of the term on
the preceding line is different from the naive super-Maxwell value.

An alternative, perhaps more elegant way to resolve the discrepancies
between the super-Maxwell transformation rules and those of the
supergravity multiplet can be found in the literature (see
e.g.~\dcite{cai1}, {}~\dcite{Gross:1987mw}, \dcite{Bellucci:1988ff}
and~\dcite{berg8}), although it is strictly bound to ten dimensions
(which is why we did not adopt it above). It consists of using a
spin connection with $H$-torsion, which is defined in the string frame
as
\begin{equation}
\Omega_{\mu\,rs} = \omega_{\mu\, rs} + \hat H_{\mu rs}
 = \omega_{\mu\,rs} + H_{\mu rs} - 3\, \bar\psi_{[\mu}\Gamma_{r}\psi_{s]}\, .
\end{equation}
While one usually considers this shift to incorporate the effects of
the field strength $H$ in the action, the fermionic terms are also
important.  The above combination, in contrast to the spin connection
itself, transforms in a nice way under supersymmetry:
\begin{equation}
\delta\Omega_{\mu}{}^{rs} = -4\bar\epsilon\Gamma_\mu\hat\psi^{rs}\, .
\end{equation}
It is easy to see that by using $W(\Omega)$ instead of $W(\omega)$ in
the bosonic part of the action, one generates upon expansion precisely
the fermionic terms with left-right mixed indices (the term linear in
the field strength $H$ drops out) and one obtains the naive
generalisation of the super-Maxwell action. The above argument does,
unfortunately, not have an obvious generalisation that applies in
arbitrary dimensions. 

In analogy with the discussion of the super-Maxwell action, let us expand 
the $\teight{r}$ on the left-moving side to arrive at the analogue of 
\eqn{e:expandedSYMstring} (all indices have been lowered for esthetical reasons):
\begin{equation}
\label{e:expandedSUGRAstring}
\begin{aligned}
%
%
(\alpha')^{-3} {\cal L}_{\Gamma^{[0]}} =\,\,
&-&\tfrac{1}{32}e\,\teight{s}\,& 
\big( (W^2)^2 - 4 W^4 \big) \\[1ex]
&+& \tfrac{1}{384}&\varepsilon^{t_1t_2r_1\cdots r_8}
\teight{s} B_{t_1t_2}
W_{r_1r_2s_1s_2}\cdots W_{r_7r_8s_7s_8}\, , \\[1ex]
%
%
(\alpha')^{-3} {\cal L}_{\Gamma^{[1]}} =\,\, &-&4e\,\teight{s} 
&(\bar\psi_{s_1s_2} \Gamma_{r_1} D_{r_2} \psi_{s_3s_4})
W_{r_1r_3s_5s_6}W_{r_3r_2s_7s_8} &\quad&  \\[1ex]
&-& \tfrac{1}{4}e\,\teight{s} &(\bar\psi_{r_1}\Gamma_{r_2}\psi_{s_7s_8}) 
W_{r_1r_2s_1s_2} W_{mns_3s_4}W_{nms_5s_6} & & \\[1ex]
&-& e\,\teight{s}& (\bar\psi_{r_1}\Gamma_{r_2}\psi_{s_7s_8}) 
W_{r_1ms_1s_2}W_{mns_3s_4}W_{nr_2s_5s_6} \\[1ex]
 &+& e\,\teight{s} &(\bar\psi_{r_1}\Gamma_{s_7}\psi_{r_2s_8}) W_{r_1r_2s_1s_2}
W_{mns_3s_4}W_{nms_5s_6} \\[1ex]
 &-& 4e\,\teight{s}&(\bar\psi_{r_1}\Gamma_{s_7}\psi_{r_2s_8}) 
W_{r_1ms_1s_2}W_{mns_3s_4}W_{nr_2s_5s_6} \\[1ex]
&+&\tfrac{1}{4} e\,\teight{s} &(\bar\psi_m\Gamma_n\psi_{ms_8}) 
W_{pqs_1s_2}W_{qps_3s_4}W_{ns_7s_5s_6} && (+\tfrac{2}{d-2}) \\[1ex]
&-& e\,\teight{s} &(\bar\psi_m\Gamma_n\psi_{ms_8}) 
W_{nps_1s_2}W_{pqs_3s_4}W_{qs_7s_5s_6} \, , && (-\tfrac{8}{d-2})\\[1ex]
%
%
(\alpha')^{-3} {\cal L}_{\Gamma^{[3]}} =\,\, &+&2e\, \teight{s} &(\bar\psi_{s_5s_6}\Gamma_{r_1r_2r_3} D_{r_4}
\psi_{s_7s_8}) W_{r_1r_2s_1s_2}W_{r_3r_4s_3s_4}  \\[1ex]
 &-& \tfrac{1}{8}e\,\teight{s} &(\bar\psi_m \Gamma_{mr_1r_2}\psi_{s_7s_8})
W_{r_1r_2s_1s_2} W_{pns_3s_4}W_{nps_5s_6} & &  \\[1ex] 
 &+& \tfrac{1}{2}e\, \teight{s} &(\bar\psi_m \Gamma_{mr_1r_2} \psi_{s_7s_8}) 
W_{r_1ps_1s_2}W_{pns_3s_4}W_{nr_2s_5s_6} \\[1ex]
 &+&  e\,\teight{s} &(\bar\psi_m \Gamma_{r_1r_2r_3}\psi_{s_7s_8}) 
W_{r_1r_2s_1s_2} W_{mn s_3s_4}W_{nr_3 s_5s_6}\, , \\[1ex]
%
%
(\alpha')^{-3} {\cal L}_{\Gamma^{[5]}} =\,\,  &+&\tfrac{1}{8}e \,
  \teight{s} &(\bar\psi_{r_6} \Gamma_{r_1\cdots r_5}\psi_{s_7s_8})
  W_{r_1r_2s_1s_2}W_{r_3r_4s_3s_4} W_{r_5r_6s_5s_6}  \, , \\[1ex]
%
%
(\alpha')^{-3} {\cal L}_{\Gamma^{[7]}} =\,\,  &+&\tfrac{1}{48}e \,
  \teight{s}&(\bar\psi_m \Gamma_{m r_1\cdots r_6}\psi_{s_7s_8})
  W_{r_1r_2s_1s_2}W_{r_3r_4s_3s_4} W_{r_5r_6s_5s_6}  \, . 
\end{aligned}
\end{equation}
Written in this form, all direct dependence of the action on the
space-time dimension can be made explicit, as we indeed have done in
the rightmost column. This makes it suitable for lifting to eleven
dimensions, an issue we will return to in section~\ref{s:lifting}.
Next we will, however, discuss the string-theory origin of the terms
in~\eqn{e:SUGRAstring}.

\end{sectionunit}

\begin{sectionunit}
\title{String-amplitude calculations and extensions to type II theories}
\maketitle

We have intentionally not started with a string analysis for the
gravity case, as there are several additional tricky elements there as
compared to the super-Maxwell situation. Nevertheless, the
action~\eqn{e:SUGRAstring} can be understood also from a string point
of view, as we will explain in the present section. In addition, this
allows us to make a few comments about possible generalisations of
the superinvariant to the type IIA and IIB theories, although we are
at this moment not able to give a detailed discussion of those cases.
Readers who are more interested in the lifting procedure to eleven
dimensions can skip this section and continue with
section~\ref{s:lifting}.

Let us first discuss the vertex operators for the gravity case.  We
will focus on the type IIA/IIB theories as those are slightly more
complicated, but the heterotic results can be extracted without too
much difficulty. Again, because the explicit forms of vertex operators
in various pictures are scattered in the literature, we list here the
vertex operators which we need. In the NS$\otimes$NS sector we have
the graviton, dilaton and two-form operators, combined into
\begin{align}
V^{(0,0)}_g(k) &=\int\!{\rm d}^2 z\, \zeta_{\mu\nu} \norder{(i\partial X^\mu + \tfrac{\alpha'}{2} k\cdot  \Psi  \,  \Psi^\mu)\times
(i\bar\partial X^\nu + \tfrac{\alpha'}{2} k\cdot \tilde \Psi  \, \tilde \Psi^\nu)
e^{ik\cdot X}} \, ,\\[1ex]
V^{(-1,-1)}_g(k) &= \int\!{\rm d}^2 z\, \zeta_{\mu\nu} \norder{\Psi^\mu \tilde \Psi^\nu e^{-\phi-\tilde
  \phi} e^{ik\cdot X}} \, .
\end{align}
The $\zeta_{\mu\nu}$ denotes the polarisation, while $X^\mu$ and
$\Psi^\mu$ are the usual world-sheet bosons and fermions.
In the R$\otimes$NS sector we have the operators for the gravitino,
\begin{align}
V^{(-1/2,0)}_{\psi}(k) &= \int\!{\rm d}^2 z\, \bar\psi_\nu \norder{S  e^{-\phi/2}
(i\bar\partial X^\nu + \tfrac{\alpha'}{2} k\cdot \tilde \Psi  \, \tilde \Psi^\nu)
e^{ik\cdot X}}\, , \\[1ex]
V^{(1/2,0)}_{\psi}(k) &= \int\!{\rm d}^2 z\, \bar\psi_\nu \Gamma_\mu
\norder{S  e^{\phi/2} (i\partial X^\mu + \tfrac{\alpha'}{2} k\cdot  \Psi  \,  \Psi^\mu)
(i\bar\partial X^\nu + \tfrac{\alpha'}{2} k\cdot \tilde \Psi  \, \tilde \Psi^\nu)
e^{ik\cdot X}} \, .
\end{align}
As with the super-Maxwell vertex operators~\eqn{e:SYMvertexoperators}, we have
ignored any overall normalisation factors.
In order for these vertex operators to have the right conformal
dimension to be primary fields, the polarisation tensors have to
satisfy the conditions $\Gamma^\mu \psi_\mu=0$ as well as $k^\mu
\zeta_{\mu\nu}=0$. 

\begin{table}
\smaller\smaller
\begin{align*}
\intertext{\bf even/even spin structure:}
\left.\begin{aligned}
{}& V_g^{(0,0)} &&\quad V_g^{(0,0)} &&\quad V_g^{(0,0)} &&\quad V_g^{(0,0)} \\
{}& \norder{k\Psi\Psi} &&\quad \norder{k\Psi\Psi} &&\quad
\norder{k\Psi\Psi} 
&&\quad \norder{k\Psi\Psi} \\[1ex]
{}& \norder{k\tilde\Psi\tilde\Psi} &&\quad \norder{k\tilde\Psi\tilde\Psi} &&\quad
\norder{k\tilde\Psi\tilde\Psi} &&\quad \norder{k\tilde\Psi\tilde\Psi}
\end{aligned}\quad\right\}& \quad t_8 t_8 W^4\\[2ex]
\left.\begin{aligned}
{}& T_F &&\quad V_g^{(0,0)} && &&\quad V_\psi^{(-1/2,0)}\!\! &&
&&\quad V_\psi^{(-1/2,0)}\!\! && &&\quad V_g^{(0,0)} \\
{}& \partial X \Psi && \quad \norder{k\Psi\Psi} && 
\stackrel{\Gamma^{[1]}S^+\Psi^2}{\longleftrightarrow} && 
\quad\quad S^+ && \stackrel{\Gamma^{[1]}\Psi}{\longleftrightarrow}
&& \quad\quad S^+ && \stackrel{\Gamma^{[1]}S^+\Psi^2}{\longleftrightarrow}
&& \quad \norder{k\Psi\Psi}  \\
{} & && \quad\norder{k\tilde\Psi\tilde\Psi} && &&\quad \norder{k\tilde\Psi\tilde\Psi}&& &&\quad
\norder{k\tilde\Psi\tilde\Psi} && &&\quad \norder{k\tilde\Psi\tilde\Psi}
\end{aligned}\quad\right\}& \quad \teight{r}\teight{s}
(\bar\psi_{ss} \Gamma_{r} \eta_{rr} D_r \psi_{ss}) W_{rrss} W_{rrss}\\[2ex]
\left.\begin{aligned}
{}& V_\psi^{(-1/2,0)} && &&\quad V_\psi^{(-1/2,0)}&&T_F &&\quad V_g^{(0,0)} &&\quad V_g^{(0,0)} \\
{}&   S^+   && \stackrel{{\Gamma^{[1]} \Psi}}{\longleftrightarrow}
&&\norder{S^+\exp(i kX)}&\leftrightarrow&\partial X \Psi && \quad\norder{k\Psi\Psi} && \quad \norder{k\Psi\Psi} \\
{}& \norder{k\tilde\Psi\tilde\Psi} && &&\quad \norder{k\tilde\Psi\tilde\Psi} && &&\quad
\norder{k\tilde\Psi\tilde\Psi} &&\quad \norder{k\tilde\Psi\tilde\Psi}
\end{aligned}\quad\right\}& \quad \teight{s}(\bar\psi_{ss} \Gamma_m D_n \psi_{ss})
W_{mpss}W_{pnss}\\[2ex]
\left.\begin{aligned}
{}& T_F && \quad V_\psi^{(-1/2,0)} && &&\quad V_\psi^{(-1/2,0)}
&&\quad V_g^{(0,0)} &&\quad V_g^{(0,0)}  &&\quad V_g^{(0,0)}\\
{}&\partial X\Psi && \quad S^+ &&
\stackrel{\Gamma^{[1]}\Psi}{\longleftrightarrow} 
&&\quad  S^+ &&\quad \norder{k\Psi \Psi} 
&&\quad \norder{k\Psi\Psi} &&\quad \norder{k\Psi\Psi}\\
{}& && \quad \bar\partial X && && \quad\norder{k\tilde\Psi\tilde\Psi} &&\quad \norder{k\tilde\Psi\tilde\Psi} &&\quad
\norder{k\tilde\Psi\tilde\Psi} &&\quad \norder{k\tilde\Psi\tilde\Psi}
\end{aligned}\quad\right\}& \quad \teight{r}\teight{s}(\bar\psi_r
\Gamma_r \psi_{ss}) W_{rrss}W_{rrss}W_{rrss}\\[3ex]
\intertext{\bf odd/even spin structure:}
\left.\begin{aligned}
{}& V_\psi^{(-1/2,0)} && &&\quad V_\psi^{(-1/2,0)} && && \quad T_F &&\quad V_g^{(0,0)}
&& \quad V_g^{(0,0)}\\
{}& S^+ && \stackrel{\Gamma^{[5]}\Psi^5}{\longleftrightarrow} && \quad
:S^+\,\exp(ikX): &&\leftrightarrow &&\, \partial X\Psi  &&\quad
\norder{k\Psi\Psi} && \quad \norder{k\Psi\Psi} \\
{}& \norder{k\tilde\Psi\tilde\Psi} && &&\quad \norder{k\tilde\Psi\tilde\Psi} && && &&\quad
\norder{k\tilde\Psi\tilde\Psi} &&\quad \norder{k\tilde\Psi\tilde\Psi}
\end{aligned}\quad\right\}& \quad \varepsilon^{(r)}_{10} \teight{s} (\bar\psi_{ss}
\Gamma_{rrrrr} D_r\psi_{ss}) W_{rrss} W_{rrss} \\[2ex]
\left.\begin{aligned}
{}& V_B^{(-1,0)} && \quad T_F &&\quad V_g^{(0,0)} &&\quad
V_g^{(0,0)} &&\quad V_g^{(0,0)}
&& \quad V_g^{(0,0)}\\
{}& \Psi && \quad {\partial X \Psi}  &&\quad \norder{k\Psi\Psi} &&\quad \norder{k\Psi\Psi}
&&\quad \norder{k\Psi\Psi} && \quad \norder{k\Psi\Psi} \\
{}& \bar\partial X && && \quad \norder{k\tilde\Psi\tilde\Psi} &&\quad \norder{k\tilde\Psi\tilde\Psi} &&\quad
\norder{k\tilde\Psi\tilde\Psi} &&\quad \norder{k\tilde\Psi\tilde\Psi}
\end{aligned}\quad\right\}& \quad \varepsilon_{10} B \teight{s} W^4  \\[2ex]
\left.\begin{aligned}
{}& T_F &&\quad V_\psi^{(-1/2,0)} && &&\quad V_\psi^{(-1/2,0)} &&\quad V_g^{(0,0)}
&& \quad V_g^{(0,0)} && \quad V_g^{(0,0)} \\
{}& {\partial X \Psi} && \quad S^+ && \stackrel{\Gamma^{[3]}\Psi^3}{\longleftrightarrow} && \quad
S^+ &&\quad \norder{k\Psi\Psi} && \quad \norder{k\Psi\Psi} &&\quad \norder{k\Psi\Psi}\\
{}& && \quad \bar\partial X && &&\quad \norder{k\tilde\Psi\tilde\Psi} &&
\quad\norder{k\tilde\Psi\tilde\Psi} &&\quad
\norder{k\tilde\Psi\tilde\Psi}  &&\quad 
\norder{k\tilde\Psi\tilde\Psi}
\end{aligned}\quad\right\}& \quad \varepsilon^{(r)}_{10} \teight{s} (\bar\psi_{r}
\Gamma_{rrr} \psi_{ss}) W_{rrss} W_{rrss} W_{rrss}\\[2ex]
\intertext{\bf odd/odd spin structure:}
\left.\begin{aligned}
{}& T_F &&\quad V_g^{(-1,-1)} &&\quad V_g^{(0,0)} &&\quad V_g^{(0,0)}
&&\quad V_g^{(0,0)} && \quad V_g^{(0,0)}\\
{}& \partial X\Psi && \quad \Psi &&\quad \norder{k\Psi\Psi} &&\quad \norder{k\Psi\Psi}
&&\quad \norder{k\Psi\Psi} && \quad \norder{k\Psi\Psi} \\
{}& \bar\partial X\tilde\Psi && \quad \tilde\Psi
&& \quad \norder{k\tilde\Psi\tilde\Psi}  &&\quad \norder{k\tilde\Psi\tilde\Psi} &&\quad
\norder{k\tilde\Psi\tilde\Psi} &&\quad \norder{k\tilde\Psi\tilde\Psi}
\end{aligned}\quad\right\} & \quad \varepsilon_{10} \varepsilon_{10} W^4
\end{align*}
\caption{The vertex operator products for the IIA string leading to
the supersymmetric completion of the $t_8 t_8 W^4$ effective action.
In the heterotic model, the $t_8$ tensors on the right-moving side
arise from $(\bar\partial X e^{ikX})^4$ instead of
$(k\tilde\Psi\tilde\Psi)^4$, while the odd/odd term is absent.}
\label{t:sugravops}
\end{table}

Using these operators, one arrives at table~\ref{t:sugravops}. Observe
that there is now one more way to produce a five-point function in the
odd/odd spin-structure sector. As shown in table~\ref{t:sugravops}, in
the odd/odd spin structure one graviton-vertex operator has to be
taken in the $(-1,-1)$ ghost picture, say the fifth state, while the 
other four are in the $(0,0)$ ghost picture. The resulting tensorial 
structure is
\begin{equation}
\label{e:epsepstrick}
\varepsilon_{10}^{\mu\kappa r_1\cdots r_8}
\varepsilon_{10}^{\nu\lambda s_1\cdots s_8} \times
\eta_{\mu\nu} \times \zeta^{(5)}_{\kappa\lambda} \times
\prod_{i=1}^4 \zeta_{r_{2i-1}s_{2i-1}}^{(i)}
k^{(i)}_{r_{2i}} k^{(i)}_{s_{2i}}\ .
\end{equation}
Here, the $\eta_{\mu\nu}$ tensor arises from the contraction of the
$\partial X^\mu$ with $\bar\partial X^\nu$ from the supercurrent.
Furthermore, $\zeta^{(5)}$ is the polarisation of the fifth graviton
and is seen as the fluctuation of the metric around the flat
background $g_{\kappa\lambda} \sim \eta_{\kappa\lambda} + \kappa_{10}
\zeta_{\kappa\lambda}$. Therefore, \eqn{e:epsepstrick} is the
linearised version of $\varepsilon_{10}^{\mu\kappa r_1\cdots r_8}
\varepsilon_{10}^{\nu\lambda s_1\cdots s_8}\,
g_{\mu\nu}g_{\kappa\lambda}\, R^4$.  We should remark that the double
epsilon term does not seem to appear in the heterotic theory, as the
relevant operator product does not exist. An alternative way to check
the absence of this term is to compute the relative normalisation of
the $t_8 t_8 R^4$ and $\varepsilon_{10} t_8 B R^4$ terms along the
lines of~\ddcite{Lerche:1987sg}{Lerche:1988qk}, and verify that it
corresponds to the combination appearing in the $I_X$ invariant.
\end{sectionunit}

\begin{sectionunit}
\title{Lifting to eleven dimensions}
\maketitle
\label{s:lifting}

After the string interlude of the previous section, let us get
back to the field theory construction of supersymmetric invariants and
discuss how to lift the results obtained in section~\ref{s:max2sugra}
to eleven dimensions. As we shall see, this is essentially straightforward 
at the level of the purely gravitational terms (i.e.~including just the 
graviton and gravitino) and the anomaly cancellation term. 

The supersymmetry transformation rules which we have used so far are
shared between the ten- and eleven-dimensional theories (differences
only come in when the gauge-field and dilaton parts of the theory are
considered). Therefore, the supersymmetry analysis performed in
section~\ref{s:max2sugra} remains valid provided one takes proper
care of explicit dimension dependence arising from contracted
Kronecker deltas. As we have already remarked below~\eqn{e:maxwellhorror}, 
one never encounters such explicit dimension factors in the variation
starting from the super-Maxwell $F^4$ action~\eqn{e:SYMstring} 
(there are, for instance, no contracted delta symbols arising from the 
expansion of gamma-matrix products).

A first attempt at constructing the eleven-dimensional version of the
invariant could be to try to lift the compact form~\eqn{e:SUGRAstring}. 
However, it is easy to see that such a naive approach is bound to fail. 
The reason is that there is a term in~\eqn{e:SUGRAstring} with index
contractions within a single $t_8$ tensor, hiding an explicit dependence
on the spacetime dimension (see~\eqn{e:t8_definition}). To be precise, 
the third line of~\eqn{e:SUGRAstring} contains the tensor $\eta_{r_2r_3}$, 
stemming from a product $[\Gamma^{[2]},\Gamma^{[1]}]$ in the string
amplitude analysis (cf.~\eqn{e:etagamma1}). When verifying supersymmetry 
invariance (which, as we have seen in section~\ref{s:maxwellsusy}, was 
done on the expanded action~\eqn{e:expandedSUGRAstring}), these trace 
terms come out with the right coefficients only in ten dimensions. 
Moreover, it is not difficult to see that it is impossible to write 
down a generalised $t_8$ tensor in eleven dimensions with the same 
symmetry properties as~\eqn{e:t8_definition} that reproduces these
coefficients.

Having made this observation, the remedy suggests itself immediately:
instead of trying to preserve the full $t_8$ structure of the
ten-dimensional action, we should lift the expanded
action~\eqn{e:expandedSUGRAstring}; this action contains left-over
$t_8$ tensors of the right-moving sector, but does not involve any
Kronecker delta contractions.  The only dimension dependence of this
action is, as we have already alluded to, related to the subtraction
of the equation-of-motion terms (cf.~the discussion
below~\eqn{e:Riemann_subst}). For these terms, however, the dimension
dependence cancels in the variation, as should be clear from their
origin.  As a result, the lifting procedure is essentially trivial in
the gravitational sector, including the replacement of the two-form in
ten dimensions with the three-form gauge field in eleven dimensions.

The net result reads (again with all indices lowered)
\begin{equation}
\label{e:elevendimI_X}
\begin{aligned}
%
%
(\alpha_M')^{-3} {\cal L}_{\Gamma^{[0]}} =\,\, 
&+&\tfrac{1}{192}&e\,\teight{r}\teight{s} W_{r_1r_2s_1s_2} W_{r_3r_4s_3s_4}
  W_{r_5r_6s_5s_6} W_{r_7r_8s_7s_8} \\[1ex]
&+& \tfrac{1}{(48)^2}&\varepsilon^{t_1t_2t_3r_1\cdots r_8}
\teight{s} B_{t_1t_2t_3}
W_{r_1r_2s_1s_2}\cdots W_{r_7r_8s_7s_8}\, , \\[1ex]
%
%
(\alpha_M')^{-3} {\cal L}_{\Gamma^{[1]}} =\,\, &-&4e\,\teight{s} 
&(\bar\psi_{s_1s_2} \Gamma_{r_1} D_{r_2} \psi_{s_3s_4})
W_{r_1ms_5s_6}W_{mr_2s_7s_8} &\quad&  \\[1ex]
&-& \tfrac{1}{4}e\,\teight{s} &(\bar\psi_{r_1}\Gamma_{r_2}\psi_{s_7s_8}) 
W_{r_1r_2s_1s_2} W_{mns_3s_4}W_{nms_5s_6} & & \\[1ex]
&-& e\,\teight{s}& (\bar\psi_{r_1}\Gamma_{r_2}\psi_{s_7s_8}) 
W_{r_1ms_1s_2}W_{mns_3s_4}W_{nr_2s_5s_6} \\[1ex]
 &+& e\,\teight{s} &(\bar\psi_{r_1}\Gamma_{s_7}\psi_{r_2s_8}) W_{r_1r_2s_1s_2}
W_{mns_3s_4}W_{nms_5s_6} \\[1ex]
 &-& 4e\,\teight{s}&(\bar\psi_{r_1}\Gamma_{s_7}\psi_{r_2s_8}) 
W_{r_1ms_1s_2}W_{mns_3s_4}W_{nr_2s_5s_6} \\[1ex]
&+&\tfrac{2}{9} e\,\teight{s} &(\bar\psi_m\Gamma_n\psi_{ms_8}) 
W_{pqs_1s_2}W_{qps_3s_4}W_{ns_7s_5s_6} \\[1ex]
&-& \tfrac{8}{9} e\,\teight{s} &(\bar\psi_m\Gamma_n\psi_{ms_8}) 
W_{nps_1s_2}W_{pqs_3s_4}W_{qs_7s_5s_6} \, , 
\end{aligned}
\end{equation}
{\renewcommand{\theequation}{\ref{e:elevendimI_X} cont.}
\addtocounter{equation}{-1}
\begin{equation}
\begin{aligned}
%
%
(\alpha_M')^{-3} {\cal L}_{\Gamma^{[3]}} =\,\, &+&2e\, \teight{s} &(\bar\psi_{s_5s_6}\Gamma_{r_1r_2r_3} D_{r_4}
\psi_{s_7s_8}) W_{r_1r_2s_1s_2}W_{r_3r_4s_3s_4}  \\[1ex]
 &-& \tfrac{1}{8}e\,\teight{s} &(\bar\psi_m \Gamma_{mr_1r_2}\psi_{s_7s_8})
W_{r_1r_2s_1s_2} W_{pns_3s_4}W_{nps_5s_6} & &  \\[1ex] 
 &+& \tfrac{1}{2}e\, \teight{s} &(\bar\psi_m \Gamma_{mr_1r_2} \psi_{s_7s_8}) 
W_{r_1ps_1s_2}W_{pns_3s_4}W_{nr_2s_5s_6} \\[1ex]
 &+&  e\,\teight{s} &(\bar\psi_m \Gamma_{r_1r_2r_3}\psi_{s_7s_8}) 
W_{r_1r_2s_1s_2} W_{r_3n s_3s_4}W_{nm s_5s_6}\, , \\[1ex]
%
%
(\alpha_M')^{-3} {\cal L}_{\Gamma^{[5]}} =\,\,  &+&\tfrac{1}{8}e \,
  \teight{s} &(\bar\psi_{r_6} \Gamma_{r_1\cdots r_5}\psi_{s_7s_8})
  W_{r_1r_2s_1s_2}W_{r_3r_4s_3s_4} W_{r_5r_6s_5s_6}  \, , \\[1ex]
%
%
(\alpha_M')^{-3} {\cal L}_{\Gamma^{[7]}} =\,\,  &+&\tfrac{1}{48}e \,
  \teight{s}&(\bar\psi_m \Gamma_{m r_1\cdots r_6}\psi_{s_7s_8})
  W_{r_1r_2s_1s_2}W_{r_3r_4s_3s_4} W_{r_5r_6s_5s_6}  \, . 
\end{aligned}
\end{equation}}
Here the dimensionful parameter $\alpha_M'$ is defined in terms of the 
eleven-dimensional Planck length by $(\alpha_M')^3=4\pi (l_P)^6$.
The coefficients $2/9$ and $-8/9$ arise from the dimension-dependent
replacement of the Riemann tensor with the Weyl tensor plus fermion
bilinears as explained below~\eqn{e:Riemann_subst}.

Perhaps one could have expected a result along these lines.
Obviously, there is no left-right splitting principle in eleven
dimensions, and much of the symmetry between the left-moving `$r$'
indices and right-moving `$s$' ones possessed by the ten-dimensional
action is indeed lost in~\eqn{e:elevendimI_X}.  It would be
interesting to see whether the tensorial structure
of~\eqn{e:elevendimI_X} has a natural explanation in terms of a
light-cone supermembrane calculation. However, we will not dwell on
this issue any further and instead continue with a discussion of the
modifications to the supersymmetry transformation rules and the
resulting supersymmetry algebra.

\end{sectionunit}

\begin{sectionunit}
\title{Superspace constraints from modified supersymmetry rules}
\maketitle
\label{s:modsusyrules}

In contrast to the higher-derivative \emph{actions}, which as we have seen  
share many structural features between the super-Maxwell and the 
supergravity models, the modifications to the \emph{transformation rules} 
are rather different in the two cases. This is due to the fact that
the formal analogy that led us to the substitution trick described in 
section~\ref{s:max2sugra} does not extend to the equations of motion,
as is evident from a comparison of \eqn{e:SYMeoms} and \eqn{e:sugraeoms}.  
For this reason, the conclusion that the supersymmetry algebra does not 
receive any modifications when the super-Maxwell $(\alpha')^2$ corrections 
are taken into account (see section~\ref{s:SYMmodsusy}) cannot simply be 
taken over to the supergravity case. Instead, these modifications have to 
be recomputed from scratch. However, before we do so in the next section, 
let us explain which of the modifications are actually of interest from 
a superspace point of view.

In order to identify the relevant modifications to the transformation 
rules, we need to examine the structure of the supersymmetry algebra.
Specifically, for the Cremmer-Julia-Scherk action, the commutator of 
two supersymmetry transformations has the schematic structure
(the ${\cal S}$ tensor is defined in~\eqn{e:calSdef})
\begin{equation}
\begin{aligned}
\label{e:11dalgebra}
{}[\delta_{\epsilon_1},\delta_{\epsilon_2}] =\,\,& 
\delta^{\text{translation}}(2\,\bar\epsilon_2\Gamma^\nu \epsilon_1) + 
\delta^{\text{susy}}(-2\,\bar\epsilon_2\Gamma^\nu \epsilon_1\,\psi_\nu) \\[1ex]
& + \delta^{\text{gauge}} ( - 4\,\bar\epsilon_2\Gamma^\sigma \epsilon_1  
B_{\sigma\nu\rho} -2\bar\epsilon_2\Gamma_{\nu\rho}\epsilon_1) \\[1ex] 
& + \delta^{\text{Lorentz}}(2\,\bar\epsilon_2\Gamma^\nu \epsilon_1\,\hat
\omega_\nu{}^r{}_t + 4\,\bar\epsilon_2 {\cal S}^{r\; \nu\rho\sigma\kappa}{}_t 
\epsilon_1 \hat H_{\nu\rho\sigma\kappa})\, .
\end{aligned}
\end{equation}
When corrections at higher order in $\alpha'_M$ to the transformations on
the left-hand side are taken into account, the parameters on the right-hand 
side can receive field-dependent modifications at the corresponding order.
In particular, we will focus on the corrections to the translation 
parameter for reasons that will become clear below.

Since our analysis allows us to compute the modifications to the 
transformation rules only to lowest order in the fermions, and since we have
not considered variations proportional to the gauge field, we are 
only able to compute the algebra on the vielbein. For this case,
the translation parameter can be identified by focussing on the part
\begin{equation}
[\delta_{\epsilon_1},\delta_{\epsilon_2}]\,e_\mu{}^r = 
  \partial_\mu\xi^\nu\,e_\nu{}^r + \cdots 
\end{equation}
of the $\alpha'_M$-corrected version of~\eqn{e:11dalgebra}. 
The expansion of the left-hand side in powers of $\alpha'_M$ reads
\begin{equation}
\Big( \tfrac{1}{2}\big[ \delta_{\epsilon_1}^{(\alpha'_M)^0},
\delta_{\epsilon_2}^{(\alpha'_M)^0}\big] 
+ \delta_{\epsilon_1}^{(\alpha'_M)^0} \delta_{\epsilon_2}^{(\alpha'_M)^3} 
- \delta_{\epsilon_2}^{(\alpha'_M)^3} \delta_{\epsilon_1}^{(\alpha'_M)^0}
- (\epsilon_1\leftrightarrow\epsilon_2) \Big)  e_\mu{}^r + {\cal O}\big((\alpha'_M)^6\big)\, .
\end{equation}
Inserting the lowest-order transformation rules, the corrections at order
$(\alpha'_M)^3$ can thus be written as
\begin{equation}
\label{e:algsimplify}
\delta_{\epsilon_2}^{(\alpha'_M)^3}e_\mu{}^r\Big|_{\begin{aligned}
&{}\scriptstyle \psi_\lambda\rightarrow D_\lambda\epsilon_1 \\
&{}\scriptstyle \psi_{mn}\rightarrow \tfrac{1}{8}\Gamma^{pq}\epsilon_1 W_{pq mn}\end{aligned}}
 - 2\bar\epsilon_1 \Gamma^r
\delta_{\epsilon_2}^{(\alpha'_M)^3}\psi_\mu - (\epsilon_1\leftrightarrow\epsilon_2)\, .
\end{equation}
All terms in this result which have the $\mu$-index sitting on either
a gamma matrix or a Weyl tensor can be interpreted as modifications to
the local Lorentz transformation parameter on the right-hand side of
the supersymmetry algebra. Instead, we only have to look for terms which have
the $\mu$-index sitting on a (covariant) derivative.\footnote{Of
  course, there could in principle be subtleties involving the Ricci
  cyclic identity so one has to be careful to also check for terms
  containing $D_.W_{\mu...}$; however, it turns out that no such terms
  appear.}  Looking at~\eqn{e:algsimplify}, these terms can arise in
three ways: from terms in $\delta^{(\alpha'_M)^3}e_\mu{}^r$ which have
the $\mu$-index on the gravitino, from terms in this transformation
rule which are proportional to $\psi_{(2)}$ and have the $\mu$-index
on the derivative, and finally from terms in
$\delta^{(\alpha'_M)^3}\psi_\mu$ where the $\mu$-index sits on the
derivative. 

Restricting to these three classes of terms, the analysis already
becomes much simpler. However, even among the terms that fall under
one of these categories, not all correspond to non-trivial
modifications of the translation parameter.  We shall explain how to
isolate those who do shortly, but first we need to make contact with
superspace.
 
Starting from our component-field results, the link to superspace is made
as follows (see also~\dcite{kas_thesis}). When embedding the component
theory in superspace, one makes a gauge choice for the lowest
components of the superfields. The most natural one is
\begin{equation}
\label{e:supervielbein_gauge}
\begin{aligned}
E_\mu{}^r &= e_\mu{}^r + {\cal O}(\theta)\, ,\\
E_\nu{}^a &= \psi_\nu{}^a + {\cal O}(\theta)\, ,\\
E_\alpha{}^a &= \delta_\alpha{}^a \, ,\\
E_\alpha{}^r &= {\cal O}(\theta)\, ,
\end{aligned}\quad\text{implying}\quad
\begin{aligned}
E_r{}^\mu    &= e_r{}^\mu + {\cal O}(\theta)\, ,\\
E_r{}^\alpha &= -\psi_r{}^\alpha + {\cal O}(\theta)\, ,\\
E_a{}^\mu    &= {\cal O}(\theta)\, ,\\
E_a{}^\alpha &= \delta_a{}^\alpha \, .
\end{aligned}
\end{equation}
The torsion is defined in terms of the supervielbeine and superconnections by
\begin{equation}
\label{e:torsiondef}
T_{AB}{}^C = (-)^{M(B+N)} E_A{}^M E_B{}^N \Big( {\cal D}_M E_N{}^C - (-)^{MN}
{\cal D}_N E_M{}^C \Big)\, ,
\end{equation}
where the local Lorentz covariant derivatives are defined as
\begin{equation}
\begin{aligned}
{\cal D}_M E_N{}^r &= \partial_M E_N{}^r + \Omega_M{}^r{}_s E_N{}^s\,
,\\
{\cal D}_M E_N{}^a &= \partial_M E_N{}^a + \tfrac{1}{4}(-)^{bN} \Omega_{Mrs}
(\Gamma^{rs})^a{}_b \, E_N{}^b\, .
\end{aligned}
\end{equation}
Imposing the gauge~\eqn{e:supervielbein_gauge}, the dimension-zero
torsion component in \eqn{e:torsiondef} reduces at $\theta=0$ to
\begin{equation}
\label{e:elvisisdead}
T_{ab}{}^r\Big|_{\text{gauge of~\eqn{e:supervielbein_gauge}}} = 
-2\, \delta_a{}^\alpha \delta_b{}^\beta \, \partial_{(\alpha}
E_{\beta)}{}^r \, .
\end{equation}
It now only remains to find out how the modified supersymmetry algebra 
determines the order-$\theta$ term of the vielbein component appearing 
in~\eqn{e:elvisisdead}.

This component of the supervielbein can be found by using a procedure
called gauge completion (used previously in e.g.~\dcite{crem3}
and~\dcite{deWit:1998tk} to determine the theta expansion of the
superfields for the standard Cremmer-Julia-Scherk theory). This
procedure relates the transformation rules of the component fields (on
the left-hand side of the equation below) to the transformation rules
of the superfields (on the right-hand side). Explicitly, one has
\begin{equation}
\label{e:vielbeincompletion}
\delta_{\text{components}} E_M{}^A = 
\Xi^N \partial_N E_M{}^A + \partial_M \Xi^N E_N{}^A 
  + \tfrac{1}{2}(-)^{BM}(\Lambda^{rs}X_{rs})^A{}_B E_M{}^B\, ,
\end{equation}
where $\tfrac{1}{2}(\Lambda^{rs}X_{rs})^m{}_n=\lambda^m{}_n$ and $\tfrac{1}{2}
(\Lambda^{rs}X_{rs})^a{}_b=\tfrac{1}{4}\lambda^{rs} (\Gamma_{rs})^a{}_b$
while the spinor-vector components vanish.
Writing out the summations in bosonic and fermionic parts, one first finds 
that the $\partial_\mu \epsilon$ pieces cancel, as expected. Denoting the
remaining part of the gravitino component transformation rule by $\hat\delta$,
one obtains the following characterisation of the order-$\theta$ components:
\begin{equation}
\label{e:gaugechoice_implications}
\left.
\begin{aligned}
(\bar\theta \hat\delta) \psi_\mu{}^a &= E_\mu{}^a\Big|_{\theta}\, ,\\
(\bar\theta \delta) e_\mu{}^r        &= E_\mu{}^r\Big|_{\theta}\, ,\\
\Xi^\beta\Big|_{\theta} &= - \Xi^\nu\Big|_{\theta} \, \psi_\nu{}^\beta\, \\
\epsilon^\beta \partial_\beta E_\alpha{}^r\Big|_{\theta} &= 
-\partial_\alpha \Xi^\nu\Big|_{\theta} e_\nu{}^r 
\end{aligned}
\right\} \text{in gauge~\eqn{e:supervielbein_gauge}}
\end{equation}
(here $(\bar\theta\delta)$ in the first two lines denote the respective
supersymmetry transformations with the supersymmetry parameter 
$\epsilon^\alpha$ substituted by $\theta^\alpha$). 
From the fourth line one observes that the problem of determining
the dimension-zero constraint is now reduced to finding the 
superspace translation parameter~$\Xi^\nu$.
This parameter is defined in superspace by the relation
\begin{equation}
\Xi_3^\nu\Big|_{\theta=0} = 
\epsilon^\alpha_2 \partial_\alpha \Xi^\nu_1\Big|_{\theta} - 
\epsilon^\alpha_1 \partial_\alpha \Xi^\nu_2\Big|_{\theta} \, .
\end{equation}
Whenever the left-hand side picks up $\alpha'_M$ corrections, the above
equation implies that these corrections will also be visible in the
first-order $\theta$ component of the transformation parameter~$\Xi^\nu$. 
Hence, we find that knowledge about the supercharge 
commutator---represented by $\Xi_3^\nu$---is sufficient to fix the 
dimension-zero component $T_{ab}{}^r$ of the supertorsion. This
component is of particular importance in the superspace formulation
of eleven-dimensional supergravity, where the constraint imposed on it 
completely determines the dynamics of the theory via the superspace
Bianchi identities; we shall have reason to come back to this point in 
the following. 

Although our primary interest, for the reason just discussed, lies in 
determining the dimension-zero component $T_{ab}{}^r$, we should mention 
that our gauge-completion approach also allows us to find explicit
manifestations of the interdependence of the various torsion components 
expressed by the superspace Bianchi identities. For instance, in the 
chosen gauge we obtain
\begin{equation}
\label{e:dimhalf_torsion}
T_{ar}{}^s = \delta_a{}^\alpha e_r{}^\mu \partial_\alpha E_\mu{}^s
  \Big|_{\theta^1} - \psi_r{}^b\,T_{ab}{}^s \,.
\end{equation}
The second equation in~\eqn{e:gaugechoice_implications} then allows us
to write
\begin{equation}
T_{ar}{}^s = e_r{}^\mu\,(\tilde\delta e_\mu{}^s)_a - \psi_r{}^b\,T_{ab}{}^s\,,
\end{equation}
where $\tilde\delta$ is defined by 
$\delta e_\mu{}^s = \epsilon^a(\tilde\delta e_\mu{}^s)_a$. For this
dimension-1/2 component of the torsion we thus see that $\alpha'_M$ 
corrections will result as a consequence of such corrections to the 
vielbein transformation rule as well as to the dimension-zero torsion
constraint.

Before we turn to the issue of how the torsion constraints are
affected by the inclusion of the higher-derivative
action~\eqn{e:elevendimI_X}, we need to discuss how to identify
corrections that cannot be set to zero by field redefinitions.  For
example, not all potential corrections to the dimension-zero torsion
component $T_{ab}{}^r$ are non-trivial, as explained
in~\ddcite{Cederwall:2000ye}{Cederwall:2000fj} (see also references
therein).  In eleven dimensions, the most general expression for this
constraint reads
\begin{equation}
\label{e:generaltorsion}
T_{ab}{}^r = 2 \big( ({\cal C}\Gamma^{r_1})_{ab} \, X^r{}_{r_1} + 
  \tfrac{1}{2!}({\cal C}\Gamma^{r_1r_2})_{ab}\, X^r{}_{r_1r_2} +
  \tfrac{1}{5!}({\cal C}\Gamma^{r_1\cdots r_5})_{ab}\, X^r{}_{r_1\cdots r_5} 
  \big)\,.
\end{equation}
As was argued in~\cite{Cederwall:2000ye,Cederwall:2000fj}, the interesting 
physics is contained in the coefficients $X^r{}_{r_1r_2}$ and 
$X^r{}_{r_1\cdots r_5}$. At this stage, the former is not relevant for us
as it is necessarily zero within our gauge-field independent analysis. 
As far as the latter coefficient is concerned, the only 
SO(10,1)-irreducible component contained therein which can not be redefined 
away is the `fish-hook' one, $\tilde{\tableau{2 1 1 1 1}}$, of dimension 
$4290$; the two other possibilities---the antisymmetric tensors 
$\tableau{1 1 1 1 1 1}$ and $\tableau{1 1 1 1}$---can be 
made to vanish by a superfield redefinition, 
which in our language corresponds to an appropriate change of the gauge
choice~\eqn{e:supervielbein_gauge} for the supervielbein.
Indeed, the symmetrised product of three Weyl-tensor representations 
$\bf1144$ ($\tilde{\tableau{2 2}}$) does contain $\bf4290$, so 
$(\alpha'_M)^3\,W^3$ corrections could in principle appear.  

The upshot of this section is that, in order to find non-trivial
modifications to the supertorsion component $T_{ab}{}^r$, we have 
to study corrections to the supertranslation parameter in the algebra 
which involve a factor $\bar\epsilon_2\Gamma^{[5]}\epsilon_1$. All other
superspace modifications are of lesser importance and can be ignored.
\end{sectionunit}

\begin{sectionunit}
\title{Absence of corrections to the supersymmetry algebra}
\maketitle
\label{s:modtorsion}

Although our method will eventually produce the modifications to the 
supertorsion constraints along the lines sketched in the previous
section, we are unfortunately rather severely restricted by the fact
that we have not yet included the gauge-field terms in the superinvariant. 
As we will show in this section, there are no non-trivial modifications 
to the supersymmetry algebra when one only considers the part of the 
superinvariant given by~\eqn{e:elevendimI_X}. We will comment on possible 
reasons for and implications of the absence of these corrections.
However, let us begin by discussing in general terms which variations 
are responsible for the equation-of-motion terms, in particular those 
of the kind that would give rise to modifications of the supersymmetry
algebra.
 
As far as the vielbein equation-of-motion terms are concerned, they
can, broadly speaking, be generated in two different ways: either by the 
split of a Riemann tensor in Weyl-tensor plus Ricci parts, or from 
additional, explicit Ricci terms that appear in the $\slashed{D}\psi_{(2)}$ 
identity~\eqn{e:Dslashedpsi2} and the $D\cdot \psi_{(2)}$ 
identity~\eqn{e:Ddotpsi2} as well as in the contracted 
Bianchi identity $D _{[\mu} R_{\nu\rho]}{}^{rs}\,e_r{}^\mu=0$
(the Ricci terms should in all cases be subsituted by their 
on-shell expressions given in \eqn{e:RfromEOM}--\eqn{e:RiccifromEOM}). 
We shall return to the role of the contracted Bianchi identity in
the variation of the action shortly and instead first discuss the
$\slashed{D}\psi_{(2)}$ and $D\cdot \psi_{(2)}$ identities, which
are also the only sources for gravitino equation-of-motion terms. 
Upon insertion of~\eqn{e:RfromEOM}--\eqn{e:RiccifromEOM} they read:
\begin{equation}
\label{e:Dslashedpsi2_rewritten}
\begin{aligned}
\slashed{D}\psi_{\mu\nu} =\,\,& \tfrac{1}{8}\Gamma^m\Gamma^{rs} W_{\mu\nu
rs} \psi_m \\[1ex]
& + \frac{1}{(d-1)(d-2)}\Gamma_{m\mu\nu}\psi^m\,\eom{e}_{\lambda}{}^{\lambda}
-\frac{1}{d-2}\Gamma_{m[\mu}{}^s\psi^m\, \eom{e}_{\nu]s} \\[1ex]
{}\quad
& +\frac{1}{d-2}\bigg(\Gamma_{[\mu}\psi^m + (d-3)\Gamma^m\psi_{[\mu}\bigg)\,
\eom{e}_{\nu]m}
+ \frac{d-3}{(d-1)(d-2)}\Gamma_{[\mu}\psi_{\nu]}\,\eom{e}_{\lambda}{}^{\lambda}
\\[1ex]
&+ D_{[\mu}\widetilde{\eom{\bar\psi}}_{\nu]} + {\cal O}(H^2) +
{\cal O}(\psi^3) \,,
\end{aligned}
\end{equation}
\begin{equation}
\label{e:Ddotpsi2_rewritten}
\begin{aligned}
D^\nu\psi_{\mu\nu} =\,\,& \tfrac{1}{8}\Gamma^{rs}\psi^\nu W_{\mu\nu rs} \\[1ex]
& - \frac{1}{2(d-2)}\Gamma_{\mu\nu}\psi^r\,\eom{e}_r{}^\nu
  - \frac{d-3}{2(d-2)}\Gamma^{rs}\psi_r\,\eom{e}_{\mu s} 
  - \frac{1}{2} \psi^r\,\eom{e}_{\mu r} \\[1ex]
& - \frac{d-3}{2(d-1)(d-2)}\Gamma_{\mu\nu}\psi^\nu\,\eom{e}_\lambda{}^\lambda
  + \frac{1}{4} D_\mu(\Gamma\cdot\widetilde{\eom{\bar\psi}})
  - \frac{1}{2} \slashed{D}\,\widetilde{\eom{\bar\psi}}_\mu  + {\cal O}(H^2) 
  + {\cal O}(\psi^3) \,.
\end{aligned}
\end{equation}
Note that in both of the above identities some of the vielbein
equation-of-motion terms come from expanded Riemann tensors
(cf.~\eqn{e:Dslashedpsi2}--\eqn{e:Ddotpsi2}).
Additionally, further vielbein equation-of-motion terms appear after
expansion of the Riemann tensor in the supersymmetry transformation rule 
for the gravitino curvature~\eqn{e:hatpsi2_trafo}:
\begin{equation}
\label{e:hatpsi2_trafo_rewritten}
\begin{aligned}
\delta \psi_{\mu\nu} =\,\,& \tfrac{1}{8}W_{\mu\nu mn}\Gamma^{mn}\epsilon + 
\frac{1}{d-2}\Big[ \Gamma^n{}_{[\mu}\epsilon\,\eom{e}_{\nu]n}
+\frac{1}{d-1} \Gamma_{\mu\nu}\epsilon\, \eom{e}_\lambda{}^\lambda\Big]
+ {\cal O}(F^2) + {\cal O}(\psi^2)\, .
\end{aligned}
\end{equation}
In the above equations one has to be careful with the index order on
$\eom{e}$; the second index corresponds to the curved index of the
vielbein (cf.~\eqn{e:graviton_eom}).

A quick inspection of the variation of the eleven-dimensional
action~\eqn{e:elevendimI_X} shows that, when we
use~\eqn{e:Dslashedpsi2_rewritten}--\eqn{e:hatpsi2_trafo_rewritten} to
isolate the equation-of-motion terms, there is a plethora of terms
which have to be cancelled by modifying the supersymmetry
transformation rules. For the gravitino one can collect these
modifications in a reasonably compact form (see~\eqn{e:modpsirule}
below), but the result for the vielbein has an order of magnitude more
terms. Fortunately, as explained in the previous section the interesting 
physics is contained in a limited subset of these terms, so we do not 
have to consider them all.

With the identities~\eqn{e:Dslashedpsi2_rewritten} 
and~\eqn{e:Ddotpsi2_rewritten} at hand, the full set of 
gravitino equation-of-motion terms in the variation of the 
action~\eqn{e:elevendimI_X} is readily determined to be\footnote{Although 
we are discussing the eleven-dimensional case, here and in the following
we keep the spacetime dimension $d$ as a parameter as this helps us to
keep track of the origin of various terms.}
\begin{equation}
\label{e:psieomterms}
\begin{aligned}
(\alpha'_M)^{-3}&\delta{\cal L}\big|_{\eom{\bar\psi}} = \,\,
  \bareom{\bar\psi}{}^\mu\,e\,\teight{s}\bigg[ 
-\tfrac{1}{24} \tilde P_{\mu s_8}\Gamma^{r_1\cdots r_6} D_{s_7} 
  (\epsilon W_{r_1r_2s_1s_2}\cdots W_{r_5r_6s_5s_6}) \\[1ex]
&+\tfrac{1}{24}\, \teight{r} \,\tilde P_{\mu s_8} \Gamma_{r_7r_8} D_{s_7}
  (\epsilon W_{r_1r_2s_1s_2}\cdots W_{r_5r_6s_5s_6}) \\[1ex]
&+\tfrac{1}{3(d-2)}\, \teight{r}\, \tilde P_{\mu s_8}\Gamma_m\Gamma_{r_8}
  \eta_{r_7s_7} D_m (\epsilon W_{r_1r_2s_1s_2}\cdots W_{r_5r_6s_5s_6}) \\[1ex]
~ &-\tfrac{1}{6(d-2)} \, \teight{r} \, \tilde P_{\mu m} \Gamma_m \Gamma_{r_8} 
  \eta_{r_7s_7} D_{s_8} (\epsilon W_{r_1r_2s_1s_2}\cdots W_{r_5r_6s_5s_6}) \\[1ex]
&+\tfrac{1}{3(d-2)}\teight{r}\big(\tilde P_{\mu r_7} \eta_{ms_7}\eta_{r_8s_8} 
    \epsilon
  -\tfrac{1}{d-2} \tilde P_{\mu m} \eta_{r_7s_7}\eta_{r_8s_8} \epsilon
  \big) D_m (\epsilon W_{r_1r_2s_1s_2}\cdots W_{r_5r_6s_5s_6})\bigg] \,.
\end{aligned}
\end{equation}
In computing the above result one has to vary the $W^4-\psi^2$ action
and cannot get away with a simple variation of the $R^4$ action only,
since the two differ by equation-of-motion terms and the variations
will similarly differ by such terms. 
The complete compensating gravitino transformation rule at order 
$(\alpha'_M)^3$ readily follows:
\begin{equation}
\label{e:modpsirule}
\begin{aligned}
\delta\psi_\mu =&\, D_\mu\epsilon + (\alpha'_M)^3\,\teight{s} \bigg[ 
  \tfrac{1}{24}\tilde P_{\mu s_8}
\Gamma^{r_1\cdots r_6} D_{s_7}\big( \epsilon\, W_{r_1r_2s_1s_2} \cdots
W_{r_5r_6s_5s_6}\big) \\[1ex]
 & -\tfrac{1}{24}\, \teight{r} \tilde P_{\mu s_8} \Gamma_{r_7r_8} D_{s_7}\big(
\epsilon\, W_{r_1r_2s_1s_2}\cdots W_{r_5r_6s_5s_6}\big)\\[1ex]
 & -\tfrac{1}{3(d-2)}\, \teight{r} \tilde P_{\mu s_8} \Gamma_m \Gamma_{r_8} 
\eta_{r_7s_7} D_{m} \big( \epsilon\, W_{r_1r_2s_1s_2}\cdots W_{r_5r_6s_5s_6}\big)\\[1ex]
 & +\tfrac{1}{6(d-2)} \,\teight{r} \tilde P_{\mu m} \Gamma_m \Gamma_{r_8}
\eta_{r_7s_7} D_{s_8} \big( \epsilon\, W_{r_1r_2s_1s_2}\cdots W_{r_5r_6s_5s_6}\big)\\[1ex]
 & +\tfrac{1}{3(d-2)} \teight{r} \Big(
\eta_{r_8s_7}\eta_{r_7m}\tilde P_{\mu s_8} \epsilon 
+
\tfrac{1}{d-1}\eta_{r_7s_7}\eta_{r_8s_8}\tilde P_{\mu m}\epsilon
\Big) D_m\big(W_{r_1r_2s_1s_2}\cdots W_{r_5r_6s_5s_6}\big)\bigg] \, .
\end{aligned}
\end{equation}
where we used a shorthand notation for (see~\eqn{e:alternative_grino_eom})
\begin{equation}
\tilde P_{mn} \defas \frac{1}{2(d-2)}\big(\Gamma_{mn} - (d-3)\eta_{mn}\big)\, .
\end{equation}
The above result can in principle be checked in an independent way by
verifying that the modified equation of motion for the gravitino is
indeed supercovariant under the transformation
rule~\eqn{e:modpsirule}, but this turns out to be considerably more
difficult than the analogous check which we performed in
section~\ref{s:SYMmodsusy} for the super-Maxwell theory (consequently,
we have only verified supercovariance for the highest-order 
gamma-matrix terms). 

Recall from the previous section that the $\delta^{(\alpha'_M)^3}\psi_\mu$
terms that modify the supersymmetry algebra are those where the $\mu$-index
sits on the covariant derivative. Although there are such terms contained
in the third and fifth lines of~\eqn{e:modpsirule}, none of these are
proportional to either $\Gamma^{[4]}$ or $\Gamma^{[6]}$, which are
the only ones that would contribute to the $\Gamma^{[5]}$ component 
of the translation parameter. Hence, \eqn{e:modpsirule} does not lead to 
any non-trivial modification to the supersymmetry algebra.

Turning to the vielbein equation-of-motion terms, we again focus on
terms that modify the $\Gamma^{[5]}$ component of the translation 
parameter. As discussed in the previous section, these correspond to
terms in $\delta^{(\alpha'_M)^3}e_\mu{}^r$ that either have 
the $\mu$-index on the gravitino or contain $\psi_{(2)}$ and have the
$\mu$-index on a covariant derivative.  
Only terms of the former kind can potentially be generated by the
equation-of-motion terms 
in~\eqn{e:Dslashedpsi2_rewritten}--\eqn{e:hatpsi2_trafo_rewritten}.
There is indeed such a term, namely
\begin{equation}
\label{e:gamma7_contrib}
(\alpha'_M)^{-3} \,\delta {\cal L}_{\Gamma[7]} \rightarrow \, 
  -\frac{1}{4(d-2)}\,e\,\teight{s}\,(\bar\epsilon\Gamma^{r_1\cdots r_5}
  \psi_\mu)\,\delta^{r_6}_{s_7}\, W_{r_1r_2s_1s_2}\cdots W_{r_5r_6s_5s_6}\,
  \eom{e}_{s_8}{}^\mu \, .
\end{equation}
Cancellation of this term requires the transformation rule to be
modified according to
\begin{equation}
\label{e:L7mod}
\delta e_\mu{}^r\Big|_{\text{from \eqn{e:gamma7_contrib}}} = +\frac{(\alpha'_M)^3}{4(d-2)}\,\teight{s}\,(\bar\epsilon
  \Gamma^{r_1\cdots r_5}\psi_\mu)\,\delta^{r_6r}_{s_7s_8}\,
  W_{r_1r_2s_1s_2}\ldots W_{r_5r_6s_5s_6} \,,
\end{equation}
which for the supersymmetry algebra, in turn, implies the contribution
\begin{equation}
\label{e:old}
{}[ \delta_{\epsilon_1},\delta_{\epsilon_2}]\, e_\mu{}^r\Big|_{\text{from \eqn{e:gamma7_contrib}}} =
+\frac{(\alpha'_M)^3}{4(d-2)} \, \teight{s} \, D_\mu\big(\bar\epsilon_2
\Gamma^{r_1\cdots r_5}\epsilon_1\big) \, \delta^{r_6 r}_{s_7s_8}\,
W_{r_1r_2s_1s_2} W_{r_3r_4s_3s_4} W_{r_5r_6s_5s_6}\, .
\end{equation}

We are not done yet, however; in the cases where the super-Maxwell variations 
produce terms proportional to ${\cal E}(A)^m=D_n F^{mn}+\cdots$, the 
analogous Weyl-tensor terms in the supergravity action give 
equation-of-motion terms for the vielbein. That this is so can
be seen by inserting the decomposition \eqn{e:Riemann_decompose} of the 
Riemann tensor in the contracted Bianchi identity 
$D_{[\mu}R_{\nu\rho]}{}^{rs}\,e_r{}^\mu=0$, and then substituting 
the on-shell expressions \eqn{e:RfromEOM}--\eqn{e:RiccifromEOM} for the 
Ricci terms. Neglecting all higher-order terms and keeping only the 
equation-of-motion terms of interest one is left with
\begin{equation}
\label{e:DdotW}
D^n W_{mnrs} = \frac{4}{d-2} D_\mu \eom{e}_{[s}{}^\mu\, \eta_{r]m} 
  + \cdots \, .
\end{equation}
Instead of scanning the variation of the eleven-dimensional $W^4$ 
action~\eqn{e:elevendimI_X} for terms of this kind, we can, alternatively,
make use of the $\eom{A}$ terms in the variation of the super-Maxwell $F^4$
action displayed 
in~\eqn{e:maxwellhorror}--\eqn{e:maxwellhorror2}.\footnote{Although the 
left-right mixing and the replacement $R\rightarrow W$ on the gravity side 
lead to modifications as compared to the super-Maxwell $F^4$ 
Lagrangian~\eqn{e:expandedSYMstring}, it is easy to see that these do not 
affect the terms can potentially contribute to the $\Gamma^{[5]}$ part of the 
translation parameter.} In this way we obtain the term
\begin{equation}
\label{e:gamma3_contrib}
(\alpha'_M)^{-3} \, \delta L_{\Gamma^{[3]}}^1 \rightarrow 
  + \tfrac{1}{8}\,e\,\teight{s}\, (\epsilon \Gamma^{r_1\cdots r_5} 
  \bar\psi_{s_7s_8}) W_{r_1r_2s_1s_2} W_{r_3r_4s_3s_4} D^n W_{r_5 n s_5s_6}\,,
\end{equation}
which arises from the fourth line of $\delta L_{\Gamma[3]}^1$ 
in~\eqn{e:maxwellhorror} after mapping to the supergravity side. 
Using~\eqn{e:DdotW}, the modification to the transformation rule is then 
found to be
\begin{equation}
\label{e:DWmod}
\delta e_\mu{}^r \Big|_{\text{from \eqn{e:gamma3_contrib}}}= + \frac{(\alpha'_M)^3}{2(d-2)}\, \teight{s} \,D_\mu
\big(\bar\epsilon\Gamma^{r_1\cdots r_5} \psi_{s_7s_8} \big)
W_{r_1r_2s_1s_2} W_{r_3r_4s_3s_4} \delta^{r_5 r}_{s_5 s_6}\, .
\end{equation}
Such a non-supercovariant term is not very elegant (we will get back
to this shortly) but we can still use it to compute the supersymmetry algebra.
The modification~\eqn{e:DWmod} produces only a $\Gamma^{[5]}$ term
because of the anti-symmetrisation on $\epsilon_1$ and $\epsilon_2$.
The result is
\begin{equation}
\begin{aligned}
\label{e:new}
{}[ \delta_{\epsilon_1},\delta_{\epsilon_2}] \, e_\mu{}^r \Big|_{\text{from \eqn{e:gamma3_contrib}}}&=
-\frac{(\alpha'_M)^3}{4(d-2)}\,\teight{s}\, D_\mu 
(\bar\epsilon_2\Gamma^{r_1\cdots r_5} \epsilon_1\big) W_{r_1r_2s_1s_2} 
W_{r_3r_4s_3s_4} W_{r_5r_6s_5s_6} \delta^{r_6 r}_{s_7 s_8} \,.
\end{aligned}
\end{equation}

Finally, adding~\eqn{e:old} and~\eqn{e:new} we arrive at the conclusion
that the $\Gamma^{[5]}$ part of the translation parameter in the 
supersymmetry algebra, and thereby also the $\Gamma^{[5]}$ part of the 
torsion component $T_{ab}{}^r$, does not receive any corrections 
proportional to $(\alpha'_M)^3\, W^3$.

In order to check the correctness of this rather unexpected result, we
can rederive it in a different way. Namely, since the non-trivial
information encoded in the supersymmetry algebra (and the superspace
constraints) should not be affected by field redefinitions in the
higher-derivative action, we should obtain the same vanishing result
for non-trivial $\Gamma^{[5]}$ corrections by using the supersymmetric
$R^4$ action instead of the $W^4$ action~\eqn{e:elevendimI_X} (recall
the discussion around~\eqn{e:Riemann_subst}). As the former is
obtained from the latter simply by replacing the Weyl tensor with the
Riemann tensor and removing the last two terms of ${\cal
  L}_{\Gamma^{[1]}}$, we can to a large extent draw on the above
analysis of the $W^4$ case. The main difference is that in the
variation of the $R^4$ action no vielbein equation-of-motion terms
arise from splitting Riemann tensors into Weyl and Ricci parts, since
such splitting is not required. But the equation-of-motion terms
generating the modifications~\eqn{e:L7mod} and~\eqn{e:DWmod} are both
of precisely this type, as can be seen from the factor $(d-2)^{-1}$ in
their coefficients (cf.~\eqn{e:Dslashedpsi2}--\eqn{e:Ddotpsi2}).
Hence, neither of these modifications occurs in the variation of the
$R^4$ action, and since no new $\Gamma^{[5]}$ corrections to the
transformation rules are generated, we can indeed again conclude that
the supersymmetry algebra is not modified at this level.

At this point, it is interesting to notice that the situation is
slightly different for the dimension-1/2 component $T_{ar}{}^s$.  For
this case, \eqn{e:dimhalf_torsion} shows that there is a direct
relation between the transformation rule of the vielbein and the
torsion component. However, the gauge
choice~\eqn{e:supervielbein_gauge} is not appropriate in case the
vielbein does not transform supercovariantly. The field transformation
that brought us from the $R^4$ action to the $W^4$ action has
introduced such non-supercovariant terms, as one can see
from~\eqn{e:DWmod}. We will not go into details here, but it suffices
to say that we can only reliably compute the dimension-1/2 component
of the torsion, under the assumptions of section~\ref{s:modsusyrules},
for the $R^4$ action. The modifications induced by~\eqn{e:L7mod}
and~\eqn{e:DWmod} are then absent, but this does not constitute the
complete result at this level; other equation-of-motion terms which
are not relevant for the discussion of the supertranslation algebra
(and therefore have not been discussed here) do nevertheless
contribute to the dimension-1/2 torsion. However, we shall not concern
ourselves with a systematic analysis of these terms here, but instead
return to discuss the dimension-zero torsion.

As hinted at above, the vanishing $(\alpha'_M)^3\,W^3$ correction to
the supersymmetry algebra is quite surprising in light of Howe's
result~\cite{howe7} that by imposing \emph{only\/} the standard
constraint \mbox{$T_{ab}{}^r = 2\, (\Gamma^r)_{ab}$} on the
supertorsion in eleven-dimensional superspace, the equations of motion
for classical Cremmer-Julia-Scherk supergravity follow upon solving
the superspace Bianchi identities. Hence, any non-trivial correction
to the latter theory requires that the constraint on this supertorsion
component be modified in a non-trivial manner. We are thus led to the
conclusion that these non-trivial corrections must be due solely to
the gauge-field terms that we have not yet included in the $t_8 t_8
W^4$ superinvariant.  Moreover---and more remarkably---when the
gauge-field strength $H_{(4)}$ is set to zero, our result, in
conjunction with Howe's, seems to tell us that the dynamics encoded in
the action~\eqn{e:elevendimI_X} is equivalent to the dynamics of the
CJS theory for configurations with vanishing gauge-field
strength!\footnote{Some gauge-field dependent contributions to the
  bosonic part of the invariant were obtained
  by~\ddcite{Deser:1998jz}{Deser:2000xz} from an analysis of
  supergravity four-point functions at tree-level.  However, their
  method is neither complete (higher-point functions are necessary as
  well), nor does it produce a result which exhibits directly the
  $t_8$~tensorial structures.} We should, however, point out that a
complete analysis of the action at order $(\alpha'_M)^3$ is likely to
require the inclusion of the $\epsilon\epsilon\, W^4$ term (together
with its associated fermi bilinears). Its presence can be deduced by
lifting the one-loop term in the IIA action~\eqn{e:L_IIA}, and it is
conceivable that this term is also required once one includes the
gauge field in the supersymmetry analysis.  At present we do, however,
not know whether this purely gravitional term will produce non-trivial
modifications to the algebra; it seems more likely that such
corrections will again arise from the gauge-field terms once they are
included.

To the best of our knowledge, the only other computation of a
supergravity commutator (where corrections could be expected) in the
presence of higher-derivative terms that is available in the
literature was done in the type IIB paper by \dcite{Green:1998by}.
These authors considered the algebra on the dilatino. As their main
result required only the information of the terms independent of the
Riemann tensor, they only had to take into account the modified
dilatino transformation rule. As a consequence their calculation did
not reveal any modifications to the supertranslation parameter either.

\end{sectionunit}

\end{sectionunit}

\begin{sectionunit}
\title{Discussion, conclusions and outlook}
\maketitle

In this paper we have presented the first part of a systematic
derivation of the modifications to the superspace torsion constraints
due to higher-order $\alpha'$ string effects. In contrast to existing
approaches, our analysis is based on the component formalism of
supergravity. Using string-amplitude calculations in combination with
supersymmetry requirements, we were able to find a compact form for
the superinvariant associated to the anomaly cancellation term in ten
dimensions. In addition, we derived the modifications to the
supersymmetry transformation rules. In the sector which we considered,
which is largely independent of the gauge field, the invariant
could be lifted to eleven dimensions. The field-dependent parameters
in the supersymmetry algebra were shown to receive no corrections from 
this `purely gravitational' part of the theory. 

Given the complexity of higher-derivative supergravity actions, it is
most encouraging that the use of string-amplitude information has
enabled us to reduce the supersymmetry analysis to a problem that can
be worked out by hand. As a result, we now have a much better
understanding of the tensorial structure of the fermion bilinears. In
addition, the compact form of our result has enabled us to lift our
results to eleven dimensions, in spite of the fact that it was based 
on string input. 

The next step of our programme is obviously to extend the analysis to
cover the gauge-field sector in our analysis and also to include the
$\epsilon\epsilon W^4$ term. Since we have found that no non-trivial
purely gravitational modifications to the dimension-zero supertorsion
constraint are generated by the higher-derivative interactions we have
derived so far, while the superspace analysis of \dcite{howe7} proves
that any non-trivial M-theory corrections must show up precisely in
this torsion component, it is clear that either one (or both) should
be responsible for these corrections. The strong link between the
presence of the gauge field and the structure of eleven-dimensional
superspace at the quantum level is perhaps not so unexpected, given
the central roles the membrane and the five-brane---both supported by
a non-vanishing gauge-field strength---play in M-theory. In addition,
there are several reasons to expect the presence of the
$\epsilon\epsilon W^4$ term in the action. First of all it is obtained
by lifting the IIA action. But more importantly, absence of corrections
to the torsion constraints generated by this term would imply, through
Howe's analysis~\cite{howe7}, that the dynamics of the purely gravitational
theory (i.e.~setting the gauge field to zero) is equivalent to that
of the Cremmer-Julia-Scherk theory. It is not clear how such a rather
strong conclusion would fit into the various duality conjectures.

The presence of the $\epsilon\epsilon\,W^4$ term in the
higher-derivative eleven-dimensional action can in principle be
studied via our string theory analysis. As our approach is based on
one-loop amplitudes, and because the heterotic string does not exhibit
this particular bosonic term at one loop, it is, however, necessary to
first extend our analysis to the maximally extended supergravity
theories in ten dimensions. Treatment of the gauge-field sector using
our methods is also possible, although one expects that the lifting
procedure to eleven dimensions will be more complicated. But perhaps
the rather compact form of our invariant will make it possible to
analyse these gauge-field dependent terms using only supersymmetry.
Work on these issues is in progress. Once the gauge-field terms have
been included in the higher-derivative action as well as in the
transformation rules, it also becomes of interest to study
$\alpha'_M$-corrected M-brane supergravity solutions as well as
applications to compactification problems.

On the sideline, many other interesting questions have appeared.  One
of them is to understand whether a superparticle (or supermembrane)
vertex operator analysis (in the space-time supersymmetric formalism)
is able to reproduce the tensorial structures of the fermionic
bilinears in our eleven-dimensional action~\eqn{e:elevendimI_X}. This,
however, requires complicated zero-mode integrals to be performed (an
alternative way to state this problem is that one has to calculate
contractions of a sixteen-dimensional spinorial epsilon tensor with a
number of gamma matrices, which is hard except for special cases like
the contraction that leads to the $t_8t_8$ structure). 

The main goal of this programme, however, remains to understand how the
higher-deriva\-tive modifications to the target-space theory are related
to similar corrections of world-volume theories of branes. For instance,
the known bosonic higher-derivative gravitational corrections to the 
Born-Infeld part of the D-brane actions (as derived by 
\dcite{Bachas:1999um}) and the ones correcting the Wess-Zumino term 
(see e.g.~\dcite{gree11} and \dcite{Cheung:1998az}) have so far not 
been incorporated in a kappa-symmetric framework, generalising the
actions of \ddcite{Cederwall:1997pv}{cede4},
\dcite{berg14} and \dcite{Aganagic:1996nn}. 
Intuitively one expects that the gravitational corrections, together 
with the modified torsion constraints and perhaps a modified
form of the kappa-symmetry projector, conspire to yield again a 
kappa-symmetric action. The superembedding formalism seems to be a very
promising tool with which to address this question.

\end{sectionunit}

\section*{Acknowledgements}

We have benefitted from discussions with Eugene Cremmer, Stanley
Deser, Jim Gates, Michael Green, Renata Kallosh, Hermann Nicolai, Jan
Plefka, Sebastian Silva, Per Sundell, Paul Townsend, Andrew Waldron,
Niclas Wyllard and Marija Zamaklar. In addition, we would like to
thank the theory group at Chalmers University in Gothenburg, and in
particular Martin Cederwall, Ulf Gran, Mikkel Nielsen and Bengt
Nilsson, for an inspiring week of discussions, for sharing with us an
early draft of~\cite{Cederwall:2000ye} and for very useful remarks
concerning our superspace results. Finally, we thank Mees de Roo for
many behind-the-scenes comments about~\cite{dero3} and for providing
us with details about their computer calculation.  \medskip

\noindent P.V. was partially supported by the TMR contract ERB FMRXCT 96-0012.
\vfill\eject

\appendix
\begin{sectionunit}
\title{Supergravities in first- and second-order formalism}
\maketitle
\label{s:sugra_appendix}
\begin{sectionunit}
\title{Normalisation issues}
\maketitle

This appendix contains various details about the standard,
non-extended supergravity theories in ten and eleven dimensions
(though some of the results apply to other supergravity theories as
well). Our main goal is to explain the origins of various
normalisation factors, to elaborate on the presence of the dilaton and
to explain some of the subtleties one encounters when dealing with the
transformation rule of the spin connection. None of the results are
new, although few accounts of higher-dimensional supergravity in the
first-order formulation have appeared previously in the literature
(exceptions are the papers by~\dcite{cast2} and \dcite{juli2}). As we
do not need higher-order fermi terms in the main text, they have been
omitted here for the sake of brevity. The original references for the
theories discussed here are \dcite{crem1} (eleven dimensions) and
\dcite{cham1} and \dcite{berg1} (ten dimensions).

The gravity supermultiplets for the theories under consideration consist
first of all of a graviton (described by the vielbein $e_\mu{}^r$), a
gravitino ($\psi_\mu$) and a bosonic gauge field (which we will denote
by $B_{\mu_1\cdots \mu_{n-1}}$). In ten dimensions, there is in
addition a dilaton ($\phi$) and dilatino ($\lambda$). We define the
bosonic and fermionic field strengths as
\begin{equation}
\begin{aligned}
H_{\mu_1\cdots\mu_n} &= n\, \partial_{[\mu_1} B_{\mu_2\cdots \mu_n]}\,
,\\[1ex]
\psi_{\mu\nu} &= D_{[\mu}\psi_{\nu]}\, .
\end{aligned}
\end{equation}
Note the perhaps somewhat unusual normalisation of the gravitino
curvature. 

Let us first discuss some normalisation issues. In eleven dimensions
the kinetic terms are unique up to normalisations. In ten dimensions,
in contrast, the presence of the dilaton forces us to make a choice of
Weyl scaling. In the present section we eliminate this choice by
requiring that the kinetic terms for the fermions are diagonal and we
will comment on the other possibilities later. With this choice, our
normalisations are fixed by
\begin{equation}
\label{e:action_ansatz}
\begin{aligned}
S_{\text{graviton}} &=& -\frac{1}{4\kappa_d^2} \int\!{\rm d}^{d} x\,& e\,
R(\omega)\, ,\\[1ex]
S_{\text{gravitino}} &=& -\frac{1}{\kappa_d^2} \int\!{\rm d}^{d}x\,&
e\,\bar\psi_\mu\Gamma^{\mu\nu\rho} \psi_{\nu\rho} \, ,\\[1ex]
S_{\text{gauge field}} &=& -\frac{1}{2\kappa_d^2}\int\!{\rm d}^dx& \,
e\,\frac{1}{n!}\,\phi^{-p}
\,H_{\mu_1\cdots\mu_n}H^{\mu_1\cdots\mu_n}\, .
\end{aligned}
\end{equation}
In eleven dimensions $n=4$, while in the ten-dimensional theory $n=3$.
The power $p$ of the dilaton in the last line depends on the dimension
and the rank of the gauge field, and is completely fixed by
supersymmetry once we have fixed the normalisation of the dilaton
kinetic term. For this field (and its superpartner) we use the
actions
\begin{equation}
\begin{aligned}    
S_{\text{dilaton}} &=& -\frac{1}{2\kappa_d^2} \int\!{\rm d}^d x\, &
e\, (\phi^{-1} \partial_\mu \phi)^2\, ,\\[1ex]
S_{\text{dilatino}} &=& -\frac{1}{\kappa_d^2} \int\!{\rm d}^d x\, &
e\, \bar\lambda \Gamma^\mu D_\mu \lambda\, .
\end{aligned}
\end{equation}
With this choice, it turns out that $p=2$ for the three-form theory
(see below).  The spinors are all minimal, i.e.~in eleven dimensions
the gravitino is Majorana while in ten dimensions the gravitino and
the dilatino are both Majorana and Weyl (though of opposite
chirality). In addition to the above terms, the complete non-linear
actions will contain three-point vertices coupling the fermions, the
gauge field and the dilaton. Finally, there will of course be
higher-order fermi terms, and in eleven dimensions also Chern-Simons
couplings of the gauge fields; we will ignore the former but comment
on the way in which the latter arise at the end of this section.

The structure of the supersymmetry transformation rules of the various
fields can be determined by analysing the kinetic terms given above
(see also \dcite{Townsend:1983kk}). Let us first state the result, which 
can be summarised by the general form
\begin{subequations}\label{e:trafo_ansatz}\begin{align}
\label{e:trafo_e}
\delta e_\mu{}^r &= 2\,\bar\epsilon \Gamma^r \psi_\mu \, ,\\[1ex]
\label{e:trafo_psi}
\delta \psi_\mu  &= D_\mu \epsilon  
+  {\cal N}_{\psi}\,\phi^{-p/2}\, \big({\cal T}_\mu{}^{\sigma_1\cdots\sigma_n} \epsilon\big)
\hat H_{\sigma_1\cdots
\sigma_n} \, ,\\[1ex]
\label{e:trafo_B}
\delta B_{\mu_1\cdots\mu_{n-1}} &= (n-1)\, {\cal N}_{\psi}\,\phi^{p/2}\,
\big(\bar\psi_{[\mu_1}\Gamma_{\mu_2\cdots\mu_{n-1}]}\epsilon\big) 
- {\cal N}_\lambda\,\phi^{p/2}\,\big(\bar\lambda\Gamma_{\mu_1\cdots\mu_{n-1}}
\epsilon\big)\, ,\\[1ex]
\label{e:trafo_phi}
\phi^{-1} \delta\phi &= \sqrt{2}\big(\bar\epsilon\lambda\big)\, ,\\[1ex]
\label{e:trafo_lambda}
\delta\lambda &= \tfrac{1}{\sqrt{2}} \big(\Gamma^\mu\epsilon\big)\,
\phi^{-1} \hat\phi_\mu -\tfrac{1}{2\cdot n!}
{\cal N}_\lambda \, \phi^{-p/2} \big(\Gamma^{\sigma_1\cdots\sigma_n}\epsilon\big)\, \hat
H_{\sigma_1\cdots\sigma_n}\, .
\end{align}
\end{subequations}
We should stress that these rules are \emph{not} the ones which we
use in the main text; instead, a super-Weyl rescaling to the string
frame has been used there. More details are given in the next section.
The spin connection appearing e.g.~in the graviton kinetic action can
be viewed as either an independent field (first-order formalism) or as
a composite one defined in terms of the vielbein (second-order
formalism). In between the first- and second-order formalisms there is
also the so-called 1.5-order formulation, where the spin-connection is
treated as an independent field for the purpose of the supersymmetry
variation only. These issues will be discussed in more detail below.

Given the transformation rules, one can define so-called
supercovariant objects, which are by definition such that they
transform without any derivatives of the supersymmetry parameter.
Notationally, we distinguish fields with this property by hats.
They are readily constructed with the result:
\begin{subequations}
\label{e:supercovariant_objects}
\begin{align}
\hat\phi_\mu    &= \partial_\mu \phi - \sqrt{2}\bar\lambda \psi_\mu\, \phi \, ,\\[1ex]
\hat\lambda_\mu &= D_\mu \lambda - \tfrac{1}{\sqrt{2}} \big(\Gamma^\nu
\psi_\mu\big) \phi^{-1}\hat\phi_\nu +
\tfrac{1}{2\cdot n!}{\cal N}_\lambda
\phi^{-p/2}\big(\Gamma^{\sigma_1\cdots\sigma_n}\psi_\mu\big)
\hat H_{\sigma_1\cdots\sigma_n}\, ,\\[1ex]
\hat\psi_{\mu\nu} &= D_{[\mu} \psi_{\nu]} + {\cal N}_\psi\, \phi^{-p/2} {\cal T}_{[\mu} \cdot \hat
H  \, \psi_{\nu]}\, ,\\[.5ex]
\hat H_{\sigma_1\cdots\sigma_n} &= n\,\partial_{[\sigma_1}
B_{\sigma_2\cdots\sigma_n]} - \tfrac{1}{2}n(n-1){\cal N}_\psi \phi^{p/2}\, \big(\bar\psi_{[\sigma_2}
\Gamma_{\sigma_3\cdots\sigma_{n}} \psi_{\sigma_1]} \big)\nonumber\\[1ex]
&\quad\quad\quad\quad\quad\quad\quad\quad+ n{\cal N}_\lambda \phi^{p/2}\,\big( \bar\lambda\Gamma_{[\sigma_2\cdots\sigma_n}\psi_{\sigma_1]}\big)\, .
\end{align}
\end{subequations}

Let us now explain how to obtain the gauge-field dependent
coefficients in~\eqn{e:trafo_ansatz} (the other ones are easily
fixed). The first thing to do is to determine the structure of the
three-point vertices in the action. The easiest way to achieve this is
to demand that the fermion equations of motion are
supercovariant. Using the definitions~\eqn{e:supercovariant_objects}
one then writes down these couplings immediately. The next step is to
fix the structure of the tensors ${\cal T}$. This can be done by
focussing on variations proportional to a derivative of the gauge
field strength, and observing that terms with a $\Gamma^{[n]}$ matrix
only come from this tensor and should therefore vanish by themselves.
Once this tensor is known, the normalisation factors ${\cal N}_\psi$
and ${\cal N}_\lambda$ follow by looking at for instance the $H^2$
terms in the variation, or by computing the algebra on the bosonic
fields.  Finally, the exponent $p$ can be fixed by considering
variations proportional to one power of the gauge-field strength and a
derivative of the dilaton.

To illustrate this procedure, let us discuss the invariance of the
eleven-dimensional action in some more detail. The three-point
vertices can all be incorporated in the action by just taking the
supercovariant gravitino curvature in~\eqn{e:action_ansatz} (such a
simple substitution does no longer work in ten dimensions, but the
rest of the logic is the same). This object, and the supercovariant 
gauge-field strength, transform as
\begin{align}
\label{e:hatpsi2_trafo}
\delta \hat\psi_{\mu\nu} &= \tfrac{1}{8}\, R_{\mu\nu mn}
\Gamma^{mn}\epsilon + {\cal N}_\psi\, D_{[\mu}\big( {\cal T}_{\nu]}\cdot \hat H\big)\epsilon
+ {\cal N}_\psi^2\, {\cal T}_{[\mu}\cdot \hat H\, {\cal T}_{\nu]}\cdot \hat H\, \epsilon
+ {\cal O}(\epsilon\psi^2)\, ,\\[1ex]
\delta \hat H_{\mu\nu\rho\sigma} &= -12{\cal N}_\psi\, \bar\epsilon
\Gamma_{[\mu\nu} \psi_{\rho\sigma]} - 12{\cal N}_\psi\,\bar\epsilon T_{[\mu\nu}{}^r
\Gamma_{|r|\rho} \psi_{\sigma]} + {\cal O}(\epsilon\psi^3)\, .
\end{align}
Employing $\delta e = e\, e_r{}^\mu\, \delta e_\mu{}^r$ as well as
$\partial_\mu e = e\, e_n{}^\lambda \partial_\mu e_\lambda{}^n$, one
finds that the three kinetic terms transform as
{\allowdisplaybreaks
\begin{subequations}
\begin{align}
\label{e:ekin}
\kappa_d^2\, \delta S_{\text{graviton}} &= e \Big( R_n{}^\mu -
\tfrac{1}{2} e_n{}^\mu  R \Big )\, \bar\epsilon\Gamma^n \psi_\mu +
\tfrac{1}{2}e \Big( \tfrac{1}{2} T_{mn}{}^\nu
 - T_{m\lambda}{}^\lambda e_n{}^\nu \Big) \delta \omega_\nu{}^{mn}\, .\\[2ex]
\label{e:gravikin}
\kappa_d^2\, \delta S_{\text{gravitino}} &=  -e \Big( R_n{}^\mu -
\tfrac{1}{2} e_n{}^\mu R \Big )\, \bar\epsilon\Gamma^n \psi_\mu
+ \tfrac{1}{4} e\, \bar\epsilon \Gamma^{\mu\nu\rho rs} \psi_\mu \, 
 R_{\nu\rho rs}   \nonumber \\[.5ex]
&\quad\quad- e\,\bar\epsilon\Big( 
 T_{\mu\lambda}{}^\nu\, 
\Gamma^{\rho\mu\lambda}
- T_{\mu\lambda}{}^\lambda\, \Gamma^{\mu\nu\rho} \Big) \hat\psi_{\nu\rho}\nonumber\\[.5ex]
& \quad\quad + e{\cal N}_\psi\,\bar\epsilon\Gamma^{\mu\nu\rho} D_\mu\Big({\cal
T}_\nu \cdot \hat H \psi_\rho\Big) - e\,{\cal N}_\psi\,\bar\epsilon\bar{\cal T}_\lambda
\cdot\hat H \Gamma^{\lambda\mu\nu} \psi_{\mu\nu} \\[.5ex]
&\quad\quad - e\,{\cal N}_\psi\,\bar\psi_\mu\Gamma^{\mu\nu\rho} D_\nu\Big( {\cal
T}_\rho \cdot\hat H \Big)\epsilon  -
\tfrac{1}{4}e\, \bar\psi_\mu\Gamma^{\mu\nu\rho} \Gamma^{rs}\psi_\rho \,
\delta\omega_{\nu rs} \nonumber \\[2ex]
& \quad\quad + e\,{\cal N}_\psi^2\,\bar\epsilon \bar{\cal T}_\mu\cdot \hat H \Gamma^{\mu\nu\rho} 
{\cal T}_\nu\cdot H \psi_\rho + {\cal O}(\epsilon\psi^3) \, .\nonumber \\[2ex]
\label{e:gfkin}
\kappa_d^2\, \delta S_{\text{gauge field}}  &= -\tfrac{1}{48} e \bar\epsilon \Gamma^\mu
\psi_\mu H^2 + \tfrac{1}{3} e\, \bar\epsilon\Gamma^\mu \psi^\nu \, 
H_{\mu\lambda\kappa\sigma} H_{\nu}{}^{\lambda\kappa\sigma} 
- \tfrac{1}{2}e\, {\cal N}_\psi\, g^{\mu\lambda}\, \partial_\mu H_{\lambda\nu\rho\sigma}\, \bar\epsilon
\Gamma^{\nu\rho} \psi^{\sigma}\nonumber\\[.5ex]
&\quad\quad + \tfrac{1}{2} e\,{\cal N}_\psi\, \bar\epsilon \Gamma_{\nu\rho}
\psi_\sigma\, H^{\mu\kappa\rho\sigma} \Big( T_{\mu\kappa}{}^\nu - T_{\mu\lambda}{}^\lambda
\delta_\kappa{}^\nu \Big) + \tfrac{1}{4}e\,{\cal N}_\psi\,
\bar\epsilon\Gamma_{\nu\rho}\psi_\sigma H^{\mu\nu\rho\lambda}
T_{\mu\lambda}{}^\sigma\nonumber\\[.5ex]
&\quad\quad - \tfrac{1}{2}e\,{\cal N}_\psi\,
\bar\epsilon\Gamma^{s_2s_3}\psi^{s_4} D_\mu\Big(
e_{s_1}{}^{\lambda_1}\cdots
e_{s_4}{}^{\lambda_4}\Big) e^{s_1\mu}  H_{\lambda_1\cdots\lambda_4} \,.
\end{align}
\end{subequations}}
We have kept $\delta\omega$ as well as the terms proportional to the
torsion\footnote{A useful relation, valid under the integral,
which isolates the appearance of the torsion terms, is
\begin{equation}
\label{e:usefulintermediate}
\begin{aligned}
e\, \delta\Big[ e_r{}^\mu e_s{}^\nu R_{\mu\nu tu} \Big] \, S^{rstu}
&= 4\, e \Big( \bar\epsilon\Gamma^\mu \psi_{[r} \Big)\, R_{s]\mu tu} S^{rstu}\\[.5ex]
&\quad + 2\, \Big( \tfrac{1}{2} T_{rs}{}^\nu S^{rstu} -
T_{r\lambda}{}^\lambda S^{r\nu tu} \Big) \delta\omega_{\nu tu}\\[.5ex]
&\quad -2\, e \Big( D_\mu S^{rstu} \Big) e_r{}^\mu e_s{}^\nu \, \delta\omega_{\nu tu}\, ,
\end{aligned}
\end{equation}
This can be used in higher-derivative actions, but of course also 
reproduces the variation of the Einstein-Hilbert term after insertion of
$S^{rstu}=\tfrac{1}{2}(\eta^{ru}\eta^{st}-\eta^{su}\eta^{rt})$
(which satisfies $D_\mu S^{rstu}=0$).}
 $T_{\mu\nu}{}^r = 2\,D_{[\mu}e_{\nu]}{}^r$ for the discussion
in section~\ref{s:omega_trafo}.  The first terms to focus on are those
with a derivative on the gauge-field strength. They only come with a
$\Gamma^{[n-2]}$ matrix in~\eqn{e:gfkin}, and by comparison
with~\eqn{e:gravikin} one finds
\begin{equation}
\label{e:T_def}
{\cal T}_{\mu}{}^{\sigma_1\cdots \sigma_n} = \frac{n-1}{2(d-2)n!}
\left(
\Gamma_\mu{}^{\sigma_1\cdots\sigma_n} - \frac{n}{n-1}(d\!-\!n\!-\!1)
\, \delta_\mu{}^{[\sigma_1}
\Gamma^{\sigma_2\cdots\sigma_n]}\right) \, .
\end{equation}
The cancellation now arises due to the fact that relative normalisation
between the two terms leads to
\begin{equation}
\Gamma_{\mu\nu}{}^{\rho} {\cal T}_\rho{}^{\sigma_1\cdots\sigma_n} =
-\frac{1}{2\cdot n!}\Big(
\Gamma_{\mu\nu}{}^{\sigma_1\cdots\sigma_n} +
 n(n-1)\, \delta_{\mu}{}^{[\sigma_1}\delta_{\nu}{}^{\sigma_2}
\Gamma^{\sigma_3\cdots\sigma_n]} \Big) \, .
\end{equation}
When inserted in the relevant terms in~\eqn{e:gravikin}, one finds
that the first gamma produces a Bianchi identity on the gauge-field
strength, while the second one exactly cancels the contribution
from~\eqn{e:gfkin}. By computing the algebra on the two bosonic fields
or by analysing the variations proportional to $H^2$, one obtains
${\cal N}_\psi=1$.

The Chern-Simons term in the eleven-dimensional action arises because
of the fact that there is a $\bar\psi \Gamma^{[9]} \epsilon H^2$ term
left over after all other terms with two powers of the gauge-field
strength have been cancelled. By dualising the gamma matrix one can
get rid of this variation by adding the term
\begin{equation}
S_{\text{Chern-Simons}} = -\frac{2}{(12)^4\, \kappa_{11}^2}\int{\rm d}^{11}x\, 
\varepsilon^{\mu_1\cdots \mu_{11}} B_{\mu _1\mu _2\mu _3}
H_{\mu _4\cdots\mu _7} H_{\mu _8\cdots\mu_{11}}\, .
\end{equation}
The story is very similar in the ten-dimensional case, where one finds
the same tensorial structure for ${\cal T}$ and in addition obtains
that ${\cal N}_\lambda=1/\sqrt{2}$ and $p=2$.
\end{sectionunit}

\begin{sectionunit}
\title{Super-Weyl transformations and the string frame}
\maketitle

In ten dimensions we have the freedom of going to a different frame by
rescaling the vielbein by a power of the dilaton. In order to maintain
the canonical transformation rule of the vielbein, an accompanying
redefinition of the fermion fields is necessary. Starting from the
transformation rules~\eqn{e:trafo_ansatz} and performing a rescaling
\begin{equation}
\begin{aligned}
 e_\mu{}^r &= \phi^{-\alpha} \tilde e_\mu{}^r\, , &\quad
 \lambda   &= \phi^{\alpha-\beta} \tilde \lambda\, ,\\[1ex]
 \psi_\mu  &= \phi^{-\beta}\tilde\psi_\mu - \gamma e_\mu{}^r
\Gamma_r \lambda\, , &\quad \epsilon &= \phi^{-\beta}\tilde \epsilon\, ,
\end{aligned}
\end{equation}
(the second column ensures that the first terms in the dilatino and
gravitino transformation rules remain independent of the dilaton) then
one finds that the new vielbein transforms as
\begin{equation}
\label{e:etilde_trafo}
\delta\tilde e_\mu{}^r = 2\,\phi^{\alpha-2\beta}
\,\big(\Tilde{\Bar\epsilon}\Gamma^r
\tilde\psi_\mu\big) + e_\mu{}^r \big(\bar\epsilon\lambda\big) \phi^\alpha \left(
\sqrt{2} \alpha - 2 \gamma\right) 
- 2\gamma\, \phi^{\alpha-2\beta} \big(\Tilde{\Bar \epsilon} 
\Gamma_{rs}\tilde \lambda\big) \tilde e_\mu{}^s\, .
\end{equation}
Imposing that this is identical to the transformation rule in the
original frame requires that the two remaining parameters are expressed in
terms of $\alpha$ as
\begin{equation}
\beta =\tfrac{1}{2}\alpha\, ,\quad \gamma = \tfrac{1}{\sqrt{2}}\alpha\, .
\end{equation}
In addition, there are some changes in the supersymmetry
transformation rules. Firstly, a Lorentz transformation has to be
added in order to accommodate the last term in~\eqn{e:etilde_trafo},
and addition there are some changes to the fermionic transformation
rules. Focussing on the $n=3$ case (we drop the tildes from now on)
one obtains:
\begin{subequations}
\label{e:trafotilde_ansatz}
\begin{align}
\label{e:trafotilde_e}
\delta e_\mu{}^r &= 2\,\bar\epsilon \Gamma^r \psi_\mu 
- \sqrt{2}\alpha\, (\bar\epsilon
\Gamma^{r}{}_s\lambda) e_\mu{}^s \, ,\\[1ex]
\label{e:trafotilde_psi}
\delta \psi_\mu  &= D_\mu \epsilon  
+  \phi^{q}\, \big({\cal
T}_\mu{}^{\sigma_1\cdots \sigma_3} \epsilon\big) \hat H_{\sigma_1\cdots
\sigma_3} - \tfrac{\alpha}{ 4\cdot n!}\,
\phi^{q}\, \big(\Gamma_\mu \Gamma^{\sigma_1\cdots \sigma_3}
\epsilon\big) \hat H_{\sigma_1\cdots\sigma_3} \\[1ex]
\label{e:trafotilde_B}
\delta B_{\mu_1\mu_{2}} &= 
\begin{gathered}[t] 2\,\phi^{-q}\,
\big(\bar\psi_{[\mu_1}\Gamma_{\mu_2]}\epsilon\big) 
+ \tfrac{(2\alpha-1)}{\sqrt{2}} \phi^{-q}\,\big(\bar\lambda\Gamma_{\mu_1\mu_2}
\epsilon\big)\, ,
\end{gathered}
\\[1ex]
\label{e:trafotilde_phi}
\phi^{-1} \delta\phi &= \sqrt{2}\big(\bar\epsilon\lambda\big)\, ,\\[1ex]
\label{e:trafotilde_lambda}
\delta\lambda &= \tfrac{1}{\sqrt{2}} \big(\Gamma^\mu\epsilon\big)\,
\phi^{-1} \hat\phi_\mu -\tfrac{1}{2\sqrt{2}\cdot n!}
\phi^{q} \big(\Gamma^{\sigma_1\cdots\sigma_3}\epsilon\big)\, \hat
H_{\sigma_1\cdots\sigma_3}\, ,
\end{align}
\end{subequations}
(here $q=\alpha(n-1)-p/2$). The rules above of course include higher order
fermi terms which we have suppressed.

Under these super-Weyl rescalings, the kinetic terms for the graviton,
the gravitino, the dilaton and the dilatino all pick up a dilaton
prefactor $\phi^{-(d-2)\alpha}$. The gauge-field kinetic term instead
will get an overall factor $\phi^{(2n-d)\alpha-p}$ where $p$ is the
factor that was already present. The frame in which all transformation
rules are independent of the dilaton, namely for which
$\alpha=p/(2n-2)$, is called the ``string frame''. In this frame one
in addition observes that all kinetic terms have the same dilaton
prefactor, while the transformation rule of the gauge field becomes
independent of the dilatino and the gravitino rule simplifies
drastically.  We refer the reader to the work of
\dcite{Kallosh:1985eh} and \dcite{Kallosh:1986cd} (the $N=1$ case) and
\dcite{bell2} (the extended supergravities) for more information.

These results can of course be formulated in superspace as well.
Following \dcite{Gates:1980ky}, \dcite{howe9} and in particular
section~6 of \dcite{gate9}), one finds that the transformed
supervielbeine are given by
\begin{equation}
\label{e:superWeyl}
\begin{aligned}
\tilde E_{M}{}^{A} &= 
\begin{pmatrix}
e^{\beta\Phi} \Big( E_\alpha{}^a - f_r{}^a E_\alpha{}^r \Big) & e^{\alpha\Phi} E_\alpha{}^r \\
e^{\beta\Phi} \Big( E_\mu{}^a - f_r{}^a E_\mu{}^r \Big) & e^{\alpha\Phi} E_\mu{}^r  \\
\end{pmatrix}
\, , \\[1ex]
\tilde E_{A}{}^{M} &= 
\begin{pmatrix}
e^{-\beta\Phi} E_a{}^\alpha  & e^{-\beta\Phi} E_a{}^\mu \\
e^{-\alpha\Phi} \Big( E_r{}^\alpha + f_r{}^a E_a{}^\alpha\Big) &
e^{-\alpha\Phi} \Big( E_r{}^\mu    + f_r{}^a E_a{}^\mu \Big) \\
\end{pmatrix}
\, .
\end{aligned}
\end{equation}
Observe that the components $E_\mu{}^r$ and $E_\alpha{}^r$ (which are
the only components appearing in e.g.~the string world-sheet action)
transform in a simple way. The $f_r{}^a$ field is a superfield and has
a non-trivial expansion in powers of theta.

\end{sectionunit}

\begin{sectionunit}
\title{Transformation of the spin connection in eleven dimensions}
\maketitle 
\label{s:omega_trafo}

We have not yet discussed how the terms proportional to the torsion
and the variation of the spin connection
in~\eqn{e:ekin}--\eqn{e:gfkin} are cancelled. In the second-order
formalism the torsion is of higher order in the fermions, while the
transformation rule of the spin connection follows from that of the
vielbein. In contrast, the first-order formalism keeps both of these
as independent objects. The 1.5-order formalism, somewhere in between
the previous two, applies only to the \emph{classical} supergravity
theories as it makes use of the fact that the equation of motion of
the spin connection is algebraic. As a result, this object does not
have to be varied in the action as it only leads to terms proportional
to its defining equation. We will exhibit in detail how the
differences arise in the case of the eleven-dimensional theory.

We first observe that the fully anti-symmetrised Riemann tensor
in~\eqn{e:gravikin} can be rewritten using
\begin{equation}
\label{e:Ricciwithtorsion}
R_{[\mu\nu\rho]}{}^s = -D_{[\mu}T_{\nu\rho]}{}^s\, .
\end{equation}
The only candidate $\delta\omega$ term that can cancel this variation
is the second line of~\eqn{e:ekin} with a non-supercovariant term in
$\delta\omega$ (the first term in \eqn{e:deltaomega} below). Part of
the transformation rule for the spin connection should therefore be
\begin{equation}
\label{e:deltaomega}
\delta_1 \omega_\nu{}^{mn} = D_\phi\Big( \bar\epsilon
\Gamma_\nu{}^{\phi\rho m n} \psi_\rho\Big) +
4\, \bar\epsilon \Gamma^{mnp}
\hat\psi_{\nu p} -  \frac{4}{d-2}\, \, \bar\epsilon \Gamma^{rs[m} 
\hat\psi_{rs} e_\nu{}^{n]}  + {\cal O}(\epsilon\psi^3) \, ,
\end{equation}
(for the cancellation of the torsion terms, there is a third possible
candidate in the variation with a $\Gamma^{[3]}$, namely $\bar\epsilon
\Gamma_\nu{}^{p[n} \psi^{m]}{}_p$, but it turns out that this term is
not needed).  However, this is not the complete story as we have
remaining terms coming from the third and fourth line of
\eqn{e:gravikin} as well as the second line of \eqn{e:gfkin}. Taken
together, they are
\begin{equation}
\label{e:remaining}
\begin{aligned}
\text{remaining} &= -e\,\Big( \bar\psi_\rho \Gamma^{\mu\nu\rho} {\cal T}_\lambda\cdot \hat H\, \epsilon
- (\epsilon\leftrightarrow\psi_\rho) \Big)\,  T_{\mu\nu}{}^\lambda
- \tfrac{1}{12}e\,
\bar\epsilon\Gamma^{\mu\rho\lambda\sigma_2\sigma_3\sigma_4}
T_{\mu\lambda}{}^{\sigma_1} H_{\sigma_1\cdots\sigma_4} \psi_\rho\\[.5ex]
&\quad\quad + \tfrac{1}{4} e\, \bar\epsilon \Gamma_{\nu\rho}
\psi_\sigma\, H^{\mu\kappa\rho\sigma} \Big( T_{\mu\kappa}{}^\nu - T_{\mu\lambda}{}^\lambda
\delta_\kappa{}^\nu \Big) + \tfrac{1}{8}e\,
\bar\epsilon\Gamma_{\nu\rho}\psi_\sigma H^{\mu\nu\rho\lambda}
T_{\mu\lambda}{}^\sigma\, .
\end{aligned}
\end{equation}
The first two terms can be reduced further, as the anti-symmetric
combination picks out only the $\Gamma^{[6]}$ and $\Gamma^{[2]}$
pieces. The fact that \eqn{e:remaining} is non-zero implies that we
need additional terms in $\delta\omega$.

Just adding $\epsilon\psi H$ terms to the transformation rule of the
spin connection does not, however, make the action invariant. The only
way out of this problem is to \emph{add} an additional term to the
action, which vanishes when the torsion is taken on-shell. Such an
addition has also appeared in \dcite{cast2} and \dcite{juli2}.  The
guiding principle to find this action will be to make
\eqn{e:deltaomega} supercovariant. This requires
\begin{equation}
\label{e:deltaomega2}
\delta_2 \omega_\nu{}^{mn} = - (D_\phi\bar\epsilon)
\Gamma_\nu{}^{\phi\rho mn}\psi_\rho + \bar\epsilon
\Gamma_\nu{}^{\kappa\lambda mn} {\cal T}_\kappa\cdot \hat H
\psi_\lambda\, .
\end{equation}
The first term produces a variation proportional to the derivative of
the supersymmetry parameter. The appropriate term in the action to 
cancel this variation is
\begin{equation}
\label{e:ST}
S_T = \frac{1}{\kappa_{11}^2} \int\!{\rm d}^{11}x\, \tfrac{1}{8}e\, \Big( T_{mn}{}^\nu - \text{``value of
$T_{mn}{}^\nu$ on-shell''}\Big)\, \bar\psi_\lambda
\Gamma_\nu{}^{\lambda\kappa mn} \psi_\kappa\, .
\end{equation}
The second term in \eqn{e:deltaomega2} now produces additional terms
proportional to the three-form field strength and so does the
variation of the gravitini in $\delta S_T$. The sum of these, even
though it involves a contraction of $\Gamma^{[5]}$ with ${\cal T}$, is
rather simple:
\begin{equation}
\label{e:delta2stuff}
\kappa_{11}^2 \Big( \delta_2 S_{\text{CJS}} + \delta S_T\Big) = 
\tfrac{1}{4}e\, T_{mn}{}^\nu \bar\epsilon 
\Gamma_\nu{}^{\kappa\sigma mn} {\cal T}_\kappa\cdot \hat H\psi_\sigma
+ (\epsilon\leftrightarrow \psi_\sigma)\, .
\end{equation}
Adding~\eqn{e:delta2stuff} to the terms~\eqn{e:remaining} which were
still remaining, and working out the $\Gamma^{[6]}$ and $\Gamma^{[2]}$
terms in the gamma products (this is a bit tedious but can be done
more easily by making use of~\eqn{e:contractedgammagamma}), one finds that many
terms cancel and one obtains the following rather elegant result
\begin{equation}
\kappa_{11}^2\,\delta S  = -\tfrac{1}{72} \Big( \bar\psi_\lambda
\Gamma^{\mu\nu}{}_{\sigma_1\cdots\sigma_4} \epsilon 
+ 24\, \bar\psi_\lambda \Gamma^{\sigma_1\sigma_2} \delta_{\sigma_3}{}^\mu
\delta_{\sigma_4}{}^\nu\epsilon \Big) \, \Big(
\tfrac{1}{2}T_{\mu\nu}{}^\lambda - T_{\mu\kappa}{}^\kappa
\delta_\nu{}^\lambda\Big) \, .
\end{equation}
These can be cancelled by one final addition to the transformation of the
spin connection,
\begin{equation}
\delta_3 \omega_\nu{}^{mn} = -4e\, \bar\epsilon{\cal
S}^{m\sigma_1\cdots\sigma_4}{}_n H_{\sigma_1\cdots\sigma_4} \psi_\nu\,,
\end{equation}
where we have defined the tensor ${\cal S}$ as
\begin{equation}
\label{e:calSdef}
{\cal S}^{m\sigma_1\cdots\sigma_4}{}_n = \tfrac{1}{288}\left( \Gamma^{m\sigma_1\cdots\sigma_4}{}_n + 
24 \eta^{m[\sigma_1} \Gamma^{\sigma_2\sigma_3} \eta^{\sigma_4]}{}_n \right) \, .
\end{equation}
It also happens to be the symmetric part of the product of a single
$\Gamma$ with ${\cal T}$, but it is as of yet unclear how to find
${\cal S}$ in the above calculation in such an elegant way (it
probably requires rewriting of the transformation rule of the
three-form, as that is the only place where ${\cal T}$ is not yet
present manifestly).

Summarising, the action~\eqn{e:action_ansatz} (plus higher-order fermi
terms) is the first-order form of  eleven-dimensional supergravity
provided we add $S_T$ as given in \eqn{e:ST}. It is invariant under
the transformations~\eqn{e:trafo_ansatz} together with the
transformation rule of the spin connection
\begin{equation}
\label{e:deltaomega_full}
\delta \omega_\nu{}^{mn} = \bar\epsilon
\Gamma_\nu{}^{\phi\rho m n} \hat\psi_{\phi\rho}
+ 4\, \bar\epsilon \Gamma^{mnp}
\hat\psi_{\nu p} -  \tfrac{4}{9}\, \, \bar\epsilon \Gamma^{rs[m} 
\hat\psi_{rs} e_\nu{}^{n]}  
- 4\,\bar\epsilon{\cal S}^{mn}\cdot \hat H \psi_\nu
+ {\cal O}(\epsilon\psi^3) +{\cal O}(T) \, .
\end{equation}
(The additional torsion term can be determined by considering the
variation of $S_T$). The first two terms above can be rewritten in
such a way that the relation to the second-order formalism becomes
more clear. Using $\Gamma^{[5]}=\Gamma^{[2]}\Gamma^{[3]}-\Gamma^{[3]}\eta$ 
and the corresponding expansion of $\Gamma^{[3]}$, one finds
\begin{equation}
\label{e:deltaomega_full_new}
\begin{aligned}
\delta \omega_\nu{}^{mn} &=
-\tfrac{1}{2}\bar\epsilon\Gamma^{mn}\eom{\bar\psi}_\nu + 
(1+\tfrac{2}{9})\,\bar\epsilon\,{\eom{\bar\psi}}{}^{[m} e_\nu{}^{n]}
+2\,\bar\epsilon\Gamma_{\nu}{}^{[m} \widetilde{\eom{\bar\psi}}{}^{n]}
-2\,\bar\epsilon\Gamma^{mn}\widetilde{\eom{\bar\psi}}_\nu \\[.5ex]
&\quad
-2\,\bar\epsilon\Gamma_\nu\hat\psi^{mn}
+4\,\bar\epsilon\Gamma^{[m} \hat\psi^{n]}{}_{\nu}
- 4\,\bar\epsilon{\cal S}^{mn}\cdot \hat H \psi_\nu
+ {\cal O}(\epsilon\psi^3) + {\cal O}(T)\, .
\end{aligned}
\end{equation}
The capital $\eom{\bar\psi}$ symbols on the first line are proportional to the
gravitino equation of motion (see \eqn{e:grino_eom} and
\eqn{e:alternative_grino_eom} below). The gauge-field independent
terms of this transformation rule also apply to the ten-dimensional theory.
\end{sectionunit}

\begin{sectionunit}
\title{Equations of motion and other identities}
\maketitle

The equations of motion associated to the eleven-dimensional action
discussed in the previous sections are as follows ($n=4$ in the following):
\begin{subequations}
\label{e:sugraeoms}
\begin{align}
\label{e:graviton_eom}
\eom{e}_r{}^\mu := \frac{\kappa_{d}^2}{e} \frac{\delta S_{d}}{\delta e_\mu{}^r} &= \tfrac{1}{2}\Big(R_r{}^\mu-\tfrac{1}{2}e_r{}^\mu R\Big)
+ \tfrac{1}{2\,n!}\Big(2n(H^2)_r{}^\mu - e_r{}^\mu H^2\Big)\nonumber\\[1ex]
&\quad\quad + 3\Big( \bar\psi_{[r} \Gamma^{\mu\nu\rho} \psi_{\nu\rho]}
- \tfrac{1}{3} e_r{}^\mu\bar\psi_\lambda \Gamma^{\lambda\nu\rho}\psi_{\nu\rho}\Big)
+ {\cal O}(\psi^2 H) +  {\cal O}(\psi^4) \, ,\\[1ex]
\label{e:grino_eom}
\eom{\bar\psi}^\mu := \frac{\kappa_{d}^2}{e}
\frac{\delta S_{d}}{\delta \bar\psi_\mu} &= 2\,\Gamma^{\mu\nu\rho} \hat\psi_{\rho\nu}
\, ,\\[1ex]
\eom{A}^{\sigma_1\cdots\sigma_{n-1}} := \frac{\kappa_{d}^2}{e}\frac{\delta S_{d}}{\delta A_{\sigma_1\cdots\sigma_{n-1}}} 
&= \frac{1}{e}\partial_\kappa\left(e
H^{\kappa\sigma_1\cdots\sigma_{n-1}}\right) + 4\, D_\kappa \Big[
 \bar\psi_\mu \Gamma^{\mu\nu\lambda} {\cal T}_\nu{}^{\kappa\mu\nu\rho}
\psi_\lambda\Big] \nonumber\\[1ex]
&\quad\quad - \frac{1}{(n!)^2\,e}\,\varepsilon^{\sigma_1\cdots\sigma_{n-1} \lambda_1\cdots\lambda_8}
H_{\lambda_1\cdots\lambda_4}
H_{\lambda_5\cdots\lambda_8}\, ,\\[1ex]
\label{e:omega_eom}
\eom{\omega}_{mn}{}^\nu := \frac{\kappa_{d}^2}{e} \frac{\delta
S_{d}}{\delta \omega_{\nu}{}^{mn}} &= 
\tfrac{1}{2}\big( \tfrac{1}{2}T_{mn}{}^\nu - T_{m\lambda}{}^\lambda
e_{n}{}^\nu\big)
- \bar\psi_\lambda\Gamma^\lambda \psi_{[m} e_{n]}{}^\nu - \tfrac{1}{2}
\bar\psi_m \Gamma^\nu \psi_n\, .
\end{align}
\end{subequations}
Here, $n$ is the field strength form degree.  In the equation of
motion of the spin connection, the $\Gamma^{[5]}$ part cancels between
the variation of the gravitino kinetic term and the variation of the
extra part $S_T$ (see \eqn{e:ST}) that was added to the action.  

At this point one can make contact between the first-order results and
the well-known second order formulations. Using the equation of motion
for the spin connection one determines the on-shell value of the
torsion,
\begin{equation}
\label{e:EMspin}
T_{\mu\nu}{}^r = 2\bar\psi_{[\mu} \Gamma^r \psi_{\nu]} \, .
\end{equation}
The transformation rule of the spin connection in the second-order
formalism is obtained by using the equation of motion for the
gravitino and inserting that in the first-order transformation rule.

The Ricci tensor and scalar can now be expressed as
\begin{align}
\label{e:RfromEOM}
R &= \frac{4}{d-2} \Big[ -\eom{e}_\lambda{}^\lambda -
\frac{1}{2n!}(d-2n) H^2 - (d-3) \bar\psi_\mu \Gamma^{\mu\nu\rho}
\psi_{\nu\rho} \Big] + {\cal O}(\psi^2 H) + {\cal O}(\psi^4)\, ,\\
\label{e:RiccifromEOM}
R_r{}^\mu &= 2\,\eom{e}_r{}^\mu - \frac{2}{d-2}
\eom{e}_\lambda{}^\lambda e_r{}^\mu
+ \frac{2}{(n-1)!}(H^2)_r{}^\mu - \frac{n-1}{d-2}\frac{2}{n!} H^2 e_r{}^\mu\nonumber\\[1ex]
&\quad\quad\quad\quad\quad- \frac{d-1}{d-2} \bar\psi_\lambda
\Gamma^{\lambda\nu\rho}\psi_{\nu\rho} e_r{}^\mu
+ 6\bar\psi_{[r} \Gamma^{\mu\nu\rho} \psi_{\nu\rho]} + {\cal O}(\psi^2 H) + {\cal O}(\psi^4)\, .
\end{align}

The gravitino equation of motion will also often appear in a slightly
different form, obtained by multiplying the field equation with a
non-singular operator,
\begin{equation}
\label{e:alternative_grino_eom}
\widetilde{\eom{\bar\psi}}_\lambda \equiv \frac{1}{2(d-2)}\Big(\Gamma_{\lambda\mu} -
(d-3)\eta_{\lambda\mu}\Big) \eom{\bar\psi}^\mu =2\,
\Gamma^{\rho}\hat\psi_{\rho\lambda}\, ,
\end{equation}
where the operator can be shown to satisfy
$\Big(\Gamma_{\lambda\mu} - (d-3)\eta_{\lambda\mu}\Big)\Big(
\Gamma^{\mu\nu} - \eta^{\mu\nu}\Big) = 2(d-2)\,\eta_\lambda{}^\nu$. We will
encounter this singly contracted gravitino curvature several times in
the variation of the higher-order corrections to the action.

We also need a few more identities to get rid of covariant derivatives
on gravitino curvatures. In particular, using
\begin{equation}
\label{e:cyclicpsi2}
D_{[\mu} \psi_{\nu\lambda]} = \tfrac{1}{8} \Gamma^{mn}\psi_{[\mu} R_{\nu\lambda]mn}\, ,
\end{equation}
one derives that
\begin{equation}
\label{e:Dslashedpsi2}
\slashed{D} \psi_{\nu\lambda} =  D_{[\nu}\widetilde{\eom{\bar\psi}}_{\lambda]} 
+\tfrac{1}{2}\,\Gamma^s\psi_{[\nu} R_{\lambda]s} + \tfrac{1}{8}
\Gamma^m\Gamma^{rs}\psi_m R_{\nu\lambda rs} + {\cal O}(\psi^3) + {\cal
O}(T \psi)\, .
\end{equation}
Further multiplication with a single gamma matrix can be used to
derive
\begin{equation}
\label{e:Ddotpsi2}
D^\nu \psi_{\lambda\nu} = -\tfrac{1}{2}\slashed{D}\widetilde{\eom{\bar\psi}}_\lambda +
\tfrac{1}{4}D_\lambda(\Gamma\cdot\widetilde{\eom{\bar\psi}}) - \tfrac{1}{8}
R^\nu{}_{\lambda rs}\Gamma^{rs}\psi_\nu + \tfrac{1}{4} R_{\lambda r}
\big( \Gamma^{rs}\psi_s - \psi^r\big) + \tfrac{1}{8}R\psi_\lambda\, .
\end{equation}

The equations in the present section can be used in the
ten-dimensional theory as well, as long as one restricts to the sector
of the theory which is independent of the gauge field, dilaton and
dilatino.

\end{sectionunit}
\end{sectionunit}


\begin{sectionunit}
\title{Conventions, notation and some $\Gamma$ algebra}
\maketitle
\label{s:conventions}
\begin{sectionunit}
\title{Indices and signs}
\maketitle

We use a `mostly plus' metric $\eta=(-,+,+,\ldots,+)$. Our index conventions
are as exhibited in the following table:
\begin{center}
\begin{tabular}{lccc}
              & spinor                & vector            & super\\
curved        & $\alpha,\beta,\ldots$ & $\mu,\nu,\ldots$  & $M,N,\ldots$ \\
flat          & $a,b,\ldots$          & $r,s,\ldots$      & $A,B,\ldots$ 
\end{tabular}
\end{center}
We denote the charge conjugation matrix by ${\cal C}$ and always write it
explicitly wherever it appears. 

The sign conventions we use for the Riemann tensor and its contractions are
\begin{align}
R(\omega)_{\mu\nu mn} &= \partial_{\mu} \omega_{\nu mn}
 + \omega_{\mu mp}\,\omega_{\nu}{}^p{}_n - (\mu\leftrightarrow\nu) \,,\\[1ex]
R(\omega)_{\mu\nu} &= R_{\mu\lambda}{}^{\lambda}{}_{\nu}\, , \qquad
R(\omega) = R_\mu{}^\mu \,.
\end{align}
The covariant derivative associated to $\omega$ acts on Lorentz vector and 
spinor indices according to
\begin{align}
D(\omega)_\mu\, V^m &= \partial_\mu V^m
  + \omega_\mu{}^m{}_n\,V^n \,,\\[1ex]
D(\omega)_\mu \epsilon &= \partial_\mu\epsilon 
  + \frac{1}{4}\omega_\mu{}^{rs} \,\Gamma_{rs}\epsilon \,.
\end{align}
With these conventions one obtains
\begin{equation}
\label{e:DDcommutator}
[D_\mu,D_\nu] \epsilon = \tfrac{1}{4}\, R_{\mu\nu rs}\,\Gamma^{rs}\epsilon\, .
\end{equation}

The decomposition of the Riemann tensor in terms of the Weyl tensor,
the Ricci tensor (or the irreducible traceless Ricci tensor $S_m{}^n =
R_m{}^n -d^{-1} \delta_m{}^n R$) and the Ricci scalar is given by
\begin{equation}
\begin{aligned}
\label{e:Riemann_decompose}
R_{mn}{}^{pq} &= W_{mn}{}^{pq} - \frac{4}{d-2}\, \delta_{[m}{}^{[p}
R_{n]}{}^{q]} + \frac{2}{(d-1)(d-2)} R \, \delta_m{}^{[p}
\delta_n{}^{q]} \\[1ex]
&= W_{mn}{}^{pq} - \frac{4}{d-2}\, \delta_{[m}{}^{[p}
S_{n]}{}^{q]} - \frac{2}{d(d-1)} R \, \delta_m{}^{[p}
\delta_n{}^{q]}\,.
\end{aligned}
\end{equation}
From the former expansion the following operator that projects onto the 
Weyl part of the Riemann tensor is readily derived:
\begin{equation}
\begin{aligned}
\label{e:Weylprojop}
P_{m_1m_2m_3m_4}{}^{n_1n_2n_3n_4} =\,\,&
\delta_{m_1m_2}^{n_1\,n_2}\, \delta_{m_3m_4}^{n_3n_4}
+ \frac{4}{d-2} \delta_{m_1m_2}^{n_1[m_3}\, \delta_{n_3\;n_4}^{m_4]n_2}
+ \frac{2}{(d-1)(d-2)} \delta_{m_1m_2}^{m_3m_4}\,\delta_{n_1n_2}^{n_3n_4}\,. 
\end{aligned}
\end{equation}
Here we used the notation $\delta_{p_1\cdots p_k}^{q_1\cdots q_k}\defas 
\delta_{p_1}{}^{[q_1}\cdots \delta_{p_k}{}^{q_k]}$ with  
antisymmetrisation with unit weight ($\delta_p{}^q$ should be read as 
$\eta_{pq}$ where appropriate).
As appropriate, the operator~\eqn{e:Weylprojop} obeys the property $P^2=P$, 
is traceless on the $m$-indices and satisfies
\begin{equation}
P_{m_1m_2m_3m_4}{}^{n_1n_2n_3n_4}R_{n_1n_2n_3n_4} = W_{m_1m_2m_3m_4}\,.
\end{equation}

\end{sectionunit}

\begin{sectionunit}
\title{Riemann tensor polynomials and the $t_8$ tensor}
\maketitle
\label{s:fullinginvariants}

For reference, we present in this section the expressions for the
various higher-derivative invariants in terms of the tensor polynomial
normal forms as analysed by \dcite{Fulling:1992vm}. Our choice of
basis has a maximal number of invariants with no mixed-type index
contractions (double indices are summed over using a metric and traces
are on Lorentz indices):
\begin{equation}
\label{e:Fulling_invariants}
\begin{aligned}
R_{41} &= {\rm tr}\big( R_{\mu\nu} R_{\nu\rho} R_{\rho\sigma}
R_{\sigma\mu} \big) && = A_6\,,\\
R_{42} &= {\rm tr}\big( R_{\mu\nu} R_{\nu\rho} R_{\mu\sigma}
R_{\sigma\rho}\big)\,, && \\[1ex]
R_{43} &= {\rm tr}\big( R_{\mu\nu} R_{\rho\sigma} \big) {\rm tr}\big(
R_{\mu\nu} R_{\rho\sigma} \big) && = A_3\,, \\
R_{44} &= {\rm tr}\big( R_{\mu\nu} R_{\mu\nu} \big) {\rm tr}\big(
R_{\rho\sigma} R_{\rho\sigma} \big) &&= A_1\,,\\
R_{45} &= {\rm tr}\big( R_{\mu\nu} R_{\nu\rho} \big) {\rm tr}\big(
R_{\rho\sigma} R_{\sigma\mu} \big) &&= A_2\,, \\
R_{46} &= {\rm tr}\big( R_{\mu\nu} R_{\rho\sigma} \big) {\rm tr}\big(
R_{\mu\rho} R_{\nu\sigma} \big) &&= A_4\,, \\[1ex]
R_{47} &= R_{mn}{}^{[mn} R_{pq}{}^{pq} R_{rs}{}^{rs} R_{tu}{}^{tu]} &&=
\tfrac{1}{2\cdot 8!}Z\, .
\end{aligned}
\end{equation}
The `$A_i$' and `$Z$' symbols are (up to an overall normalisation
$2\cdot 8!$ of the latter) the notation of \dcite{dero3} and
\dcite{Fulling:1992vm}, who do not employ $R_{42}$ but instead use
\begin{equation}
R^{pqrs} R_{pq}{}^{tu} R_r{}^v{}_t{}^w R_{svuw}=A_5 ,\quad   R^{pqrs} R_p{}^
t{}_r{}^uR_t{}^v{}_q{}^w R_{uvsw}=A_7\, .
\end{equation}
The relation between their choice and ours is given by 
\begin{equation}
\label{e:Fulling2us}
A_7 - A_5 = R_{42} - \tfrac{1}{4} R_{46}\, ,
\end{equation}
which can be verified by repeated use of the cyclic Ricci identity.

In string theory it is more useful to employ the specific tensorial
structures that appear in string-amplitude calculations. The $t_8$
tensor has four index pairs and is defined as
\begin{equation}
\label{e:t8_definition}
\begin{aligned}
t_8^{m_1m_2 n_1n_2 p_1p_2 q_1q_2} =\,\,& -2 \Big(
 \delta^{m_1n_2}\delta^{m_2n_1}\delta^{p_1q_2}\delta^{p_2q_1}
+\delta^{n_1p_2}\delta^{n_2p_1}\delta^{m_1q_2}\delta^{m_2q_1}
+\delta^{m_1p_2}\delta^{m_2p_1}\delta^{n_1q_2}\delta^{n_2q_1}\Big)\\[1.5ex]
&+8\,\Big( 
 \delta^{m_1q_2}\delta^{m_2n_1}\delta^{n_2p_1}\delta^{p_2q_1}
+\delta^{m_1q_2}\delta^{m_2p_1}\delta^{p_2n_1}\delta^{n_2q_1}
+\delta^{m_1n_2}\delta^{m_2p_1}\delta^{p_2q_1}\delta^{q_2n_1}
\Big)\\[1.5ex]
&+\text{anti-symmetrisation of each index pair, with total weight one}\, .
\end{aligned}
\end{equation}
The following special case (for an anti-symmetric tensor $N$) is useful:
\begin{equation}
t_8^{r_1\cdots r_8} M_{r_1r_2} N_{r_3r_4}\cdots N_{r_7r_8}
= -6\, M_{t_1t_2} N_{t_2t_1} N_{mn}N_{nm} + 24\,
M_{t_1t_2} N_{t_2m} N_{mn} N_{nt_1}\, .
\end{equation}
The invariants formed from the $t_8$ tensor and the usual
$\varepsilon_{10}$ can be reduced to the seven fundamental invariants
of~\eqn{e:Fulling_invariants} as follows:
\begin{multline}
\begin{aligned}
X\defas && t_8 t_8 R^4 &=&\, 192\, R_{41} &&\, + 384\, R_{42} &&\, + 24\, R_{43} &&\, +
12\, R_{44} &&\, - 192\, R_{45} &\, - 96 \, R_{46} \, , \\[.5ex]
\tfrac{1}{8} Z\defas && -\tfrac{1}{8}\varepsilon_{10}\varepsilon_{10} R^4 &=&
\,  192\, R_{41} &&\, + 384\, R_{42} &&\, + 24\,R_{43} &&\, +
12\, R_{44} &&\, -192\, R_{45}  &\, + 96\, R_{46} 
\end{aligned} \\[1ex]
 - 768\, A_7  + \text{Ricci terms}\, .
\end{multline}
This leads to 
\begin{equation}
\label{e:XminusZ}
X-\tfrac{1}{8} Z = -192 R_{46} + 768 A_7 + \text{Ricci terms}\, .
\end{equation}
In addition, the two Yang-Mills-like invariants are
\begin{equation}
\begin{aligned}
t_8 Y_2 \defas && t_8 {\rm tr} R^4       &= &\, 8\,R_{41} &&\,+16\,R_{42} &&\, &&\, &&\, - 4\, R_{45}&\, - 2\,
R_{46} + \cdots\, , \\[.5ex]
t_8 Y_1 \defas &&\, t_8 ( {\rm tr} R^2 )^2 &= &\,  &&\,  &&\, -4\,R_{43} &&\, - 2\, R_{44} &&\,
+16\, R_{45} &\, + 8\, R_{46} + \cdots\, .
\end{aligned}
\end{equation}
Moreover, there is the relation
\begin{equation}
t_8 X_8 = t_8 t_8 R^4 = 24\, t_8 {\rm tr} R^4 - 6\, t_8 ({\rm tr}
R^2)^2 \, .
\end{equation}

We should stress that the $t_8t_8$ tensor does \emph{not} automatically
project on the Weyl part of the Riemann tensor, as can easily be verified 
by e.g.\ computing $t_8t_8\,R^4$ taking only the Ricci parts in the
expansion~\eqn{e:Riemann_decompose} to be non-zero. In a similar way,
trace terms appear in the $t_{16}$ tensor of~\dcite{Green:1999by}.  So
far, the presence of these trace terms has not played any role in the
literature because they are proportional to the equations of motion
when the gauge fields are ignored. For clarity we will always write
Weyl tensors explicitly.

In this context, let us also mention that for the $N=1$ case in ten
dimensions a supersymmetric extension of $t_8t_8R^4$ written as an
integral over the sixteen spinorial superspace coordinates has been 
given by~\dcite{nils2} and \dcite{Kallosh:1987mb}. 
Indeed, we have verified by an explicit covariant calculation that 
the trace-free version of~\eqn{e:XminusZ} can be obtained as
\begin{equation}
X-\tfrac{1}{8}Z\big|_{R\rightarrow W} = \tfrac{3}{2}\,
\int\!{\rm d}^{16}\, \theta \, \left(\theta {\cal C}\Gamma^{mnr}\theta
\,  \theta {\cal C}\Gamma_r{}^{pq}\theta\, R_{mnpq}\right)^4 \ ,
\end{equation}
The $\theta$-contracted Riemann tensor above---appearing at fourth
order in the $\theta$ expansion of the scalar superfield of $N=1$
supergravity in ten dimensions, as formulated by~\dcite{nils1}---is
manifestly free from trace terms.  This is an immediate consequence of
the group-theoretical fact that the fully anti-symmetrised product of
four SO(1,9) spinor representations~$\mathbf{16}$ contains the
irrep~$\mathbf{770}$ ($\tilde{\tableau{2 2}}$), but neither~$\bf54$
($\tilde{\tableau{2}}$) nor~$\mathbf{1}$ \cite{nils2}.  In principle
it is possible to construct the linearised superinvariant along these
lines by considering also the other terms in the expansion of the
scalar superfield, and obtain the $t_8$ and $\epsilon_{10}$ structures
by going to the light-cone gauge before doing the fermionic integral.
However, apart from the fact that this is technically probably still
rather complicated (the integrals are not as simple as the highly
symmetric one above), it also does not go beyond the linearised level,
and therefore does not provide an alternative to the methods used in
the main text.
\end{sectionunit}

\begin{sectionunit}
\title{Some useful $\Gamma$-matrix identities}
\maketitle
\label{s:gammastuff}

The matrices $\Gamma_\mu\, (\mu=0\,\cdots,d-1)$ are taken to satisfy
the Clifford algebra
\begin{equation}
\label{e:clifford}
\{\Gamma_r,\Gamma_s\} = 2 \eta_{rs} \, .
\end{equation}
In $d=10$, the matrix $\Gamma^\#:=\Gamma^0\Gamma^1\cdots\Gamma^9$
squares to the identity and anti-commutes with $\Gamma^\mu$, and can
therefore be used to define the Weyl projectors
$\mathcal{P}_\pm=\tfrac{1}{2}(\mathbf{1}\pm \Gamma^\#)$.

From \eqn{e:clifford} we can derive the commutators
\begin{equation}
\begin{aligned}
{}[ \Gamma^{rs}, \Gamma^t ] &= 4\, \Gamma^{[r}\delta^{s]}_t\, ,\\[.5ex]
\label{e:g2g2}
{}[ \Gamma^{rs}, \Gamma_{tu} ] &= 8\, \Gamma^{[r}{}_{[u}
\delta^{s]}{}_{t]}\, ,\\[.5ex]
{}[ \Gamma^{rs}, \Gamma_{tuv} ] &= 12\, \Gamma^{[r}{}_{[tu}
\delta^{s]}{}_{v]}\, .
\end{aligned}
\end{equation}
These, in turn, immediately lead to the following commutators of covariant 
derivatives and gamma matrices with curved indices
:\begin{equation}
\begin{aligned}
{}[ D_\mu , \Gamma^{\nu_1\cdots\nu_n} ] &= 
 n\, \partial_\mu e_r{}^{[\nu_1} \Gamma^{|r| \nu_2\cdots\nu_n]} 
 + \tfrac{1}{4}\omega_{\mu}{}^{rs} [ \Gamma_{rs}, \Gamma^{\nu_1\cdots\nu_n} ] \, \\[.5ex]
& = n\, D_\mu e_r{}^{[\nu_1} \Gamma^{|r| \nu_2\cdots\nu_n]} \, .
\end{aligned}
\end{equation}
The following commutators and anti-commutators of higher products of gamma
matrices are also useful:
\begin{equation}
\begin{aligned}
{}[ \Gamma^{r_1\cdots r_5}, \Gamma_{s_1 s_2 s_3} ] &= 
 2\, \Gamma^{r_1\cdots r_5}{}_{s_1 s_2 s_3} -
 240\, \Gamma^{[r_1r_2r_3}{}_{[s_3} \delta^{r_4}{}_{s_1}
 \delta^{r_5]}{}_{s_2]} \, ,\\[.5ex]
{}\{ \Gamma^{r_1\cdots r_5}, \Gamma_{s_1 s_2 s_3} \} &=
 30\, \Gamma^{[r_1\cdots r_4}{}_{[s_2s_3} \delta^{r_5]}{}_{s_1]} -
 120\, \Gamma^{[r_1r_2}
 \delta^{r_3}{}_{s_1}\delta^{r_4}{}_{s_2}\delta^{r_5]}{}_{s_3}\, .
\end{aligned}
\end{equation}
Furthermore, we have the contraction identity
\begin{multline}
\label{e:contractedgammagamma}
\delta^{p}{}_q \Gamma^{r_1\cdots r_n q} \Gamma_{p s_m\cdots s_1}
 = \\[1ex]\sum_{k=0}^{\text{min}(m,n)}
(d-m-n+k) \frac{m!\,n!}{k!\,(m-k)!\,(n-k)!}\Gamma^{[r_1\cdots r_{n-k}}{}_{s_{m-k}\cdots s_1}
 \delta^{r_{n-k+1}\cdots r_n]}_{s_{m-k+1}\cdots s_m]} \, .
\end{multline}
Finally, we have made frequent use of the famous fermi flip property
\begin{equation}
\label{e:fermiflip}
\left(\psi_1 {\cal C} \Gamma^{\mu _1\cdots\mu _n} \psi_2\right) =
(-1)^{n(n+1)/2}\left(\psi_2 \,{\cal C} \Gamma^{\mu _1\cdots\mu _n} \,
\psi_1\right)
\,, 
\end{equation}
in this form valid for arbitrary spinors; for Majorana spinors we can of 
course write $\psi^T {\cal C}= \bar\psi$.

\end{sectionunit}
\end{sectionunit}

\bibliography{componentpaper}
\end{document}